\documentclass[11pt,twoside,notitlepage]{article}
\usepackage{one-column}

\begin{document}

	%
	%

	\title{%
		Arbitrary Lagrangian--Eulerian finite\\[4pt]%
		element method for lipid membranes%
	}

	\shorttitle{ALE finite element method for lipid membranes}

	\author{Amaresh Sahu\thanks{\href{mailto:asahu@che.utexas.edu}{\texttt{asahu@che.utexas.edu}}}}
	\affil{%
		McKetta Department of Chemical Engineering, University of Texas,\\%
		Austin, Texas 78712, USA%
	}

	\shortauthor{A.\ Sahu}

	\date{\today}

	%
	%

	\maketitle

	%
	%

	\normalsize

%
%

\section{Introduction} \label{sec_intro}

In this paper, we present an arbitrary Lagrangian--Eulerian (ALE) finite element
method and open-source \julia\ code \cite{mem-ale-fem} to simulate the dynamics
of curved and deforming lipid membranes.
Our developments build on ALE theories of biological membranes where the
surface parametrization is independent of the in-plane flow of lipids
\cite{hu-pre-2007, torres-jfm-2019, sahu-jcp-2020}.
In our simulations, we arbitrarily specify the dynamics of the discretized
surface and avoid highly distorted mesh elements---all while not altering
membrane dynamics.
The utility of our implementation is demonstrated by analyzing a biologically-%
motivated scenario that is difficult to simulate with existing numerical
methods: tether formation, elongation, and subsequent translation.%
\setcounter{footnote}{1}\footnote{%
	This manuscript bears some resemblance to the proposed second part of a prior
	publication \cite{sahu-jcp-2020}.
	However, the present ALE analysis, numerical implementation, code, and
	applications are not those planned in the aforementioned effort.
	We accordingly release this work as a stand-alone publication and source code.
}

Biological membranes are two-dimensional (2D) materials, comprised of lipids and
proteins, which make up the boundary of the cell and many of its internal
organelles.
The lipids and proteins flow in-plane as a 2D fluid, while the membrane bends
out-of-plane as an elastic shell.
Such membranes play a fundamental role in many cellular processes, including
endocytosis \cite{mcmahon-cell-2002, mcmahon-nrmcb-2011}, cell migration
\cite{lauffenburger-c-1996}, and tether network dynamics within the cell
\cite{terasaki-jcb-1986, sciaky-jcb-1997}.
Lipid membranes often undergo dramatic shape changes in which their in-plane and
out-of-plane dynamics are coupled.
Consequently, comprehensive models and advanced numerical techniques are needed
to describe membrane behavior.
In the early 1970's, the seminal contributions of \textsc{P.\ Canham}
\cite{canham-jtb-1970}, \textsc{W.\ Helfrich} \cite{helfrich-znc-1973}, and
\textsc{E.A.\ Evans} \cite{evans-bpj-1974}---all of which can be viewed as
extensions of \textsc{P.M.\ Naghdi}'s fundamental contributions to shell theory
\cite{naghdi-1973}---laid the foundation for theoretical developments
\cite{evans-1978, waxman-sam-1984, zhong-can-pra-1989, steigmann-mms-1998,
steigmann-arma-1999, capovilla-jpa-2002, guven-jpa-2004, pollard-arxiv-2024}
and analysis
\cite{seifert-pra-1991, seifert-epl-1993, safran-prl-1995, fournier-prl-1996,
goldstein-jp-1996, seifert-long, powers-pre-2002, du-jcp-2004,
vlahovska-pre-2007, agrawal-bmmb-2008, stone-jfm-2010, agrawal-zamp-2011,
rahimi-sm-2013, ramaswamy-prl-2014, narsimhan-jfm-2015, stone-jfm-2015,
vlahovska-2016, sabass-prl-2016, al-izzi-prl-2020, fonda-prl-2020, lin-jfm-2021,
faizi-pnas-2024, yu-arxiv-2023, reboucas-sm-2024, venkatesh-jfm-2025,
al-izzi-arxiv-2024, venkatesh-arxiv-2024, dharmavaram-arxiv-2024}.
However, the general, coupled nonlinear equations governing the in-plane and
out-of-plane dynamics of a single-component membrane were not obtained until
2007 \cite{hu-pre-2007}.
These governing equations were subsequently obtained via other techniques
\cite{arroyo-pre-2009, mandadapu-bmmb-2012, sahu-pre-2017}, and extended to
describe the dynamics of multi-component phase separation and chemical reactions
with proteins in the surrounding fluid \cite{sahu-pre-2017}.
For our detailed perspective on the development of lipid membrane theories, see
Chap.\ IV of Ref.\ \cite{sahu-thesis}.

The equations governing membrane dynamics are highly nonlinear partial
differential equations written on a surface which is itself arbitrarily curved
and deforming over time.
One cannot in general analytically solve for the time evolution of lipid flows
and membrane shape changes, which are intricately coupled.
However, it is also difficult to solve the full membrane equations numerically,
as standard solution methods from fluid and solid mechanics struggle to capture
the membrane's in-plane fluidity and out-of-plane elasticity.
Many numerical studies accordingly investigated specific aspects of membrane
behavior.
For example, several works captured the hydrodynamics of lipid flows on vesicles
with a prescribed geometry, including the effects of the surrounding fluid as
well as embedded proteins and other inclusions \cite{oppenheimer-pre-2010,
oppenheimer-prl-2011, sigurdsson-sm-2016, gross-jcp-2018, samanta-pf-2021}.
These developments were recently extended with fluctuating hydrodynamics to
incorporate phase separation and the discrete motion of proteins
\cite{rower-jcp-2022, tran-pre-2022}.
Orthogonal efforts described elastic membrane deformations either ($i$) in the
limit of no in-plane viscosity \cite{feng-jcp-2006, barrett-jcp-2008,
barrett-siam-2008, ma-jcp-2008, dziuk-nm-2008, elliott-jcp-2010,
bonito-jcp-2010, mercker-siam-2013}, possibly with the dynamics of the
surrounding fluid \cite{narsimhan-jfm-2015}, or ($i\mkn i$) with the in-plane
fluidity replaced by viscoelasticity \cite{rangarajan-jcp-2015, zhu-br-2022}.
In both cases, the dynamic coupling between in-plane viscous flows and shape
changes is not reflected \cite{sahu-pre-2020}.
Still other works incorporated all of the aforementioned membrane complexities,
but restricted their investigations to specific geometries
\cite{mandadapu-bmmb-2012, mandadapu-bpj-2014, walani-pnas-2015,
hassinger-pnas-2017}---for which the membrane equations are simplified.

%
%
%

We reiterate that many challenges arise when developing a general numerical
method that truly captures in-plane viscous lipid flows, out-of-plane membrane
bending, and their coupling---all on an arbitrarily curved and deforming
surface.
Several studies \cite{rodrigues-jcp-2015, barrett-pre-2015, mandadapu-jcp-2017,
omar-bpj-2020} took a Lagrangian approach, where the surface is discretized and
the resulting mesh is convected with the physical, material velocity.
Lagrangian implementations successfully capture membrane dynamics, but struggle
to resolve in-plane flows as they lead to highly distorted elements.
A remeshing procedure was used to maintain element aspect ratios, though it led
to unphysical oscillations in the membrane curvature \cite{rodrigues-jcp-2015}.
An alternative approach is for the mesh to only move normal to the surface,
so that it is unaffected by in-plane flows.
The mesh motion is then out-of-plane Lagrangian, as it tracks the material
surface, and in-plane Eulerian.
Such an approach, which we refer to as Eulerian, was implemented in Ref.\ 
\cite{reuther-jfm-2020}---though the membrane geometry was updated explicitly
in a piecemeal manner.
Given the importance of geometry to membrane dynamics \cite{sahu-pre-2020},
numerical methods relying on explicit mesh updates could suffer from issues
known to affect explicit algorithms in the study of fluids
\cite{zienkiewicz-ijnmf-1995, zienkiewicz-taylor-fem}.
Moreover, even if a fully implicit Eulerian mesh motion was implemented as in
our prior work \cite{sahu-jcp-2020}, scenarios arise where the method will fail
\cite{torres-jfm-2019}.
We show one such example in \ref{sec_sim_pull}.

Since neither Lagrangian nor Eulerian approaches can capture commonly-observed
membrane behaviors, a more general mesh motion is required.
The ALE theory underlying a general mesh motion was independently derived by
both others \cite{torres-jfm-2019} and ourselves \cite{sahu-jcp-2020}.
A new mesh motion was implemented in Ref.\ \cite{torres-jfm-2019}, in which the
mesh moved only in the direction normal to a known prior configuration of the
membrane.
In practice, such a choice was again not sufficiently general and required
remeshing steps, which introduced errors in the numerical solution.
Since the aforementioned approach resembles an Eulerian mesh motion, we
hypothesize that it may not work in certain situations---such as the tether
pulling scenario discussed in \ref{sec_sim_pull}.
There is thus still a need for numerical implementations of more general mesh
motions, which can be specified by the user when solving for membrane dynamics
in a particular scenario.

The aforementioned limitations of numerical techniques motivate our development
and implementation of a fully implicit ALE finite element method for lipid
membranes.
Rather than prescribing the mesh motion directly, we choose for the mesh
velocity to satisfy a set of partial differential equations as if the mesh were
itself another material.
We then supply appropriate boundary conditions to the mesh velocity.
In addition, the mesh and membrane are constrained to coincide via a Lagrange
multiplier---which is understood as the mesh analog of a normal force per unit
area.
The presence of a scalar Lagrange multiplier coupled to vector velocities leads
to a numerical instability reminiscent of that identified by
\textsc{O.A.\ Ladyzhenskaya} \cite{ladyzhenskaya-1969}, \textsc{I.\ Babu\v{s}ka}
\cite{babuska-nm-1973}, and \textsc{F.\ Brezzi} \cite{brezzi-rfai-1974},
hereafter referred to as the LBB condition.
We suppress the instability with the method of \textsc{C.R.\ Dohrmann} and
\textsc{P.B.\ Bochev} \cite{dohrmann-ijnmf-2004}, where the Lagrange
multiplier is projected onto a space of discontinuous, piecewise linear
functions.

The remainder of the manuscript is organized as follows.
In \ref{sec_gov}, we present the strong and weak formulations of all equations
governing the membrane and mesh.
A summary of the corresponding finite element formulation is provided in
\ref{sec_fem}; additional details can be found in Appendix \ref{sec_a_fem}.
All numerical results from Lagrangian, Eulerian, and ALE simulations are
presented in \ref{sec_sim}.
We close with conclusions and pathways for future work in \ref{sec_concl}.
Our source code is provided in the \julia\ package \memalefem\
\cite{mem-ale-fem}.

%
%

\section{The governing equations: Strong and weak formulations}
\label{sec_gov}

The membrane is treated as a single 2D differentiable manifold embedded in the
Euclidean space $ \mathbb{R}^3 $, for which we implicitly assume no slip between
the two bilayer leaflets.
In our ALE formulation, the membrane surface is parametrized by coordinates that
need not follow material flows, as we detailed in Ref.\ \cite{sahu-jcp-2020}.
Accordingly, the so-called mesh velocity $ \bmvm $ resulting from the
parametrization will in general not equal the membrane velocity $ \bmv $ of
material points.
The dynamics of both the membrane and mesh are detailed in what follows; many of
the results were derived in our prior work \cite{sahu-jcp-2020, sahu-thesis}.
The phospholipid bilayer is treated as an area-incompressible material---for
which the areal density is constant and the surface tension, $ \lambda $, is a
Lagrange multiplier field to be solved for.
Our goal is to determine the fundamental unknowns $ \bmv $, $ \bmvm $, and
$ \lambda $, as well as the membrane position $ \bmx $, over time.

%
%

\subsection{Surface parametrization, geometry, and kinematics}
\label{sec_gov_geodyn}

We begin by reviewing the framework with which lipid membranes are described.
Only the most relevant features are presented here, as Ref.\ \cite{sahu-thesis}
details our understanding of the membrane geometry and kinematics, while Ref.\
\cite{sahu-jcp-2020} provides the ALE theory.

\begin{figure}[!b]
	\centering
	\includegraphics[width=0.75\textwidth]{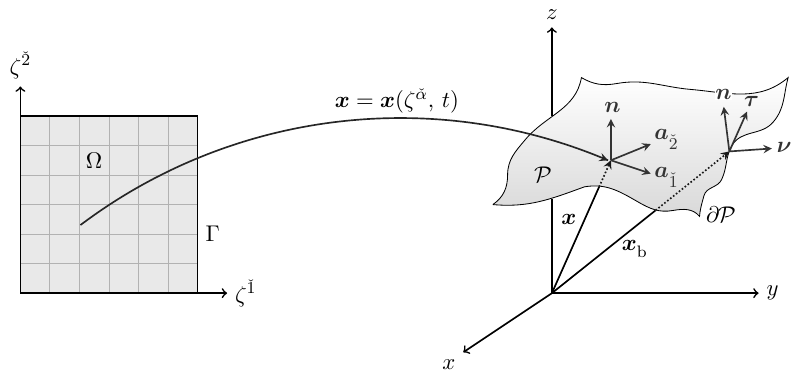}
	\caption{%
		Surface geometry.
		A schematic of the mapping
		$ \bmx = \bmx (\zeta^\calpha, t) $,
		at a single instant in time, between the parametric domain $ \Omega $ and
		the membrane patch $ \mcp $.
		The in-plane basis vectors $ \bma_\calpha $ and unit normal vector $ \bmn $
		are shown at a point $ \bmx $ on the patch, as are the in-plane unit normal
		$ \bmnu $ and unit tangent $ \bmtau $ at a point $ \bmx^{}_{\tb} $ at the
		patch boundary $ \partial \mkn \mcp $.%
	}
	\label{fig_gov_geo}
\end{figure}

Consider an arbitrarily curved and deforming membrane surface $ \mcs $, of which
we examine a patch
$ \mcp \subset \mcs $.
At any time $ t $, the surface position $ \bmx $ is parametrized by two
coordinates: $ \zetaone $ and $ \zetatwo $.
Here, as in Ref.\ \cite{sahu-jcp-2020}, the `check' accent
$ ( \, \check{\bmcdot} \, ) $ indicates the parametrization can be arbitrarily
specified.
From now on, Greek indices span the set $ \{ 1, 2 \} $, such that
$ \bmx (\zeta^\calpha, t) \in \mathbb{R}^3 $
denotes the position of a point on the membrane surface.
As shown in Fig.\ \ref{fig_gov_geo}, there is a mapping from the parametric
domain $ \Omega $ in the $ \zetaone $--$ \zetatwo $ plane to the membrane
patch $ \mcp $.
Area integrals over the patch are thus evaluated as
\begin{equation} \label{eq_gov_map_area}
	\int_\mcp \big( \ldots \big) ~\td a
	\ = \, \int_\Omega \big( \ldots \big) \, \JO ~\td \Omega
	~,
\end{equation}
where $ \JO $ is the Jacobian of the mapping.
Here $ \JO $ has dimensions of area and its
functional form is provided below.
At any point $ \bmx (\zeta^\calpha, t) $, the vectors
$ \bmaa := \partial \bmx / \partial \zeta^\calpha $
form a basis of the tangent plane to the surface, which has unit normal
$
	\bmn
	= (\bma_\cone \times \bma_\ctwo) / \lvert \bma_\cone \times \bma_\ctwo \rvert
$.
Covariant components of the metric and curvature tensors are respectively
$ a_{\calpha \cbeta} := \bma_\calpha \bmcdot \bma_\cbeta $
and
$
	b_{\calpha \cbeta}
	:= \bmn \bmcdot \bmx_{, \calpha \cbeta}
	= \bmn \bmcdot \bmx_{; \calpha \cbeta}
$.
Here
$ \pdp_{, \calpha} := \partial \pdp / \partial \zeta^\calpha $
and
$ \pdp_{; \calpha} $ are respectively the partial and covariant derivatives with
respect to $ \zeta^\calpha $.
The Jacobian $ \JO $ appearing in Eq.\ \eqref{eq_gov_map_area} is expressed as
$
	\JO
	= \sqrt{ \rule{0mm}{1.50ex} } \overline{
		\det \, a_{\calpha \cbeta}
	\,} \hspace{-0.5pt} \raisebox{1.60ex}{\vrule width 0.5pt height 0.35ex}
$.
The mean curvature $ H $ is calculated as
$
	H
	= \tfrac{1}{2} \mk \auabl b_{\calpha \cbeta}
$,
where $ \auabl $ is the contravariant metric and is calculated as
the matrix inverse of $ a_{\calpha \cbeta} $.
In the present work, the Einstein summation convention is employed, in which
repeated raised and lowered indices are summed over.
The Gaussian curvature $ K $ is computed as
$ K = \det (b_{\calpha \cbeta}) / \det (a_{\calpha \cbeta}) $.
Finally, at points $ \bmx^{}_{\tb} $ on the patch boundary
$ \partial \mkn \mcp $, additional in-plane basis vectors are defined:
$ \bmnu = \nu^\calpha \mk \bmaa = \nu_\calpha \, \bma^\calpha $
is the in-plane unit normal and
$ \bmtau = \tau^\calpha \mk \bmaa = \tau_\calpha \, \bma^\calpha $
is the in-plane unit tangent (see Fig.\ \ref{fig_gov_geo}).
Line integrals over the patch boundary are calculated as
\begin{equation} \label{eq_gov_map_line}
	\int_{\partial \mkn \mcp} \! \big( \ldots \big) ~\td s
	\ = \, \int_\Gamma \big( \ldots \big) \, \JG ~\td \Gamma
	~,
\end{equation}
where $ s $ is the arclength parametrization of $ \partial \mkn \mcp $ and
$ \Gamma := \partial \Omega $
is the boundary of the parametric domain.
In Eq.\ \eqref{eq_gov_map_line},
$ \JG = [( \tau^\cone )^2 + ( \tau^\ctwo )^2]^{-1/2} $
is the Jacobian of the mapping from $ \Gamma $ to $ \partial \mkn \mcp $.

When solving for the membrane state at a particular time, we seek the surface
tension, material velocity, and mesh velocity fields on the surface.
To this end, it is useful to define relevant function spaces over the parametric
domain $ \Omega $.
The space of square-integrable functions, $ L^2 (\Omega) $, is expressed as
\begin{equation} \label{eq_gov_geodyn_L2}
	L^2 ( \Omega )
	\, := \, \Big\{\,
		u(\zetaalpha) \, : ~ \Omega \rightarrow \mathbb{R}
		\quad
		\text{such that}
		\quad
		\Big(\textstyle
			\int_\Omega u^2 \, \td \Omega
		\Big)^{\! 1/2}
		< \, \infty
	\,\Big\}
	~.
\end{equation}
The Sobolev space of order $ k $, $ H^k (\Omega) $, consists of functions that
are square-integrable and also have up to $ k $ partial derivatives that are
square-integrable:
\begin{equation} \label{eq_gov_geodyn_Hk}
	H^k ( \Omega )
	\, := \, \Big\{\,
		u(\zetaalpha) \, : ~ \Omega \rightarrow \mathbb{R}
		\quad
		\text{such that}
		\,
		\hspace{5pt}\underbrace{%
			\hspace{-5pt}u_{, \calpha \cbeta \ldots \cgamma}%
		}_{m \text{ derivatives}}
		\hspace{-6pt} \in \, L^2 (\Omega)
		\quad
		\text{for } ~
		0 \le m \le k
	\,\Big\}
	~.
\end{equation}
From the definitions in Eqs.\ \eqref{eq_gov_geodyn_L2} and
\eqref{eq_gov_geodyn_Hk}, it is evident that
$ H^0 (\Omega) = L^2 (\Omega) $.
Moreover, since the material and mesh velocities are elements of
$ \mathbb{R}^3 $, we also define
\begin{equation} \label{eq_gov_geodyn_bmHk}
	\bmH^k ( \Omega )
	\, := \, \Big\{\,
		\bmu(\zetaalpha) \, : ~ \Omega \rightarrow \mathbb{R}^3
		\quad
		\text{such that}
		\quad
		u^{}_j
		\mk \in H^k (\Omega)
		\quad
		\text{for }
		j = 1, 2, 3
	\,\Big\}
\end{equation}
as the space of functions in which each Cartesian component is an element of
$ H^k (\Omega) $.
In Eq.\ \eqref{eq_gov_geodyn_bmHk}, we denote
$ u^{}_j := \bmu \bmcdot \bmej $,
where
$ \{ \bmej \}_{j = 1, 2, 3} $ is the canonical Cartesian basis in
$ \mathbb{R}^3 $.

We close with a discussion of membrane kinematics.
The membrane velocity
$ \bmv = \td \bmx / \td t = \bm{\dot{x}} $
is the material time derivative of the position, and is expanded in the
$ \{ \bma_\calpha, \bmn \}$ basis as
$ \bmv = v^\calpha \bma_\calpha + v \bmn $.
The mesh velocity, which is treated as a fundamental unknown independent of the
material velocity, is defined to be the rate of change of position when
$\zeta^\calpha$ is held constant---expressed as
$ \bmvm = ( \partial \check{\bmx} / \partial t ) \rvert_{\zeta^\calpha} $.
In order for the material and mesh velocities to correspond to the same surface,
kinematics require
\begin{equation} \label{eq_gov_geodyn_kinematic_constraint}
	\bmvm \bmcdot \bmn
	\, = \, \bmv \bmcdot \bmn
\end{equation}
such that the mesh motion is always out-of-plane Lagrangian
\cite{sahu-jcp-2020}.
Importantly, the in-plane material velocity components $v^\calpha$ and mesh
velocity components
$ \vam = \bmvm \bmcdot \bma^\calpha $
need not coincide, such that within our ALE framework one can dictate how the
mesh evolves within the membrane surface.
In this work, Eq.\ \eqref{eq_gov_geodyn_kinematic_constraint} is referred to as
the ALE kinematic constraint.

%
%

\subsection{The balance of mass: Material incompressibility} \label{sec_gov_inc}

As lipid membranes only stretch 2--3\% before tearing
\cite{evans-skalak, nichol-jp-1996}, we treat them as area-incompressible 2D
materials.
The local form of the balance of mass, also referred to as the continuity
equation and incompressibility constraint, is given by
\begin{equation} \label{eq_gov_inc_strong}
	\bma^\calpha (\zetagamma, t) \, \bmcdot \, \bmv_{, \calpha} (\zetagamma, t)
	\, = \, 0
	\qquad
	\forall
	\quad
	\zetagamma \mkn \in \Omega
	~,
	~ ~
	t \mk \in \, [\mk 0, \ttf \mk]
	~,
\end{equation}
where $ \ttf $ is the end of the time interval under consideration.
For notational simplicity, the functional dependence of all quantities will be
suppressed, as will the domains of $ \zetagamma $ and $ t $.
The incompressibility constraint
$ \bma^\calpha \bmcdot \bmv_{, \calpha} = 0 $
is equivalently expressed as
$ v^\calpha_{; \calpha} - 2 \mk v H = 0 $,
and is enforced by the Lagrange multiplier field
$ \lambda = \lambda (\zetaalpha, t) $---%
which has dimension of energy$\mk / \mk$area and acts as the membrane surface tension
\cite[Ch.V\,\S6(a)--(c)]{sahu-thesis}.

The weak formulation of Eq.\ \eqref{eq_gov_inc_strong} is obtained by
multiplying it with an arbitrary surface tension variation $ \delta \lambda $
and integrating over the membrane area.
The tension variation is assumed to be square-integrable, for which the weak
form of the continuity equation is expressed as
\begin{equation} \label{eq_gov_inc_weak_unstable}
	\int_\Omega
		\delta \lambda \, \big(
			\bma^\calpha \bmcdot \bmv_{, \calpha}
		\big)
	\, \JO ~ \td \Omega
	\, = \, 0
	\qquad
	\forall
	\quad
	\delta \lambda \in L^2 (\Omega)
	~.
\end{equation}
Care must be taken when discretizing Eq.\ \eqref{eq_gov_inc_weak_unstable} in
the course of finite element analysis, as one could violate the LBB condition
\cite{ladyzhenskaya-1969, babuska-nm-1973, brezzi-rfai-1974}
and observe spurious surface tension oscillations in the numerical solution.
A variety of techniques were developed to prevent such oscillations and
numerically stabilize the system \cite{malkus-cmame-1978, brezzi-1984,
zienkiewicz-cs-1984, linder-ijnme-2018}.
We choose to employ the Dohrmann--Bochev method \cite{dohrmann-ijnmf-2004}, as
it is based on an underlying theory and is straightforward to implement
numerically \cite{zienkiewicz-taylor-fem}.

%
%

\subsubsection{The Dohrmann--Bochev method: Numerical stabilization}
\label{sec_gov_inc_db}

The Dohrmann--Bochev method prevents spurious surface tension oscillations by
projecting the surface tension onto a space of discontinuous, piecewise linear
functions.
The function space, denoted $ \brL $, is defined in Eq.\ \eqref{eq_fem_disc_brL}
following a discussion of the surface discretization.
The $ L^2 $-projection of a given surface tension field
$ \lambda \in L^2 (\Omega) $
onto $ \brL $ is denoted $ \brlambda $ and is defined through the relation
\cite{dohrmann-ijnmf-2004}
\begin{equation} \label{eq_gov_inc_db_projection}
	\int_\Omega
		\delta \brlambda \, \big(
			\lambda
			\, - \, \brlambda
		\big)
	~\td \Omega
	\ = \ 0
	\qquad
	\forall
	\quad
	\delta \brlambda \in \brL
	~.
\end{equation}
In practice, the Dohrmann--Bochev method \cite{dohrmann-ijnmf-2004} is
implemented by quadratically penalizing surface tension deviations from the
space $ \brL $, for which the quantity
\begin{equation} \label{eq_gov_inc_db_weak}
	\dfrac{\alphaDB}{\zeta} \int_\Omega
		\big(
			\delta \lambda
			\, - \, \delta \brlambda
		\big)
		\big(
			\lambda
			\, - \, \brlambda
		\big)
	~ \td \Omega
\end{equation}
is subtracted from Eq.\ \eqref{eq_gov_inc_weak_unstable}.
Here, $ \zeta $ is the 2D intramembrane viscosity: a material property with
dimensions of energy$ \mk \cdot \mk $time$ \mk / \mk $area.
In addition, $ \alphaDB $ is a user-chosen parameter with dimensions of area,
such that Eqs.\ \eqref{eq_gov_inc_weak_unstable} and \eqref{eq_gov_inc_db_weak}
have the same dimensions---though the value of $ \alphaDB $ is not observed to
affect simulation results.
The numerically stabilized weak formulation of the incompressibility constraint
is written as
\begin{equation} \label{eq_gov_inc_weak_stable}
	\mcgl
	\, = \, 0
	\qquad
	\forall
	\quad
	\delta \lambda \in L^2 (\Omega)
	~,
\end{equation}
where $ \mcgl $ is the surface tension contribution to the direct Galerkin
expression \cite{zienkiewicz-taylor-solids} given by
\begin{equation} \label{eq_gov_inc_mcgl}
	\mcgl
	\ := \mk \int_\Omega
		\delta \lambda \, \big(
			\bma^\calpha \bmcdot \bmv_{, \calpha}
		\big)
	\, \JO ~ \td \Omega
	\ \, - \, \ \dfrac{\alphaDB}{\zeta} \int_\Omega
		\big(
			\delta \lambda
			\, - \, \delta \brlambda
		\big)
		\big(
			\lambda
			\, - \, \brlambda
		\big)
	~ \td \Omega
	~.
\end{equation}
In Eq.\ \eqref{eq_gov_inc_mcgl}, $ \brlambda $ and $ \delta \brlambda $ are
understood to be the $ L^2 $-projections of their respective counterparts
$ \lambda $ and $ \delta \lambda $ onto $ \brL $, according to Eq.\
\eqref{eq_gov_inc_db_projection}.

%
%

\subsection{The balance of linear momentum: Membrane dynamics}
\label{sec_gov_mem}

Consider a general, arbitrarily curved and deforming 2D material for which
inertial effects are negligible.
The local form of the balance of linear momentum for such a material is given by
\begin{equation} \label{eq_gov_mem_lin_mom}
	\bmTa_{; \mk \calpha}
	\, + \, \bmf
	\, = \, \bm{0}
	~,
\end{equation}
where $ \bmTa $ is the internal traction along a curve of constant $\zetaalpha $
on the surface and $ \bmf $ is the net body force per unit area on the material
by its surroundings.
The balance of angular momentum additionally requires that the stress vectors be
expressed as \cite{sahu-thesis}
\begin{equation} \label{eq_gov_mem_stress_vector}
	\bmTa
	\, = \, \sigmaab \mk \bmab
	\, - \, \big(
		M^{\cbeta \calpha} \mk \bmn
	\big)_{; \cbeta}
	~.
\end{equation}
Here, $ \sigmaab $ contains the couple-free, in-plane stress components and
$ \Mab $ contains the couple-stress components.
If the constitutive form of $ \sigmaab $ and $ \Mab $ are known, then Eqs.\
\eqref{eq_gov_mem_lin_mom} and \eqref{eq_gov_mem_stress_vector} govern the
dynamics of a general 2D material.

In lipid bilayers, the constitutive relations for $ \sigmaab $ and $ \Mab $ are
well-known, and depend on three material parameters.
The first is the intramembrane viscosity $ \zeta $, which characterizes the
irreversibility of in-plane flows.
The other parameters are the mean and Gaussian bending moduli, denoted $ \kb $
and $ \kg $, which have dimensions of energy and respectively penalize nonzero
mean and Gaussian curvatures \cite{canham-jtb-1970, helfrich-znc-1973,
evans-bpj-1974}.
We previously determined the membrane stresses and couple-stresses within the
framework of irreversible thermodynamics, and found \cite{sahu-thesis}
\begin{gather}
	\Mab
	\, = \, \kb \mk H \mk a^{\calpha \cbeta}
	\mk + \, \kg \mk \Big(
		2 \mk H \mk a^{\calpha \cbeta}
		\mk - \mk b^{\calpha \cbeta}
	\Big)
	\label{eq_gov_mem_Mab}
	\shortintertext{and}
	\sigmaab
	\, = \, \kb \mk \Big(
		H^2 \mk a^{\calpha \cbeta}
		\mk - \mk 2 \mk H \, b^{\calpha \cbeta}
	\Big)
	\, - \, \kg \mk K \mk a^{\calpha \cbeta}
	\mk + \, \lambda \, a^{\calpha \cbeta}
	\mk + \, \piab
	~,
	\label{eq_gov_mem_sigmaab}
	\shortintertext{where}
	\piab
	\, = \, \zeta \, \bmv_{, \cmu} \bmcdot \Big(
		\bma^\calpha \mk a^{\cmu \cbeta}
		\, + \, \bma^\cbeta \mk a^{\cmu \calpha}
	\Big)
	\label{eq_gov_mem_piab}
\end{gather}
are the in-plane viscous stress components.
The couple-stresses \eqref{eq_gov_mem_Mab} involve only $ \kb $ and $ \kg $, and
thus are purely elastic, while $ \sigmaabl $ contains bending, tensile, and
viscous contributions \eqref{eq_gov_mem_sigmaab}.

The in-plane and out-of-plane equations governing lipid membrane dynamics are
obtained by substituting Eqs.\ \eqref{eq_gov_mem_stress_vector}--%
\eqref{eq_gov_mem_piab} into Eq.\ \eqref{eq_gov_mem_lin_mom} and contracting the
result with $ \bmaa $ and $ \bmn $---which yields \cite{sahu-thesis}
\begin{gather}
	\zeta \, \Big(
		\Deltas v_{\calpha}
		\, + \, K \mk v_\calpha
		\, + \, 2 \mk v_{, \calpha} \mk H
		\, - \, 2 \mk v_{, \cbeta} \mk b^\cbeta_\calpha
		\, - \, 2 \mk v \mk H_{, \calpha}
	\Big)
	\, + \, f_\calpha
	\, + \, \lambda_{, \calpha}
	\ = \ 0
	\label{eq_gov_mem_lin_mom_in_plane}
	\shortintertext{and}
	f
	\, + \, 2 \mk \lambda \mk H
	\, + \, \zeta \, \Big(
		2 \mk b^{\calpha \cbeta} \mk v_{\calpha; \cbeta}
		\, - \, 8 \mk v \mk H^2
		\, + \, 4 \mk v \mk K
	\Big)
	\, - \, \kb \Big(
		2 \mk H^3
		\mk - \mk 2 \mk H \mk K
		\, + \, \Deltas H
	\Big)
	\ = \ 0
	~.
	\label{eq_gov_mem_lin_mom_shape}
\end{gather}
Equations \eqref{eq_gov_mem_lin_mom_in_plane} and
\eqref{eq_gov_mem_lin_mom_shape} are respectively referred to as the in-plane
and shape equations, and were independently obtained via different approaches
\cite{hu-pre-2007, arroyo-pre-2009, mandadapu-bmmb-2012, sahu-pre-2017}.
Here, the body force per unit area $ \bmf $ is decomposed as
$ \bmf = \bma^\calpha f_\calpha + f \bmn $,
and the operator $ \Deltas $ acts on an arbitrary quantity $ \pdp $ as
$ \Deltas \pdp := a^{\cmu \cnu} \pdp_{; \cmu \cnu} $.
Note that while the membrane bends elastically, the in-plane viscosity $ \zeta $
enters the shape equation \eqref{eq_gov_mem_lin_mom_shape} due to a coupling
between in-plane stresses and curvature \cite{sahu-pre-2020}.
The surface tension $ \lambda $ and surface curvatures $ H $ and $ K $ also
enter both the in-plane and shape equations, leading to nontrivial couplings
between in-plane and out-of-plane dynamics.

%
%

\subsubsection{The boundary conditions} \label{sec_gov_mem_bc}

One cannot determine a well-posed set of boundary conditions to Eqs.\
\eqref{eq_gov_mem_lin_mom_in_plane} and \eqref{eq_gov_mem_lin_mom_shape} by
inspection.
A series of systematic developments for elastic shells \cite{green-naghdi-1968,
steigmann-mms-1998, steigmann-arma-1999} underlie the formulation of the general
lipid membrane boundary conditions \cite{capovilla-pre-2002, arroyo-pre-2009,
mandadapu-bmmb-2012, sahu-pre-2017, duong-mms-2017}.
In what follows, we highlight possible boundary conditions and provide their
physical justification.
Details of our own derivations, which reproduce earlier results, are provided in
Ch.V\,\S5(d) of Ref.\ \cite{sahu-thesis}.

The in-plane equations governing lipid flows \eqref{eq_gov_mem_lin_mom_in_plane}
are identical to those governing a two-dimensional Newtonian fluid
\cite{scriven-ces-1960}.
We thus expect the boundary conditions to be similar to those of a fluid, in
which one specifies either the velocity $ \bmv $ or (for a surface) the force
per length $ \bmF $ on the boundary.
For general 2D materials, the force per length on the patch boundary is given by
\begin{equation} \label{eq_gov_mem_bc_force}
	\bmF
	\, = \, \bmTa \nu^{}_\calpha
	\, - \, \big(
		\Mab \mk \nu^{}_\calpha \mk \tau^{}_\cbeta \, \bmn
	\big)_{\! , \cmu} \, \tau^\cmu
	~,
\end{equation}
which---for the case of lipid membranes---has bending, tensile, and viscous
contributions (recall $ \bmnu $ and $ \bmtau $ are boundary basis vectors;
see Fig.\ \ref{fig_gov_geo}).
In our numerical implementation, on each edge of the membrane patch we specify
a component of either the velocity $ \bmv $ or the force per length $ \bmF $ in
each of the three Cartesian directions.
We denote $ \GammaFj $ and $ \Gammavj $ as the respective portions of the
boundary where
$ \Fj := \bmF \bmcdot \bmej $
and
$ \vj := \bmv \bmcdot \bmej $
are specified.
For
$ j \in \{ 1, 2, 3 \} $, the intersection
$ \Gammavj \cap \GammaFj = \varnothing $
and the closure of the union
$
	\overline{\rule{0mm}{1.67ex} \hspace{32pt} ~} \hspace{-35pt}
	\Gamma_{\! v}^{\mk \jmath} \cup \Gamma_{\!\! F}^{\mk \jmath}
	= \Gamma
$.
The boundary conditions are expressed as
\begin{equation} \label{eq_gov_mem_bc_in_plane}
	\vj
	\, = \, \barvj
	\quad
	\text{on}
	\quad
	\Gammavj
	\qquad
	\text{and}
	\qquad
	\Fj
	\, = \, \barFj
	\quad
	\text{on}
	\quad
	\GammaFj
	~,
\end{equation}
where $ \barvj $ and $ \barFj $ are known quantities that we prescribe.

The boundary conditions in Eq.\ \eqref{eq_gov_mem_bc_in_plane} are necessary but
not sufficient for a mathematically well-posed scenario.
To see why, note that the membrane bending energy gives rise to the
$ \Deltas H = a^{\cmu \cnu} \mk H_{; \cmu \cnu} $
term in the shape equation \eqref{eq_gov_mem_lin_mom_shape}.
Since the mean curvature contains two spatial derivatives of the surface
position through the curvature components
$ b_{\calpha \cbeta} = \bmn \bmcdot \bmx_{, \calpha \cbeta} $,
the $ \Deltas H $ term contains four derivatives of the surface position.
Following canonical developments in the theory of beam bending
\cite{timoshenko-pm-1921, timoshenko-pm-1922}, we expect to specify two
conditions along the entire boundary: either the out-of-plane velocity or force
per length, as well as either the slope of the surface or the boundary moment
\begin{equation} \label{eq_gov_mem_bc_moment}
	M
	\, := \, \Mab \mk \nu^{}_\calpha \mk \nu^{}_\cbeta
	~.
\end{equation}
Since the velocity and force boundary conditions are already contained in Eq.\
\eqref{eq_gov_mem_bc_in_plane}, we need only specify one of the latter pair.
More precisely, we partition the boundary into the disjoint sets $ \Gamman $
and $ \GammaM $, with
$ \Gamman \cap \GammaM = \varnothing $
and
$ \overline{ \mk \Gamman \cup \GammaM } = \Gamma $,
and prescribe
\begin{equation} \label{eq_gov_mem_bc_shape}
	\bmn \bmcdot \bmv_{, \calpha} \, \nu^{\calpha}
	\, = \, \barv_\nu
	\quad
	\text{on}
	\quad
	\Gamman
	\qquad
	\text{and}
	\qquad
	M
	\, = \, \barM
	\quad
	\text{on}
	\quad
	\GammaM
	~,
\end{equation}
where $ \barv_\nu $ and $ \barM $ are prescribed quantities.
With Eqs.\ \eqref{eq_gov_mem_bc_force}--\eqref{eq_gov_mem_bc_shape}, we have a
mathematically well-posed set of boundary conditions for the governing
equations.

%
%

\subsubsection{The weak formulation} \label{sec_gov_mem_weak}

The weak formulation of the balance of linear momentum is obtained by first
contracting Eq.\ \eqref{eq_gov_mem_lin_mom} with an arbitrary velocity variation
$ \delta \bmv $.
At this point, we recognize the four spatial derivatives contained in the shape
equation \eqref{eq_gov_mem_lin_mom_shape} through the $ \Deltas H $ term will,
in the subsequent weak form, yield two spatial derivatives of both the velocity
variation and the surface position.
Since $ \bmv $ and $ \bmx $ are assumed to lie in the same space of functions,
both are elements of $ \bmH^2 (\Omega) $---for which higher-order basis
functions, with continuous first derivatives, are required in the numerical
implementation.
In accordance with the boundary conditions in Eqs.\
\eqref{eq_gov_mem_bc_in_plane} and \eqref{eq_gov_mem_bc_shape}, the space of
admissible material velocity variations $ \bmmcvz $ is expressed as
\begin{equation} \label{eq_gov_mem_weak_velocity_0_space}
	\bmmcvz
	\, := \, \Big\{\,
		\bmu (\zetaalpha) : \Omega \rightarrow \mathbb{R}^3
		\quad
		\text{such that}
		\quad
		\bmu \in \bmH^2 (\Omega),
		\quad
		u^{}_j \big\rvert_{\Gammavj}
		\, = \, 0,
		\quad
		\big(
			\bmn \bmcdot \bmu_{, \calpha} \, \nu^\calpha
		\big) \big\rvert_{\Gamman} = \, 0
	\,\Big\}
\end{equation}
for
$ j \in \{ 1, 2, 3 \} $.
We integrate the result of the contraction over the membrane area, and
substitute the general form of the stress vectors
\eqref{eq_gov_mem_stress_vector} to obtain
\begin{equation} \label{eq_gov_mem_weak_contraction}
	\int_\Omega
		\delta \bmv \bmcdot \Big[\,
			\sigmaab \mk \bmab
			\, - \, \big(
				M^{\cbeta \calpha} \mk \bmn
			\big)_{; \cbeta}
		\,\Big]_{; \alpha}
	\, \JO ~ \td \Omega
	\ + \, \int_\Omega
		\delta \bmv \bmcdot \bmf
	\, \JO ~ \td \Omega
	\ = \ 0
	\qquad
	\forall ~ \delta \bmv \in \bmmcvz
	~.
\end{equation}
Starting with Eq.\ \eqref{eq_gov_mem_weak_contraction}, a series of algebraic
manipulations are required to determine the weak form of the linear momentum
balance.
The calculations rely on developments by Naghdi, \textsc{A.E.\ Green}, and
\textsc{D.J.\ Steigmann} \cite{green-naghdi-1968, steigmann-mms-1998}, and can
be found in Refs.\ \cite{mandadapu-jcp-2017, duong-mms-2017}.
Following the notation and development of our prior efforts
\cite[Ch.V$\mk$\S5(d)]{sahu-thesis},
the weak formulation of the balance of linear momentum is expressed as
\begin{equation} \label{eq_gov_mem_weak}
	\mcgv
	\, = \, 0
	\qquad
	\forall
	\quad
	\delta \bmv \in \bmmcvz
	~,
\end{equation}
where%
\footnote{%
	Jump forces can arise at corners of the membrane patch and enter Eq.\
	\eqref{eq_gov_mem_mcgv}.
	They are assumed here to be zero.%
}
\begin{equation} \label{eq_gov_mem_mcgv}
	\begin{split}
		\mcgv
		\  &:= \, \int_\Omega
			\dfrac{1}{2} \, \Big(
				\delta \bmv_{, \calpha} \bmcdot \bmab
				\, + \, \delta \bmv_{, \cbeta} \bmcdot \bmaa
			\Big) \, \sigmaab
		\, \JO ~ \td \Omega
		\  + \, \int_\Omega
			\dfrac{1}{2} \, \Big(
				\delta \bmv_{; \calpha \cbeta}
				\, + \, \delta \bmv_{; \cbeta \calpha}
			\Big) \bmcdot \bmn \, \Mab
		\, \JO ~ \td \Omega
		\\[5pt]
		&\hspace{20pt}
		- \, \int_\Omega \delta \bmv \bmcdot \bmf \, \JO ~ \td \Omega
		\ - \ \sum_{j = 1}^3 \mk \int_{\GammaFj}
			\delta v_j \, \barFj
		\, \JG ~ \td \Gamma
		\ - \, \int_{\GammaM}
			\delta \bmv_{, \calpha} \, \nu^\calpha \bmcdot \bmn \, \barM
		\, \JG ~ \td \Gamma
		~.
	\end{split}
\end{equation}
In Eq.\ \eqref{eq_gov_mem_mcgv},
$ \delta v_j := \delta \bmv \bmcdot \bmej $
is the $j^{\text{th}}$ Cartesian component of the arbitrary velocity variation.
It is useful to recognize that Eq.\ \eqref{eq_gov_mem_mcgv} is general to any
material whose stress vectors can be expressed as in Eq.\
\eqref{eq_gov_mem_stress_vector}, with the corresponding boundary conditions in
Eqs.\ \eqref{eq_gov_mem_bc_in_plane} and \eqref{eq_gov_mem_bc_shape}.

%
%

\subsection{The dynamics of the mesh} \label{sec_gov_mesh}

In our ALE formulation, the material velocity $ \bmv $ and the mesh velocity
$ \bmvm $ are independent quantities.
The evolution of the position $ \bmx (\zetaalpha, t) $ of the membrane surface
is dictated by the mesh velocity via the relations
\begin{gather}
	\bmvm (\zetaalpha, t)
	\, = \, \pp{}{t} \Big(
		\bmx (\zetaalpha, t)
	\Big)
	\label{eq_gov_mesh_velocity}
	\shortintertext{and}
	\bmx (\zetaalpha, \mk t)
	\, = \, \bmx (\zetaalpha, \mk 0)
	\, + \, \int_0^t \bmvm (\zetaalpha, \mk t') ~\td t'
	~,
	\label{eq_gov_mesh_position}
\end{gather}
which both apply over all parametric coordinates
$ \zetaalpha \mkn \in \Omega $
and for all times
$ t \in [\mk 0, \ttf \mk] $
under consideration (see Ref.\ \cite[\S2]{sahu-jcp-2020} for additional
details).
As we previously recognized
$ \bmx \in \bmH^2 (\Omega) $,
Eqs.\ \eqref{eq_gov_mesh_velocity} and \eqref{eq_gov_mesh_position} reveal
$ \bmvm \in \bmH^2 (\Omega) $
as well.
In what follows, we discuss three mesh velocity schemes of increasing
complexity, all of which are consistent with the kinematic constraint
\eqref{eq_gov_geodyn_kinematic_constraint} such that the mesh and material
always overlap.
We begin with Lagrangian and Eulerian schemes, which were previously employed
by both ourselves and others, and are known to suffer from limitations
\cite{rodrigues-jcp-2015, barrett-pre-2015, mandadapu-jcp-2017, omar-bpj-2020,
reuther-jfm-2020, sahu-jcp-2020}.
We then discuss a new class of ALE schemes in which the mesh velocity itself
satisfies dynamical equations similar to those that govern the material
velocity.

%
%

\subsection{The Lagrangian mesh motion} \label{sec_gov_mesh_lag}

We begin with the simplest membrane motion, namely a Lagrangian scheme.
In this case, the mesh velocity is prescribed to be equal to the material
velocity:
\begin{equation} \label{eq_gov_mesh_lag_vm}
	\bmvm
	\mk = \, \bmv
	~.
\end{equation}
In our numerical implementation, mesh velocity degrees of freedom are mapped to
their material velocity counterparts such that Eq.\ \eqref{eq_gov_mesh_lag_vm}
is satisfied identically over the entire surface.%
\footnote{%
	One could instead treat the membrane and mesh velocities separately, and
	directly modify the resultant tangent diffusion matrix to satisfy Eq.\
	\eqref{eq_gov_mesh_lag_vm}.
	Details of such an approach are provided in Appendix B.7 of Ref.\
	\cite{sahu-jcp-2020}.%
}
The direct Galerkin expression in the case of a Lagrangian mesh motion is then
given by
\begin{equation} \label{eq_gov_mesh_lag_mcg}
	\mcgL
	\, := \,\mk \mcgl
	\, + \, \mcgv
	\  = \ \mk  0
	\qquad
	\forall
	\quad
	\delta \lambda \in L^2 (\Omega)
	~,
	~ ~
	\delta \bmv \in \bmmcvz
	~.
\end{equation}
In Eq.\ \eqref{eq_gov_mesh_lag_mcg}, the superscript `\texttt{L}' denotes a
Lagrangian scheme, where
$ \bmvm = \bmv $
throughout and mesh positions are updated according to Eq.\
\eqref{eq_gov_mesh_position}.
In addition, $ \mcgl $ and $ \mcgv $ are respectively provided in Eqs.\
\eqref{eq_gov_inc_mcgl} and \eqref{eq_gov_mem_mcgv}.

%
%

\subsection{The Eulerian mesh motion} \label{sec_gov_mesh_eul}

The second scheme we implement is in-plane Eulerian, with
$ \bmvm \bmcdot \bmaa = 0 $,
and out-of-plane Lagrangian, for which
$ \bmvm \bmcdot \bmn = \bmv \bmcdot \bmn $.
The in-plane and out-of-plane conditions are expressed as the single equation
\begin{equation} \label{eq_gov_mesh_eul_vm_strong}
	\bmvm
	\, = \, \big(
		\bmn \otimes \bmn
	\big) \mk \bmv
	~,
\end{equation}
where `$ \otimes $' denotes the dyadic product.
As there are no spatial derivatives in Eq.\ \eqref{eq_gov_mesh_eul_vm_strong},
no mesh velocity boundary conditions are to be prescribed for an Eulerian mesh
motion.

The weak form of Eq.\ \eqref{eq_gov_mesh_eul_vm_strong} is obtained by
contracting it with an arbitrary mesh velocity variation
$ \delta \bmvm \in \bmH^2 (\Omega) $,
integrating over the membrane surface, and multiplying the result by a user-%
specified constant $ \alphamE $.
The weak formulation of the Eulerian mesh velocity equation is then given by
\cite{sahu-jcp-2020}
\begin{equation} \label{eq_gov_mesh_eul_vm_weak}
	\mcgmE
	\ := \ \alphamE \int_\Omega
		\delta \bmvm \bmcdot \Big[\,
			\bmvm
			\, - \, \big(
				\bmn \otimes \bmn
			\big) \mk \bmv
		\,\Big]
	\, \JO ~ \td \Omega
	\ = \ 0
	\qquad
	\forall
	\quad
	\delta \bmvm \in \bmH^2 (\Omega)
	~.
\end{equation}
In Eq.\ \eqref{eq_gov_mesh_eul_vm_weak}, the superscript `\texttt{E}' signifies
the scheme is Eulerian, and the constant $ \alphamE $ has dimensions of
energy$ \, \cdot \, $time$ \mk / \mk $length$ ^4 $ such that $ \mcgmE $ has the
same dimensions as $ \mcgl $ and $ \mcgv $.
In this study,
$ \alphamE = \zeta / \ell^2 $
for all results presented, as varying $ \alphamE $ was not observed to affect
membrane behavior ($ \ell $ is a chosen characteristic length).
The direct Galerkin expression for a scenario with an Eulerian mesh motion is
expressed as
\begin{equation} \label{eq_gov_mesh_eul_mcg}
	\mcgE
	\, := \,\mk \mcgl
	\, + \, \mcgv
	\, + \, \mcgmE
	\  = \ \mk  0
	\qquad
	\forall
	\quad
	\delta \lambda \in L^2 (\Omega)
	~,
	~ ~
	\delta \bmv \in \bmmcvz
	~,
	~ ~
	\delta \bmvm \in \bmH^2 (\Omega)
	~.
\end{equation}

%
%

\subsection{The arbitrary Lagrangian--Eulerian mesh motion}
\label{sec_gov_mesh_ale}

The final scheme we consider is neither in-plane Lagrangian nor in-plane
Eulerian.
Instead, the mesh is modeled as a separate 2D material with its own constitutive
relations and associated dynamical equations.
The mesh analogs of the stress vectors are then expressed as [cf.\ Eq.\
\eqref{eq_gov_mem_stress_vector}]
\begin{equation} \label{eq_gov_mesh_visc_stress_vector}
	\bmTam
	\, = \, \sigmaabm \mk \bmab
	\, - \, \big(
		M^{\cbeta \calpha}_\tm \mk \bmn
	\big)_{; \cbeta}
	~,
\end{equation}
where $ \sigmaabm $ and $ \Mabm $ are the mesh analogs of $ \sigmaab $ and
$ \Mab $.
Here $ \sigmaabm $ and $ \Mabm $ can be specified arbitrarily without altering
the dynamics of the membrane, and so the mesh dynamics can be that of an
elastic, viscous, or viscoelastic material.
When the mesh velocity $ \bmvm $ is determined by solving an arbitrary set of
governing equations, it will no longer satisfy the kinematic constraint
\eqref{eq_gov_geodyn_kinematic_constraint} across the membrane surface.
We enforce Eq.\ \eqref{eq_gov_geodyn_kinematic_constraint} with an additional
Lagrange multiplier field, which physically acts on the mesh as an external body
force per unit area in the normal direction---hereafter referred to as the mesh
pressure $ \ptm $.
Following the developments of \ref{sec_gov_mem}, the dynamical equation
governing the mesh dynamics is given by
\begin{equation} \label{eq_gov_mesh_visc_lin_mom}
	\bmTa_{\tm \mk ; \mk\mk \calpha}
	\, + \, \ptm \mk \bmn
	\, = \, \bm{0}
	~.
\end{equation}
Upon specification of appropriate boundary conditions, Eq.\
\eqref{eq_gov_geodyn_kinematic_constraint} and the three components of Eq.\
\eqref{eq_gov_mesh_visc_lin_mom} uniquely determine the mesh pressure $ \ptm $
and the three components of the mesh velocity $ \bmvm $.
The ability to specify mesh boundary conditions can prevent undesirable mesh
distortions, and is an advantage of our ALE method.
Two choices of the mesh motion, for which $ \sigmaabm $ and $ \Mabm $ are
prescribed, are discussed subsequently.
In the first case, the mesh is area-compressible and purely viscous; in the
second case it also resists bending.
In both cases, the simplest boundary conditions for the mesh velocity are
chosen.
The investigation of more complex boundary conditions, as well as more involved
mesh behaviors, is left to a future study.

%
%

\subsubsection{The weak formulation of the kinematic constraint}
\label{sec_gov_mesh_ale_kinematic_constraint}

The mesh pressure is an unknown Lagrange multiplier field enforcing the ALE
kinematic constraint in Eq.\ \eqref{eq_gov_geodyn_kinematic_constraint}.
The weak form of Eq.\ \eqref{eq_gov_geodyn_kinematic_constraint} is obtained by
multiplying it with an arbitrary variation
$ \delta \ptm \in L^2 (\Omega) $,
and integrating the result over the membrane surface to obtain
\begin{equation} \label{eq_gov_mesh_visc_weak_unstable}
	\int_\Omega
		\delta \ptm \, \bmn \bmcdot \big(
			\bmvm
			- \, \bmv
		\big)
	\, \JO ~ \td \Omega
	\, = \, 0
	\qquad
	\forall
	\quad
	\delta \ptm \in L^2 (\Omega)
	~.
\end{equation}
Equation \eqref{eq_gov_mesh_visc_weak_unstable} bears some resemblance to Eq.\
\eqref{eq_gov_inc_weak_unstable}, which was obtained from the incompressibility
constraint.
In both equations, the variation of a scalar Lagrange multiplier interacts with
a vector velocity.
Moreover, when Eq.\ \eqref{eq_gov_mesh_visc_weak_unstable} is discretized, our
numerical simulations exhibit an instability reminiscent of the checkerboarding
that arises when the LBB condition is violated.
We thus hypothesize that both numerical instabilities are similar, and attempt
to stabilize the mesh pressure with the Dohrmann--Bochev method.
Following the results of \ref{sec_gov_inc_db}, we express the weak formulation
of the kinematic constraint \eqref{eq_gov_geodyn_kinematic_constraint} as
\begin{equation} \label{eq_gov_mesh_visc_weak_stable}
	\mcgp
	\, = \, 0
	\qquad
	\forall
	\quad
	\delta \ptm \in L^2 (\Omega)
	~,
\end{equation}
where the numerically stabilized mesh pressure contribution to the direct
Galerkin expression is given by [cf.\ Eqs.\ \eqref{eq_gov_inc_mcgl},
\eqref{eq_gov_mesh_visc_weak_unstable}]
\begin{equation} \label{eq_gov_mesh_visc_mcgp}
	\mcgp
	\ := \ - \int_\Omega
		\delta \ptm \, \bmn \bmcdot \big(
			\bmvm
			- \, \bmv
		\big)
	\, \JO ~ \td \Omega
	\ \, - \, \ \dfrac{\alphaDB \ell^2}{\zeta} \int_\Omega
		\big(
			\delta \ptm
			\, - \, \delta \brpm
		\big)
		\big(
			\ptm
			\, - \, \brpm
		\big)
	~ \td \Omega
	~.
\end{equation}
In Eq.\ \eqref{eq_gov_mesh_visc_mcgp}, $ \brpm $ and $ \delta \brpm $ are the
$ L^2 $-projections of $ \ptm $ and $ \delta \ptm $ onto the space $ \brL $, as
defined through Eq.\ \eqref{eq_gov_inc_db_projection}.
The factor of $ \ell^2 $, for a characteristic length $ \ell $, is required for
dimensional consistency.
Our choice to employ the Dohrmann--Bochev method to stabilize
$ \ptm $ does not yet sit on a firm theoretical footing, and currently can only
be justified with empirical success.
We hope to investigate the mathematical nature of the stabilization of $ \ptm $
in a future manuscript.

%
%

\subsubsection{The case where the mesh dynamics are purely viscous}
\label{sec_gov_mesh_ale_visc}

For the case of purely viscous mesh dynamics, quantities are denoted with a
superscript or subscript `\texttt{v}.'
The scheme is referred to as `ALE-viscous,' with the shorthand `\texttt{Av}' or
`\alev.'
Since the mesh velocity field is area-compressible, the mesh motion resists both
shearing and dilation.

\bigskip

\noindent
\textbf{Strong formulation.---}%
We prescribe for the mesh analog of the in-plane stresses and couple-stresses to
be [cf.\ Eqs.\ \eqref{eq_gov_mem_Mab}--\eqref{eq_gov_mem_piab}]
\begin{equation} \label{eq_gov_mesh_visc_sigmaab}
	\sigmaabmv
	\mk = \, \piabm
	\mk := \, \zetam \, \bmvm_{\,\, , \mk \cmu} \bmcdot \Big(
		\bma^\calpha \mk a^{\cmu \cbeta}
		\, + \, \bma^\cbeta \mk a^{\cmu \calpha}
	\Big)
	\qquad
	\text{and}
	\qquad
	\Mabmv
	\mk = \, 0
	~,
\end{equation}
where $ \zetam $ is a user-specified parameter that we refer to as the mesh
viscosity.
Importantly, since the mesh velocity is area-compressible, our choice in Eq.\
\eqref{eq_gov_mesh_visc_sigmaab} resists both shear and dilation of the mesh.
By substituting Eqs.\ \eqref{eq_gov_mesh_visc_stress_vector} and
\eqref{eq_gov_mesh_visc_sigmaab} into Eq.\ \eqref{eq_gov_mesh_visc_lin_mom}, the
dynamical equations governing the mesh are found to be
\cite[Ch.V$\mk$\S5(c)]{sahu-thesis}
\begin{gather}
	\zetam \, \Big(
		\Deltas \vm_{\calpha}
		\, + \, K \mk \vm_\calpha
		\, + \, 2 \mk (\vm)_{, \calpha} \mk H
		\, - \, 2 \mk (\vm)_{, \cbeta} \, b^\cbeta_\calpha
		\, - \, 2 \mk \vm H_{, \calpha}
		\, + \, \big(
			a^{\cgamma \cbeta} \mk (\vm_\cgamma)_{; \cbeta}
			\, - \, 2 \mk \vm H
		\big)_{\! ; \calpha}
	\Big)
	\ = \ 0
	\label{eq_gov_mesh_visc_lin_mom_in_plane}
	\shortintertext{and}
	\ptm
	\, + \, \zetam \, \Big(
		2 \mk b^{\calpha \cbeta} \mk (\vm_{\calpha})_{; \cbeta}
		\, - \, 8 \mk \vm H^2
		\, + \, 4 \mk \vm K
	\Big)
	\ = \ 0
	~,
	\label{eq_gov_mesh_visc_lin_mom_shape}
\end{gather}
where
$ \bmvm = \vm_{\calpha} \mk \bma^\calpha + \vm \mk \bmn $.
The solution to Eq.\ \eqref{eq_gov_mesh_visc_lin_mom_in_plane} is
independent of $ \zetam $, and the mesh pressure in Eq.\
\eqref{eq_gov_mesh_visc_lin_mom_shape}  will adjust so as to satisfy the
kinematic constraint \eqref{eq_gov_geodyn_kinematic_constraint}.
The mesh dynamics accordingly do not depend on the choice of $ \zetam $.
In all numerical results presented, we choose
$ \zetam = \zeta $
for simplicity.

\bigskip

\noindent
\textbf{Boundary conditions.---}%
Since the mesh velocity satisfies differential equations [see Eqs.\
\eqref{eq_gov_mesh_visc_lin_mom_in_plane} and
\eqref{eq_gov_mesh_visc_lin_mom_shape}]
rather than an algebraic equation [cf.\ Eqs.\ \eqref{eq_gov_mesh_lag_vm},%
\eqref{eq_gov_mesh_eul_vm_strong}], boundary conditions for the mesh velocity
are required.
To this end, we introduce the mesh analog of the force per length on the
boundary:
\begin{equation} \label{eq_gov_mesh_visc_bc_force}
	\bmFmv
	\, = \, \bmTamv \, \nu^{}_\calpha
	\, = \, \sigmaabmv \, \nu^{}_\calpha \mk \bmab
	~.
\end{equation}
Recalling $ \Mabmv = 0 $ according to Eq.\ \eqref{eq_gov_mesh_visc_sigmaab}, the
purely viscous mesh dynamics cannot sustain moments---as is the case for a fluid
film \cite{sahu-jcp-2020}.
Consequently, neither slope nor moment boundary conditions are required on the
boundary, since the mesh analog of the moment
$ M^\tm = 0 $
identically throughout.
With this understanding, the parametric boundary $ \Gamma $ is partitioned into
the portions $ \Gammavmj $ and $ \GammaFmj $, on which
$ \vmj := \bmvm \bmcdot \bmej $
and
$ \Fmj := \bmFm \bmcdot \bmej $
are respectively specified.
For simplicity in our numerical formulation, the mesh is assumed to be either
static or force-free on its entire boundary, for which [cf.\ Eq.\
\eqref{eq_gov_mem_bc_in_plane}]
\begin{equation} \label{eq_gov_mesh_visc_bc_in_plane}
	\vmj
	\, = \, 0
	\quad
	\text{on}
	\quad
	\Gammavmj
	\qquad
	\text{and}
	\qquad
	\Fmj
	\, = \, 0
	\quad
	\text{on}
	\quad
	\GammaFmj
	~.
\end{equation}

\bigskip

\noindent
\textbf{Weak formulation.---}%
The weak formulation of the purely viscous ALE mesh behavior is obtained by
drawing analogy to the developments in \ref{sec_gov_mem_weak}.
First, the space of arbitrary mesh velocity variations is defined as
\begin{equation} \label{eq_gov_mesh_weak_velocity_0_space}
	\bmmcvmzv
	\, := \, \Big\{\,
		\bmu (\zetaalpha) : \Omega \rightarrow \mathbb{R}^3
		\quad
		\text{such that}
		\quad
		\bmu \in \bmH^2 (\Omega),
		\quad
		u^{}_j \big\rvert_{\Gammavmj}
		\, = \, 0
	\,\Big\}
	~.
\end{equation}
By contracting Eq.\ \eqref{eq_gov_mesh_visc_lin_mom} with an arbitrary mesh
velocity variation
$ \delta \bmvm \in \bmmcvmzv $
and repeating the steps of \ref{sec_gov_mem_weak}, the weak formulation of the
ALE-viscous mesh motion is found to be [cf.\ Eqs.\ \eqref{eq_gov_mem_weak} and
\eqref{eq_gov_mem_mcgv}]
\begin{equation} \label{eq_gov_mesh_visc_weak}
	\mcgmAv
	\, = \, 0
	\qquad
	\forall
	\quad
	\delta \bmvm \in \bmmcvmzv
	~,
\end{equation}
where
\begin{equation} \label{eq_gov_mesh_visc_mcgmA}
	\mcgmAv
	\  := \, \int_\Omega
		\dfrac{1}{2} \, \Big(
			\delta \bmvm_{, \calpha} \bmcdot \bmab
			\, + \, \delta \bmvm_{, \cbeta} \bmcdot \bmaa
		\Big) \, \sigmaabmv
	\, \JO ~ \td \Omega
	\,\  - \, \int_\Omega
		\Big(
			\delta \bmvm \bmcdot \ptm \mk \bmn
		\Big)
	\, \JO ~ \td \Omega
	~.
\end{equation}
In Eqs.\ \eqref{eq_gov_mesh_visc_weak} and \eqref{eq_gov_mesh_visc_mcgmA}, the
superscript `\texttt{Av}' refers to the choice of a purely viscous ALE mesh
motion.
The direct Galerkin expression for this motion is then written as
\begin{equation} \label{eq_gov_mesh_visc_mcg}
	\begin{split}
		&
		\mcgAv
		\, := \,\mk \mcgl
		\, + \, \mcgv
		\, + \, \mcgmAv
		\, + \, \mcgp
		\  = \ \mk  0
		\\[5pt]
		&\quad
		\forall
		\quad
		\delta \lambda \in L^2 (\Omega)
		~,
		~ ~
		\delta \bmv \in \bmmcvz
		~,
		~ ~
		\delta \bmvm \in \bmmcvmzv
		~,
		~ ~
		\delta \ptm \in L^2 (\Omega)
		~.
	\end{split}
\end{equation}

%
%

\subsubsection{The case where the mesh dynamics are viscous and resist bending}
\label{sec_gov_mesh_ale_bend}

We now consider the case where the mesh dynamics are viscous in response to
in-plane shears and area dilations, and additionally resist bending.
The mesh motion is referred to as `ALE-viscous-bending,' with shorthand
`\texttt{Avb}' or `\alevb.'
Corresponding quantities are denoted with a subscript or superscript
`\texttt{vb}.'

\bigskip

\noindent
\textbf{Strong formulation.---}%
We choose for the bending resistance of the mesh to arise in the same manner
as the membrane itself, with the viscous resistance identical to that of the
ALE-viscous scenario.
To this end, we choose [cf.\ Eqs.\ \eqref{eq_gov_mem_Mab}--%
\eqref{eq_gov_mem_piab} and \eqref{eq_gov_mesh_visc_sigmaab}]
\begin{gather}
	\Mabmvb
	\, = \, \kbm \mk H \mk a^{\calpha \cbeta}
	\mk + \, \kgm \mk \Big(
		2 \mk H \mk a^{\calpha \cbeta}
		\mk - \mk b^{\calpha \cbeta}
	\Big)
	\label{eq_gov_mesh_vb_Mab}
	\shortintertext{and}
	\sigmaabmvb
	\, = \, \kbm \mk \Big(
		H^2 \mk a^{\calpha \cbeta}
		\mk - \mk 2 \mk H \, b^{\calpha \cbeta}
	\Big)
	\, - \, \kgm \mk K \mk a^{\calpha \cbeta}
	\mk + \, \piabm
	~,
	\label{eq_gov_mesh_vb_sigmaab}
\end{gather}
where $ \piabm $ is defined in Eq.\ \eqref{eq_gov_mesh_visc_sigmaab}.
In Eqs.\ \eqref{eq_gov_mesh_vb_Mab} and \eqref{eq_gov_mesh_vb_sigmaab}, $ \kbm $
and $ \kgm $ are user-specified parameters; we always choose
$ \kbm = \kb $
and
$ \kgm = \kg $
for simplicity.
It is well-known that bending terms from the Helfrich free energy do not enter
the in-plane equations \cite{mandadapu-bmmb-2012}, and so the in-plane dynamical
mesh equations in the \alevb\ case are given by Eq.\
\eqref{eq_gov_mesh_visc_lin_mom_in_plane}.
The out-of-plane dynamical equation is expressed as [cf.\ Eqs.\
\eqref{eq_gov_mem_lin_mom_shape} and \eqref{eq_gov_mesh_visc_lin_mom_shape}]
\begin{equation} \label{eq_gov_mesh_vb_lin_mom_shape}
	\ptm
	\, + \, \zetam \, \Big(
		2 \mk b^{\calpha \cbeta} \mk v_{\calpha; \cbeta}
		\, - \, 8 \mk v \mk H^2
		\, + \, 4 \mk v \mk K
	\Big)
	\, - \, \kbm \Big(
		2 \mk H^3
		\mk - \mk 2 \mk H \mk K
		\, + \, \Deltas H
	\Big)
	\ = \ 0
	~.
\end{equation}
Contrary to the \alev\ scenario \eqref{eq_gov_mesh_visc_lin_mom_shape},
here the relative magnitude of $ \zetam $ and $ \kbm $ do affect the resultant
mesh pressure $ \ptm $.
Investigating the relationship between $ \ptm $, $ \zetam $, and $ \kbm $---as well as the
ensuing mesh dynamics---is left to a later study.

\bigskip

\noindent
\textbf{Boundary conditions.---}%
We once again introduce the mesh analog of the force per length on the boundary,
which in this case is written as [cf.\ Eqs.\ \eqref{eq_gov_mem_bc_force} and
\eqref{eq_gov_mesh_visc_bc_force}]
\begin{equation} \label{eq_gov_mesh_vb_bc_force}
	\bmFmvb
	\, = \, \bmTamvb \, \nu^{}_\calpha
	\, - \, \big(
		\Mabmvb \, \nu^{}_\calpha \mk \tau^{}_\cbeta \, \bmn
	\big)_{\! , \cmu} \, \tau^\cmu
	~.
\end{equation}
In principle, one could specify boundary forces for the mesh dynamics.
However, as in the ALE-viscous case, we choose simple boundary conditions where
the mesh is either static or force-free on the entire boundary
\eqref{eq_gov_mesh_visc_bc_in_plane}.
In addition, since the mesh couple-stresses are nonzero, either slope or moment
boundary conditions are required for $ \bmvm $---where the analog of the
boundary moment is calculated as
$ \Mm = M^{\calpha \cmu}_\tm \mk \nu^{}_\calpha \mk \nu^{}_\cmu $
[cf.\ Eq.\ \eqref{eq_gov_mem_bc_moment}].
At present, we take the simpler approach and specify zero-slope conditions on
the entire boundary [cf.\ Eq.\ \eqref{eq_gov_mem_bc_shape}]:
\begin{equation} \label{eq_gov_mesh_vb_bc_shape}
	\bmn \bmcdot \bmvm_{, \calpha} \, \nu^{\calpha}
	\, = \, 0
	\quad
	\text{on}
	\quad
	\Gamma
	~.
\end{equation}
The ability to specify the slope of the mesh at a boundary will be relevant in
the tether pulling scenario of \ref{sec_sim_pull}.

\bigskip

\noindent
\textbf{Weak formulation.---}%
At this point, it is straightforward to write the weak formulation of the mesh
dynamics.
The space of arbitrary mesh variations is given by
\begin{equation} \label{eq_gov_mesh_vb_weak_velocity_0_space}
	\bmmcvmzvb
	\, := \, \Big\{\,
		\bmu (\zetaalpha) : \Omega \rightarrow \mathbb{R}^3
		\quad
		\text{such that}
		\quad
		\bmu \in \bmH^2 (\Omega),
		\quad
		u^{}_j \big\rvert_{\Gammavmj}
		\, = \, 0,
		\quad
		\big(
			\bmn \bmcdot \bmu_{, \calpha} \, \nu^\calpha
		\big) \big\rvert_{\Gamma} = \, 0
	\,\Big\}
	~.
\end{equation}
The weak formulation is then expressed as
\begin{equation} \label{eq_gov_mesh_vb_visc_weak}
	\mcgmAvb
	\, = \, 0
	\qquad
	\forall
	\quad
	\delta \bmvm \in \bmmcvmzvb
	~,
\end{equation}
where we define [cf.\ Eq.\ \eqref{eq_gov_mem_mcgv}]
\begin{equation} \label{eq_gov_mesh_vb_mcgmA}
	\begin{split}
		\mcgmAvb
		\  :=& \ \int_\Omega
			\dfrac{1}{2} \, \Big(
				\delta \bmvm_{, \calpha} \bmcdot \bmab
				\, + \, \delta \bmvm_{, \cbeta} \bmcdot \bmaa
			\Big) \, \sigmaabmvb
		\, \JO ~ \td \Omega
		\\[5pt]
		& \hspace{10pt} + \, \int_\Omega
			\dfrac{1}{2} \, \Big(
				\delta \bmvm_{; \calpha \cbeta}
				\, + \, \delta \bmvm_{; \cbeta \calpha}
			\Big) \bmcdot \bmn \ \Mabmvb
		\, \JO ~ \td \Omega
		\  - \, \int_\Omega
			\Big(
				\delta \bmvm \bmcdot \ptm \mk \bmn
			\Big)
		\, \JO ~ \td \Omega
		~.
	\end{split}
\end{equation}
The direct Galerkin expression for the scenario with the \alevb\ mesh motion is
written as
\begin{equation} \label{eq_gov_mesh_vb_mcg}
	\begin{split}
		&
		\mcgAvb
		\, := \,\mk \mcgl
		\, + \, \mcgv
		\, + \, \mcgmAvb
		\, + \, \mcgp
		\  = \ \mk  0
		\\[5pt]
		&\quad
		\forall
		\quad
		\delta \lambda \in L^2 (\Omega)
		~,
		~ ~
		\delta \bmv \in \bmmcvz
		~,
		~ ~
		\delta \bmvm \in \bmmcvmzvb
		~,
		~ ~
		\delta \ptm \in L^2 (\Omega)
		~.
	\end{split}
\end{equation}

\section{The finite element formulation} \label{sec_fem}

With the Lagrangian, Eulerian, and ALE weak formulations, our goal is to solve
for the state of the membrane over time.
We seek to determine ($i$) the membrane velocity $ \bmv $, ($i \mkn i$) the mesh
velocity $ \bmvm $, and ($i \mkn i \mkn i$) the membrane tension $ \lambda $---%
for all parametric points
$ \zetaalpha \in \Omega $
and times
$ t \in [\mk 0, \ttf \mk] $.
In doing so, the surface position is obtained from the mesh velocity through
Eq.\ \eqref{eq_gov_mesh_position}.
We cannot solve for the highly nonlinear membrane behavior exactly, and turn to
the finite element method to numerically calculate the approximate solutions
$ \bmvh $, $ \bmvmh $, and $ \lambdah $.
An overview of our solution method is presented in what follows.
Further specifics can be found in Appendix \ref{sec_a_fem} and in the
\memalefem\ documentation \cite{mem-ale-fem}.

%
%

\subsection{The surface discretization and space of solutions}
\label{sec_fem_disc}

Let us begin by assuming there is a discretization $ \mcth $ of the parametric
domain $ \Omega $ into \nel\ (\underline{n}umber of \underline{el}ements)
non-overlapping finite elements $ \Omegae $ of characteristic length $ h $:
\begin{equation} \label{eq_fem_disc_mcth}
	\mcth
	\, := \, \big\{
		\Omega^1, \,
		\Omega^2, \,
		\ldots,   \,
		\Omega^{\nel}
	\big\}
	~.
\end{equation}
Our task is to choose the set of basis functions over each element
$ \Omegae \in \mcth $%
---denoted $ \{ N^e_k (\zetaalpha) \} $, where the index $ k $ ranges from $ 1 $
to $ \nen $ (\underline{n}umber of \underline{e}lemental \underline{n}odes)---%
such that the resultant solution spaces are consistent with their infinite-%
dimensional counterparts.
A complexity arises because mesh and membrane velocities belong to the space
$ \bmH^2 (\Omega) $.
Accordingly, $ \bmvh $ and $ \bmvmh $ are required to be
continuous and have continuous first derivatives across elements.
We satisfy the continuity criteria with a so-called tensor product of quadratic
B-spline basis functions in each of the $ \zetaone $ and $ \zetatwo $
directions \cite{piegl-tiller, cottrell}.
For our purposes, the parametric domain $ \Omega $ is partitioned into a
rectangular grid of finite elements, though more advanced discretizations
are now established \cite{toshniwal-cmame-2017, wei-cmame-2018, paul-cm-2020,
koh-cmame-2022}.
The resultant scalar and vector finite-dimensional spaces $ \mcuh $ and
$ \bmmcuh $ are respectively given by
\begin{equation} \label{eq_fem_disc_mcuh}
	\mcuh
	\, := \, \Big\{\,
		u (\zetaalpha) : \Omega \rightarrow \mathbb{R}
		\quad
		\text{such that}
		\quad
		u \in C^1 (\Omega) \cap H^2 (\Omega),
		\quad
		u \big\rvert_{\Omegae}
		\in \mathbb{Q}_2 (\Omegae)
		~ ~ \forall ~ \Omegae \in \mcth
	\,\Big\}
\end{equation}
and
\begin{equation} \label{eq_fem_disc_bmmcuh}
	\bmmcuh
	\, := \, \Big\{\,
		\bmu (\zetaalpha) : \Omega \rightarrow \mathbb{R}^3
		\quad
		\text{such that}
		\quad
		u_j \in \, \mcuh
	\,\Big\}
	~,
\end{equation}
for which
\begin{equation} \label{eq_fem_disc_v_vm_space}
	\bmvh \in \, \bmmcuh
	\qquad
	\text{and}
	\qquad
	\bmvmh \in \, \bmmcuh
\end{equation}
by construction.
In Eq.\ \eqref{eq_fem_disc_mcuh}, $ C^m (\Omega) $ denotes the space of scalar
functions on $ \Omega $ with $ m $ continuous derivatives, and
$ \mathbb{Q}_{n} (\Omegae) $ denotes the tensor-product space of
$ n^{\text{th}} $-order polynomials in the $ \zetaone $ and $ \zetatwo $
directions on the finite element $ \Omegae $.
The arbitrary variations $ \delta \bmvh $ and $ \delta \bmvmh $ are also assumed
to lie in $\bmmcuh$, and are respectively elements of the spaces
\begin{equation} \label{eq_fem_disc_delta_v_vm_space}
	\bmmcvzh
	\, := \, \bmmcvz \cap \mk\mk \bmmcuh
	\qquad
	\text{and}
	\qquad
	\bmmcvmzh
	\, := \, \bmmcvmz \cap \mk\mk \bmmcuh
	~.
\end{equation}

%
%

\subsubsection{The membrane tension, mesh pressure, and Dohrmann--Bochev
projection} \label{sec_fem_disc_db}

In general, the Lagrange multipliers $ \lambda $ and $ \ptm $ are elements of
$ L^2 (\Omega) $, and their finite-dimensional counterparts $ \lambdah $ and
$ \pmh $ need not be restricted to lie in $ \mcuh $.
However, for convenience in the numerical implementation, we do in fact choose
\begin{equation} \label{eq_fem_disc_lambda_pm_space}
	\lambdah \in \, \mcuh
	\qquad
	\text{and}
	\qquad
	\pmh \in \, \mcuh
	~.
\end{equation}
Since the same basis functions are used for the velocity and surface tension,
the well-known inf--sup condition is violated and the LBB instability arises
\cite{ladyzhenskaya-1969, babuska-nm-1973, brezzi-rfai-1974}.
We apply the Dohrmann--Bochev method \cite{dohrmann-ijnmf-2004} to stabilize
unphysical oscillations in both the surface tension and mesh pressure.
To do so, $ \lambdah $ and $ \pmh $ are projected onto the space
\begin{equation} \label{eq_fem_disc_brL}
	\brL
	\, := \, \Big\{\,
		u (\zetaalpha) : \Omega \rightarrow \mathbb{R}
		\quad
		\text{such that}
		\quad
		u \big\rvert_{\Omegae}
		\in \mathbb{P}_1 (\Omegae)
		~ ~ \forall ~ \Omegae \in \mcth
	\,\Big\}
	~,
\end{equation}
where $ \mathbb{P}_n (\Omegae) $ denotes the space of polynomials of order
$ n $ on $ \Omegae $.
Since functions in $ \brL $ form a plane over each rectangular element
$ \Omegae $, they are discontinuous across finite elements.

%
%

\subsection{The method of numerical solution} \label{sec_fem_summary}

An overview of our numerical implementation is presented below.
Additional details are provided in Appendix \ref{sec_a_fem} and the \memalefem\
package repository \cite{mem-ale-fem}.

Once the parametric domain is discretized into finite elements, all membrane
unknowns are expressed as the sum of B-spline basis functions multiplied by
membrane degrees of freedom.
The degrees of freedom are collected into a column vector and are generically
denoted $ \mbu $; they are also referred to as nodal values and their dependence
on time is implied.
Variations of membrane unknowns are similarly decomposed, with
degree-of-freedom variations compatible with all boundary conditions collected
into the column vector $ \mdeltau $.
Since the direct Galerkin expression $ \mcg $ is linear in the arbitrary
variations, its discretized counterpart $ \mcgh $ is expressed as
\begin{equation} \label{eq_fem_sum_mcgh}
	\mcgh \big( \mdeltau, \mbu \big)
	\, = \, \mdeltau^\tT \mrmu
	\, = \, 0
	\qquad
	\forall ~ \mdeltau
	~.
\end{equation}
In Eq.\ \eqref{eq_fem_sum_mcgh}, the residual vector $ \mr $ is defined to be
\begin{equation} \label{eq_fem_sum_residual}
	\mrmu
	\, := \, \pp{\mcgh \big( \mdeltau, \mbu \big)}{\mdeltau}
	~.
\end{equation}
Since the variations $ \mdeltau $ in Eq.\ \eqref{eq_fem_sum_residual} are
arbitrary, the membrane dynamics at any time
$ t \in [\mk 0, \ttf \mk] $
satisfy the nonlinear equation
\begin{equation} \label{eq_fem_sum_res_zero}
	\mrmu
	\, = \, \mzero
	~.
\end{equation}
Equation \eqref{eq_fem_sum_res_zero} is solved with the Newton--Raphson method,
where a sequence of progressively better estimates of the membrane state
$ \mbu $ is generated.
Given the $ j^{\text{th}} $ estimate $ \mbu_j $, the $ (j + 1)^{\text{th}} $
estimate is calculated according to
\begin{equation} \label{eq_fem_sum_newton_raphson}
	\mbu_{j + 1}
	\, = \, \mbu_j
	\, - \, \mK^{-1}_j \mrmuj
	~,
\end{equation}
where the global tangent diffusion matrix at the $ j^{\text{th}} $ iteration is
defined as
\begin{equation} \label{eq_fem_sum_tangent}
	\mK_j
	\, := \, \pp{\mrmu}{\mbu} \bigg\rvert_{\mbu_j}
	~.
\end{equation}

In prior studies \cite{mandadapu-jcp-2017, sahu-jcp-2020}, the tangent diffusion
matrix was calculated analytically: an involved task.
In the present work, however, $ \mK $ is calculated by numerically
differentiating $ \mr $ with respect to each entry of $ \mbu $.
In particular, the tangent matrix \eqref{eq_fem_sum_tangent} is calculated to
machine precision by extending $ \mr $ and $ \mbu $ into the complex plane,
following the general developments of \textsc{J.N.\ Lyness} and
\textsc{C.B.\ Moler} \cite{lyness-siam-1967, lyness-mc-1968} (see also e.g.\
Ref.\ \cite{tanaka-cmame-2014}).
We note that when calculating derivatives with respect to mesh velocity degrees
of freedom, one must also perturb the surface positions, as the two are related
through Eqs.\ \eqref{eq_gov_mesh_velocity} and \eqref{eq_gov_mesh_position}.

%
%

\section{The results of numerical simulations} \label{sec_sim}

We now present results from numerical simulations, where our finite element
implementation---including the various mesh motions discussed previously---was
used to simulate the dynamics of lipid membranes.
We first validate our code against a standard numerical benchmark, namely pure
bending of a lipid membrane patch.
We then simulate the drawing of a tube from a membrane sheet: a process known to
be important in various biological phenomena, including dynamic rearrangement of
the endoplasmic reticulum \cite{terasaki-jcb-1986, terasaki-cell-2013} and
tether formation by proteins traveling along microtubules
\cite{leduc-pnas-2004}.
The well-known quasi-static tether pulling behavior \cite{powers-pre-2002} is
confirmed, and the effect of pull speed on pull force is quantified.
We close by laterally translating the tether across the membrane surface with
our ALE scheme---a result which cannot be obtained with established Lagrangian
or Eulerian methods.

\begin{table}[p]
	\centering
		\caption{%
			Moments and forces on the boundary of a portion of a cylindrical membrane
			with radius $ \rc $, as shown in Fig.\ \ref{fig_sim_bend} and calculated
			in Appendix \ref{sec_a_cyl}.
			The constant surface tension $ \lambdaz $ is set by the net body force per
			area in the normal direction,
			$ f := \bmf \bmcdot \bmn $.
			In Fig.\ \ref{fig_sim_bend_boundary}, the cylindrical basis vector
			$ \bmetheta $ is tangent to the black curve and points to the left.%
		}
		\setlength{\tabcolsep}{14.5pt}
		\renewcommand{\arraystretch}{1.5}
		\begin{tabular}{c|c c c c}
			\hline
			\hline
			~
			&
			\texttt{TOP}
			&
			\texttt{BOTTOM}
			&
			\texttt{LEFT}
			&
			\texttt{RIGHT}
			\\
			\hline
			~ & ~ & ~ & ~
			\\[-13pt]
			$ M $
			&
			$
				\dfrac{\kb}{2 \mk \rc}
				\, + \, \dfrac{\kg}{\rc}
			$
			&
			$
				\dfrac{\kb}{2 \mk \rc}
				\, + \, \dfrac{\kg}{\rc}
			$
			&
			$
				\dfrac{\kb}{2 \mk \rc}
			$
			&
			$
				\dfrac{\kb}{2 \mk \rc}
			$
			\\[15pt]
			$ \bmF $
			&
			$
				\bigg[
					\dfrac{\kb}{4 \mk \rc^{\, 2}}
					\, + \, \lambdaz
				\bigg] \bmey
			$
			&
			$
				- \bigg[
					\dfrac{\kb}{4 \mk \rc^{\, 2}}
					\, + \, \lambdaz
				\bigg] \bmey
			$
			&
			$
				\bigg[
					\lambdaz
					\, - \, \dfrac{\kb}{4 \mk \rc^{\, 2}}
				\bigg] \bmetheta
			$
			&
			$
				\bigg[
					\dfrac{\kb}{4 \mk \rc^{\, 2}}
					\, - \, \lambdaz
				\bigg] \bmetheta
			$
			\\[12pt]
		\hline
		\hline
		\end{tabular}
		\label{tab_sim_bend_force_moment}
\end{table}

\begin{table}[p]
	\centering
		\caption{%
			Boundary conditions prescribed in the numerical implementation.
			We choose to set
			$ \kg = - \kb / 2 $
			and
			$ \bmf = \bm{0} $;
			the latter yields
			$ \lambdaz = \kb / (4 \mk \rc^{\, 2}) $.
			The prescribed moment $ \barM(t) $ is given by Eq.\
			\eqref{eq_sim_bend_moment_time}.
			The first three rows are repeated for $ \barvjm $ and $ \barFjm $ in the
			case of a viscous ALE mesh motion.%
		}
		\setlength{\tabcolsep}{24pt}
		\renewcommand{\arraystretch}{1.2}
		\begin{tabular}{c c c c}
			\hline
			\hline
			\texttt{TOP}
			&
			\texttt{BOTTOM}
			&
			\texttt{LEFT}
			&
			\texttt{RIGHT}
			\\
			\hline
			\\[-10pt]
			$ \barF_x = 0 $ &
			$ \barF_x = 0 $ &
			$ \barv_x = 0 $ &
			$ \barF_x = 0 $
			\\[10pt]
			$ \barv_y = 0 $ &
			$ \barv_y = 0 $ &
			$ \barv_y = 0 $ &
			$ \barF_y = 0 $
			\\[10pt]
			$ \barF_z = 0 $ &
			$ \barF_z = 0 $ &
			$ \barv_z = 0 $ &
			$ \barv_z = 0 $
			\\[10pt]
			$ \barM   = 0 $ &
			$ \barM   = 0 $ &
			$ \barM   = \barM (t) $ &
			$ \barM   = \barM (t) $
			\\[4pt]
		\hline
		\hline
		\end{tabular}
		\label{tab_sim_bend_bcs}
\end{table}

\begin{figure}[p]
	\centering
	\begin{subfigure}[b]{0.32\columnwidth}
		\centering
		\includegraphics[width=0.85\textwidth]{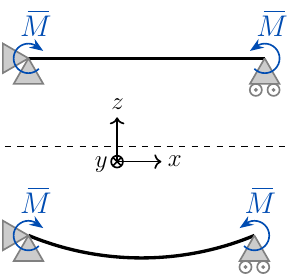}
		\caption{boundary conditions}
		\label{fig_sim_bend_boundary}
	\end{subfigure}
	\begin{subfigure}[b]{0.32\columnwidth}
		\centering
		\includegraphics[width=0.90\textwidth]{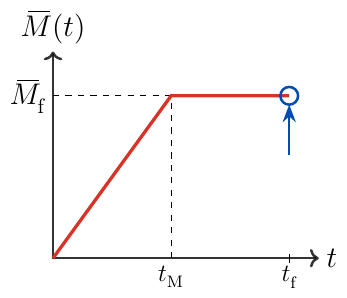}
		\caption{time dependence}
		\label{fig_sim_bend_ramp}
	\end{subfigure}
	\begin{subfigure}[b]{0.32\columnwidth}
		\centering
		\includegraphics[width=0.90\textwidth]{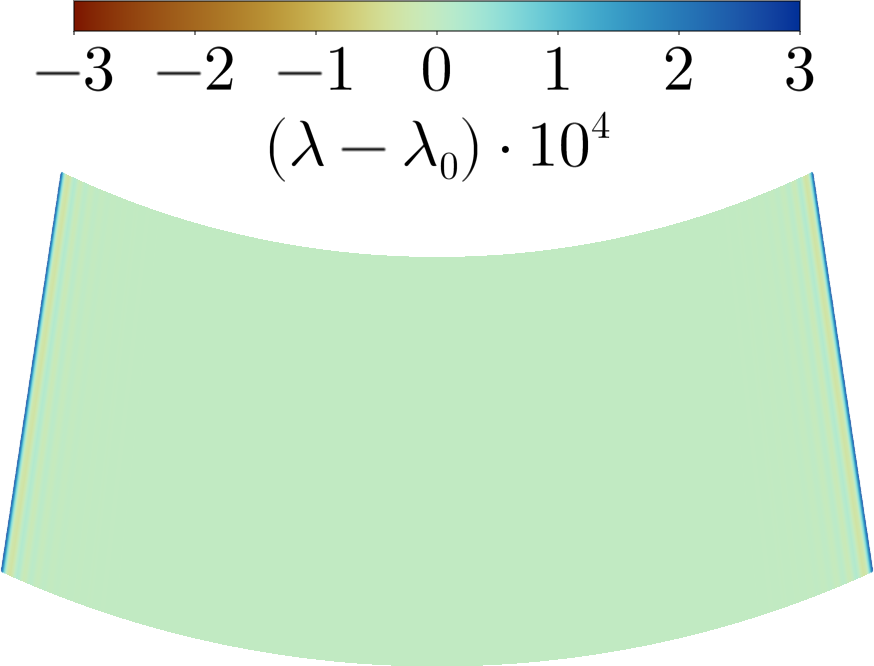}
		\\[-8pt] ~
		\caption{numerical result}
		\label{fig_sim_bend_result}
	\end{subfigure}
	\caption{%
		Pure bending scenario.
		(a) Boundary moments are applied to an initially flat patch in the $x$--$y$
		plane (top).
		The right edge is free to move in the $x$- and $y$-directions, but is
		constrained in the $z$-direction.
		At equilibrium, the membrane forms a portion of a cylinder (bottom).
		(b) Time dependence of the prescribed boundary moment, according to Eq.\
		\eqref{eq_sim_bend_moment_time}.
		Errors are calculated at the final time $ t_\tf $, as indicated by
		the blue arrow and circle.
		(c) Result of a numerical simulation on a $ 128 \times 128 $ mesh.
		The calculated surface tension differs from the analytical solution by less
		than three parts in 10$,$000, as shown by the color bar.%
	}
	\label{fig_sim_bend}
\end{figure}

%
%

\subsection{The pure bending of a flat patch} \label{sec_sim_bend}

We begin with a simple scenario: starting with a flat membrane patch, boundary
moments are applied to the left and right edges to bend the membrane, as shown
schematically in Fig.\ \ref{fig_sim_bend_boundary}.
For a given applied moment, the equilibrium configuration is known analytically
to be a portion of a cylinder \cite{mandadapu-jcp-2017, duong-mms-2017}.
The case of pure bending is thus used to validate our numerical implementation,
including the three mesh motions.
For a given observable, errors in the numerical result $ u_h $ with respect to
the known analytical solution $ u $ are calculated as
\begin{equation} \label{eq_sim_error}
	E^{\mk u}_h
	\ := \ \bigg(
		\int_\Omega \big\lvert
			u
			\, - \, u^{}_h
		\big\rvert^2
		~\td \Omega
	\bigg)^{\! 1/2}
	~,
\end{equation}
where $ h $ denotes the characteristic length of a finite element in the
parametric domain.

%
%

\subsubsection{The problem set-up} \label{sec_sim_bend_setup}

In our code, the membrane patch is initially a square of side length $ \ell $ in
the $x$--$y$ plane.
The initial velocity and position are respectively given by
\begin{equation} \label{eq_sim_bend_setup_ic_1}
	\bmv (\zetaalpha, t \le 0)
	\, = \, \bm{0}
	\qquad
	\text{and}
	\qquad
	\bmx (\zetaalpha, t \le 0)
	\, = \, x \mk \bmex + y \mk \bmey
	~,
\end{equation}
where
$ \zetaalpha \in \Omega := [\mk 0, 1] \times [\mk 0, 1] $
and the real-space coordinates $x$ and $y$ are parametrized as
\begin{equation} \label{eq_sim_bend_setup_ic_2}
	x (\zetaalpha, t \le 0)
	\, = \, \ell \, \zetaone
	\qquad
	\text{and}
	\qquad
	y (\zetaalpha, t \le 0)
	\, = \, \ell \, \zetatwo
	~.
\end{equation}
The magnitude of the applied moment, denoted $ \barM (t) $, is linearly ramped
up over time until
$ t = \ttM $%
---after which the boundary moment is held constant at the final value $\barMf$
[see Fig.\ \ref{fig_sim_bend_ramp}]:
\begin{equation} \label{eq_sim_bend_moment_time}
	\barM (t)
	\, = \, \begin{cases}
		~ \hspace{22pt}  0 &
		~                t \le 0
		\\[2pt]
		~                (t / \ttM) \, \barMf &
		~                0 < t < \ttM
		\\[6pt]
		~ \hspace{17pt}  \barMf &
		~                t \ge \ttM
		~.
	\end{cases}
\end{equation}
For times
$ t > \ttM $,
the applied boundary moment is constant in time and the membrane is bent into
a portion of a cylinder, as shown in Fig.\ \ref{fig_sim_bend_result}.
The boundary forces and moments of this stationary solution are calculated in
Appendix \ref{sec_a_cyl} and summarized in Table
\ref{tab_sim_bend_force_moment}.
Here, $ \rc $ and $ \lambdaz $ are respectively the cylinder radius and surface
tension, the latter of which is constant in the equilibrium configuration.
We choose the Gaussian bending modulus
$ \kg = - \kb / 2 $
such that
$ M = 0 $
on the top and bottom edges,
and no body force
($ \bmf = \bm{0} $)
such that
$ \lambdaz = \kb / (4 \mk \rc^{\, 2}) $
and
$ \bmF = \bm{0} $
on the left and right edges (see Table \ref{tab_sim_bend_force_moment}).
In this case, the cylinder radius $ \rc $ is related to the final boundary
moment $ \barMf $ according to
\begin{equation} \label{eq_sim_bend_moment_radius}
	\rc
	\, = \, \dfrac{\kb}{2 \mk \barMf}
	~.
\end{equation}
All of the boundary conditions prescribed in our numerical implementation for
Lagrangian and Eulerian mesh motions are provided in Table
\ref{tab_sim_bend_bcs}.
For the ALE simulations, we choose the \alev\ mesh motion discussed in
\ref{sec_gov_mesh_ale_visc}.
The first three rows of Table \ref{tab_sim_bend_bcs} are then repeated for the
mesh counterparts $ \barvjm $ and $ \barFjm $.
Recall
$ \Mm = 0 $
by construction, so there is no mesh analog of the moment boundary condition.
As a result, the membrane deforms only due to the physically applied moment
$ \barM (t) $---which is also the case for the Lagrangian and Eulerian
simulations.

%
%

\subsubsection{Non-dimensionalization} \label{sec_sim_bend_nondim}

For the case of pure bending, there are five membrane parameters: the bending
modulus $ \kb $, final bending moment $ \barMf $, ramp-up time $ \ttM $, patch
length $ \ell $, and intramembrane viscosity $ \zeta $.
These parameters all dimensionally consist of mass, length, and time.
The pure bending scenario is thus completely described by two dimensionless
quantities.
The first is the \FvK\ number, which compares surface tension to bending forces
\cite{sahu-pre-2020}.
For a cylinder with
$ \bmf  = \bm{0} $,
the membrane surface tension is constant:
$ \lambda = \lambdaz := \kb / (4 \mk \rc^{\, 2}) = \barMf^2 / \kb $,
where Eq.\ \eqref{eq_sim_bend_moment_radius} was substituted in the last
equality \cite{evans-cpl-1994}.
The \FvK\ number $ \itGamma $ is then given by
\begin{equation} \label{eq_sim_bend_fvk}
	\itGamma
	\, = \, \dfrac{\lambdaz \, \ell^2}{\kb}
	\, = \, \bigg(
		\dfrac{\barMf \, \ell}{\kb}
	\bigg)^{\! 2}
	~.
\end{equation}
The second dimensionless quantity is the Scriven--Love number $ \SL $, which
captures dynamical effects and in this scenario is set by $ \ttM $
\cite{sahu-pre-2020}.
As the characteristic velocity scale of membrane deformations is given by
$ \ell / \ttM $, the Scriven--Love number is expressed as
\begin{equation} \label{eq_sim_bend_sl}
	\SL
	\, = \, \dfrac{\ell^2 \mk \zeta}{\kb \, \ttM}
	~,
\end{equation}
which can be understood as a ratio between the fundamental membrane timescale
$ \ell^2 \zeta / \kb $
and the ramp-up time $ \ttM $.
For the results presented, we run simulations with
$ \zeta = 1.0 $,
$ \kb   = 1.0 $,
and
$ \ell  = 1.0 $.
The \FvK\ and Scriven--Love numbers are then set by choosing $ \barMf $ and
$ \ttM $, respectively.

\begin{figure}[!t]
	\centering
	\begin{subfigure}[b]{0.48\columnwidth}
		\centering
		\includegraphics[width=0.90\textwidth]{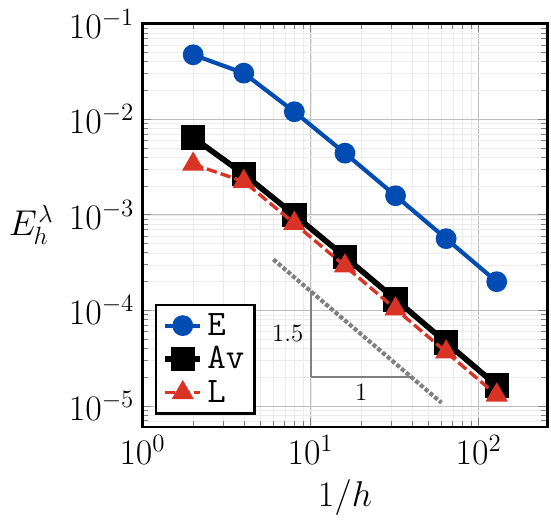}
		\caption{surface tension error}
		\label{fig_sim_bend_error_lambda}
	\end{subfigure}
	\begin{subfigure}[b]{0.48\columnwidth}
		\centering
		\includegraphics[width=0.90\textwidth]{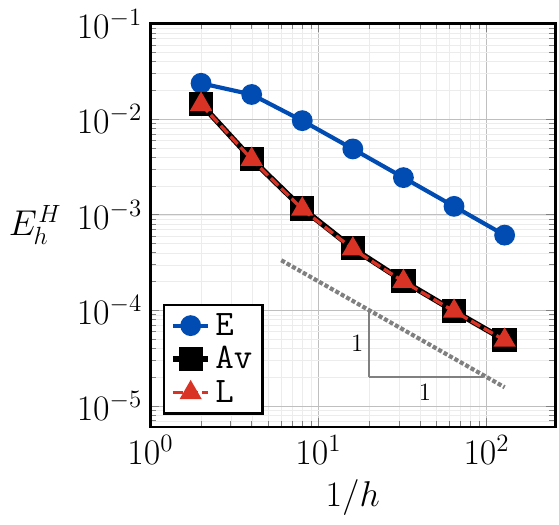}
		\caption{mean curvature error}
		\label{fig_sim_bend_error_H}
	\end{subfigure}
	\caption{%
		Errors in the (a) surface tension and (b) mean curvature at the final time
		$ \ttf $, according to Eq.\ \eqref{eq_sim_error}, when the membrane patch is
		subjected to pure bending boundary conditions.
		The mesh consists of
		$ \nel = 1 / h^2 $
		parametric area elements, each of which is a square with side length $ h $.
		Here, $ 1 / h $ ranges from $ 2^1 $ to $ 2^7 $ in powers of two.
		The labels `\texttt{E}', `\texttt{Av}', and `\texttt{L}' refer to Eulerian,
		ALE-viscous, and Lagrangian mesh motions.
		In all cases, the convergence of the error confirms our numerical
		implementation is working as expected.
		Relevant parameters are specified in \ref{sec_sim_bend_results}; we also
		choose
		$ \zeta = 1.0 $,
		$ \kb = 1.0 $,
		and
		$ \ell = 1.0 $.%
	}
	\label{fig_sim_bend_error}
\end{figure}

%
%

\subsubsection{Results} \label{sec_sim_bend_results}

We simulate pure-bending scenarios where
$ \itGamma = 0.25 $
and
$ \SL = 0.5 $,
for which
$ \barMf = \kb / (2 \mk \ell) $
and
$ \ttM = 2 \mk \ell^2 \zeta / \kb $
[see Eqs.\ \eqref{eq_sim_bend_fvk} and \eqref{eq_sim_bend_sl}].
Lagrangian, Eulerian, and ALE-viscous simulations are carried out on meshes
ranging from $ 2 \times 2 $ to $ 128 \times 128 $.
Other relevant parameters are
$ \alphaDB = \ell^2 $,
$ \alphamE = \zeta / \ell^2 $,
$ \Delta t = 0.1 \mk \ell^2 \zeta / \kb $,
and
$ \ttf = 4 \mk \ttM $.
The final configuration from one such simulation, at time $ \ttf $, is shown in
Fig.\ \ref{fig_sim_bend_result}.
The $L^2$-error in surface tension and mean curvature, relative to the
analytical solution and calculated at time $ \ttf $ according to Eq.\
\eqref{eq_sim_error}, is plotted as a function of mesh size in Fig.\
\ref{fig_sim_bend_error}.
Note that the analytical solution is valid only for a static patch; we thus
deform the membrane slowly relative to the intrinsic timescale
$ \ell^2 \zeta / \kb $
and calculate errors after it has relaxed sufficiently.
All three mesh motions converge towards the analytical solution upon mesh
refinement, indicating our numerical implementation is working as expected.
We believe the error from the Eulerian mesh motion is larger than that of
the other two schemes because boundary conditions are not prescribed for the
mesh velocity.
Instead, material velocity boundary conditions are prescribed and weakly
communicated to the mesh velocity through Eq.\ \eqref{eq_gov_mesh_eul_vm_weak};
edges are also the location where errors are the largest [see Fig.\
\ref{fig_sim_bend_result}].

%
%

\subsection{The pulling of a tether from a flat sheet} \label{sec_sim_pull}

When a single point on the membrane is displaced in the direction $ \bmn $
normal to the surface, a tent-like shape forms when deformations are small.
As the point continues to be displaced, the bilayer undergoes a nontrivial
morphological transition and forms a cylindrical tether \cite{evans-cpl-1994,
dai-bpj-1995, fygenson-prl-1997, raucher-bpj-1999}.
Tethers are known to arise in biological settings, including in the endoplasmic
reticulum \cite{terasaki-jcb-1986, scott-pnas-2024} and as the junction between
cells \cite{sowinski-ncb-2008, dubey-db-2016, imachi-n-2020}.
In addition, tethers form in \textit{in vitro} settings with optical tweezers
\cite{evans-cpl-1994, cuvelier-bpj-2005} and through the polymerization of
microtubules \cite{fygenson-prl-1997}---possibly with the use of molecular
motors \cite{roux-pnas-2002, koster-pnas-2003}.
Tether pulling is also commonly used in assays to probe both static and dynamic
membrane properties, as one can measure the force $ \bmmcfp $ required to pull
the tether \cite{cuvelier-bpj-2005, koster-prl-2005, shi-cell-2018}.
An analysis of membrane tether pulling is thus relevant to both biological and
\textit{in vitro} scenarios.
In what follows, we focus on the simplest case, in which a tether is pulled
from an initially flat sheet.


%
%

\subsubsection{The problem set-up} \label{sec_sim_pull_setup}

In numerical simulations, the membrane patch is initially a square of side
length $ \ell $ in the $x$--$y$ plane, centered at the origin.
We set
$ v_z = 0 $
on the entire patch boundary, and also pin the center node of each edge
($ \bmv = \bm{0} $)
to remove rigid body rotations and translations.
In addition, zero-slope boundary conditions are enforced by constraining
$ v_z = 0 $
for nodes adjacent to the boundary---hereafter referred to as inner boundary
nodes.
Finally, an in-plane force$/$length
$ \bmFbar_\parallel = \lambdaz \, \bmnu $
is applied on the boundary, where $ \bmnu $ is the in-plane unit normal at the
patch boundary and $ \lambdaz $ is the static surface tension---which we choose.
All boundary conditions are summarized in Table \ref{tab_sim_pull_bcs},
including those for the Eulerian and ALE schemes where $ \bmvm $ is also a
fundamental unknown.

\begin{table}[!t]
	\centering
		\caption{%
			Boundary conditions prescribed in our numerical implementation to pull a
			tether, with Lagrangian (\lag), Eulerian (\eul), and ALE-viscous-bending
			(\alevb) mesh motions.
			Here, \texttt{BDRY} refers to nodes on the boundary, \texttt{INNER} refers
			to inner nodes adjacent to the boundary, \texttt{PULL} corresponds to
			nodes associated with the central element $ \Omegaep $, and \texttt{PIN}
			refers to nodes at the center of each edge.
			The latter conditions are required to prevent rigid body rotations and
			translations.
			The \texttt{INNER} column enforces zero-slope boundary conditions.%
		}
		\setlength{\tabcolsep}{24pt}
		\renewcommand{\arraystretch}{1.2}
		\begin{tabular}{c c c c c}
			\hline
			\hline
			~
			&
			\texttt{BDRY}
			&
			\texttt{INNER}
			&
			\texttt{PULL}
			&
			\texttt{PIN}
			\\
			\hline
			\\[-10pt]
			\parbox[t]{0mm}{\multirow{2}{*}{\rotatebox[origin=c]{90}{%
				\textcolor{gray}{\lag}%
			}}} &
			$ v_z = 0 $ \hspace{-0.2pt} ~ &
			$ v_z = 0 $ &
			$ \bmv = \bmvp (t) $ &
			$ v_x = v_y = 0 $
			\\[5pt]
			&
			$ \bmFbar_\parallel = \lambdaz \mk \bmnu $ &
			~ &
			~ &
			~
			\\[8pt]
			\arrayrulecolor{gray!60}\hline
			\\[-10pt]
			\parbox[t]{0mm}{\multirow{2}{*}{\rotatebox[origin=c]{90}{%
				\textcolor{gray}{\eul}%
			}}} &
			$ v_z = 0 $ \hspace{-0.2pt} ~ &
			$ v_z = 0 $ &
			$ \bmv = \bmvp (t) $ &
			$ v_x = v_y = 0 $
			\\[8pt]
			&
			$ \bmFbar_\parallel = \lambdaz \mk \bmnu $ &
			~ &
			~ &
			$ \vm_x = \vm_y = 0 $
			\\[5pt]
			\hline
			\\[-10pt]
			\parbox[t]{0mm}{\multirow{4}{*}{\rotatebox[origin=c]{90}{%
				\textcolor{gray}{\alevb}%
			}}} &
			$ v_z = 0 $ \hspace{-0.2pt} ~ &
			$ v_z = 0 $ &
			$ \bmv = \bmvp (t) $ &
			$ v_x = v_y = 0 $
			\\[8pt]
			&
			$ \bmvm = \bm{0} $ \hspace{3.1pt} ~ &
			$ \vm_z = 0 $ \hspace{-4pt}~ &
			$ \bmvm = \bmvp (t) $ \hspace{-3.1pt} ~ &
			$ \vm_x = \vm_y = 0 $
			\\[8pt]
			&
			$ \bmFbar_\parallel = \lambdaz \mk \bmnu $ &
			~ &
			~ &
			~
			\\[5pt]
		\arrayrulecolor{black}
		\hline
		\hline
		\end{tabular}
		\label{tab_sim_pull_bcs}
\end{table}

The initial membrane state is given by
\begin{equation} \label{eq_sim_pull_setup_ic_1}
	\bmv (\zetaalpha, t \le 0)
	\, = \, \bm{0}
	\qquad
	\text{and}
	\qquad
	\lambda (\zetaalpha, t \le 0)
	\, = \, \lambdaz
	~,
\end{equation}
where
$ \zetaalpha \in \Omega := [\mk 0, 1] \times [\mk 0, 1] $.
The corresponding membrane position is expressed as
\begin{equation} \label{eq_sim_pull_setup_ic_2}
	\bmx (\zetaalpha, t \le 0)
	\, = \, x \mk \bmex + y \mk \bmey
	~,
\end{equation}
with
\begin{equation} \label{eq_sim_pull_setup_ic_3}
	x (\zetaalpha, t \le 0)
	\, = \, \big(
		\zetaone
		\, - \, 0.5
	\big) \, \ell
	\qquad
	\text{and}
	\qquad
	y (\zetaalpha, t \le 0)
	\, = \, \big(
		\zetatwo
		\, - \, 0.5
	\big) \, \ell
	~.
\end{equation}
At time
$ t = 0 $,
we seek to vertically displace the center of the membrane patch, where
$ x = 0 $
and
$ y = 0 $.
However, as shown in Appendix \ref{sec_a_pull_num}, there is no unique way to
specify the velocity at a single point given our use of B-spline basis
functions, as they are not interpolatory.
Instead, we vertically displace the portion of the membrane corresponding to the
entire parametric element $ \Omegaep $ containing the point
$ \zetaalphap := (0.5, 0.5) $
at the center of the parametric domain:
\begin{equation} \label{eq_sim_pull_constraint}
	\bmv (\zetaalpha, t \ge 0)
	\, = \, \bmvp (t)
	\qquad
	\forall ~ \zetaalpha \in \Omegaep
	~.
\end{equation}
In Eq.\ \eqref{eq_sim_pull_constraint}, $ \bmvp (t) $ is a known function of
time.
It is given by $ \vp \mk \bmez $, for constant $ \vp $, unless otherwise
specified.
Following the derivations in Appendix \ref{sec_a_pull_num}, Eq.\
\eqref{eq_sim_pull_constraint} is enforced by setting all nodal velocity degrees
of freedom associated with $ \Omegaep $ to $ \bmvp (t) $.
The resultant pull force $ \bmmcfp (t) $ is calculated as the sum of the
corresponding components of the residual vector.

%
%

\subsubsection{Non-dimensionalization} \label{sec_sim_pull_nondim}

In the scenario under consideration, there are five membrane parameters: the
bending modulus $ \kb $, equilibrium surface tension $ \lambdaz $,
two-dimensional intramembrane viscosity $ \zeta $, patch length $ \ell $, and
speed of tube drawing $ \vp $.%
\footnote{%
	We hope to include the hydrodynamics of the surrounding fluid, including
	membrane permeability \cite{alkadri-arxiv-2024}, and the effects of
	cytoskeletal contacts \cite{shi-cell-2018} in a future effort---and better
	compare our results with experiments
	\cite{cuvelier-bpj-2005, brochard-pnas-2006}.%
}
Since the five quantities dimensionally involve only mass, length, and time,
two dimensionless numbers once again determine the evolution of the system.
The \FvK\ number $ \itGamma $ is given by \cite{sahu-pre-2020}
\begin{equation} \label{eq_sim_pull_fvk}
	\itGamma
	\, := \, \dfrac{\lambdaz \, \ell^2}{\kb}
	~,
\end{equation}
and can be interpreted in two ways given the morphological changes that occur
when pulling a tether.
First, when the membrane is nearly planar and height deflections are small,
gradients in membrane shape occur over the length scale $ \ell $; $ \itGamma $
then quantifies the relative magnitude of tensile and bending forces.
In our simulations, $ \itGamma^{-1} $ is small, and consequently a boundary
layer of characteristic width
$ \sqrtl{ \kb / \lambdaz } \ll \ell $
develops at the point of application of the pull force \cite{powers-pre-2002}.
As the membrane is pulled further, a tether grows from the boundary layer
region.
The tether is close in shape to a cylinder, and the radius of a cylindrical
membrane is well-known to be
\cite{zhong-can-pra-1989}
\begin{equation} \label{eq_sim_pull_cylinder_radius}
	\rc
	\, := \, \sqrt{ \dfrac{\kb}{4 \mk \lambdaz} \, }
	~.
\end{equation}
Equation \eqref{eq_sim_pull_cylinder_radius} can be obtained from energetic
arguments alone, or equivalently from a balance of bending and tensile forces.%
\footnote{%
	In situations where there is a pressure jump across the membrane due to the
	surrounding fluid and
	$ f = \bmf \bmcdot \bmn \ne 0 $,
	Eqs.\ \eqref{eq_sim_pull_cylinder_radius} and \eqref{eq_sim_pull_force_eq} are
	not valid.
	See the discussion in Ch.\ IX, \S1(a) of Ref.\ \cite{sahu-thesis}.
}
With Eq.\ \eqref{eq_sim_pull_cylinder_radius}, the \FvK\ number
\eqref{eq_sim_pull_fvk} is understood as the ratio
\begin{equation} \label{eq_sim_pull_fvk_radius}
	\itGamma
	\, = \, \bigg(
		\dfrac{\ell}{2 \mk \rc}
	\bigg)^{\! 2}
	~.
\end{equation}
Equations \eqref{eq_sim_pull_fvk}--\eqref{eq_sim_pull_fvk_radius} confirm that
the \FvK\ number describes membrane energetics, as only lengths and the
parameters $ \kb $ and $ \lambdaz $---which respectively have dimensions of
energy and energy per area---are involved.

The second dimensionless quantity is the Scriven--Love number $ \SL $, which
compares the magnitude of viscous and bending forces in shaping the membrane
\cite{sahu-pre-2020}.
In this scenario, we define the Scriven--Love number as
\begin{equation} \label{eq_sim_pull_sl}
	\SL
	\, := \, \dfrac{\rc \, \zeta \mk \vp}{\kb}
	~.
\end{equation}
We chose to include $ \rc $---rather than $ \ell $---in
Eq.\ \eqref{eq_sim_pull_sl} because we are primarily concerned with the behavior
of the tether, rather than the entire patch.
Upon substituting Eq.\ \eqref{eq_sim_pull_cylinder_radius} into
Eq.\ \eqref{eq_sim_pull_sl}, rearranging terms, and recognizing
$ \zeta / \lambdaz $ is the fundamental timescale of lipid flows
\cite{sahu-pre-2020}, we find the Scriven--Love number can be equivalently
expressed as
\begin{equation} \label{eq_sim_pull_sl_speed}
	\SL
	\, := \, \dfrac{\vp}{4 \mk \rc \mk ( \zeta / \lambdaz )^{-1} }
	~,
\end{equation}
i.e.\ a ratio of the speed at which the tether is pulled to the natural velocity
scale of lipid flows in the tube \cite{sahu-pre-2020}.
Thus, as $ \SL $ tends to zero, we approach the quasi-static limit.

In our source code, when running tether pulling simulations, we choose
$ \zeta = 1.0 $,
$ \kb = 1.0 $,
and
$ \lambdaz = 0.25 $.
Our choice yields a tube radius
$ \rc = 1.0 $
according to Eq.\ \eqref{eq_sim_pull_cylinder_radius}.
The \FvK\ number $ \itGamma $ is altered by varying the patch size $ \ell $,
while the Scriven--Love number $ \SL $ is modified by changing the pull speed
$ \vp $.

%
%

\subsubsection{The comparison of different mesh motions}
\label{sec_sim_pull_compare}

\begin{figure}[p]
	\centering
	\includegraphics[width=0.99\textwidth]{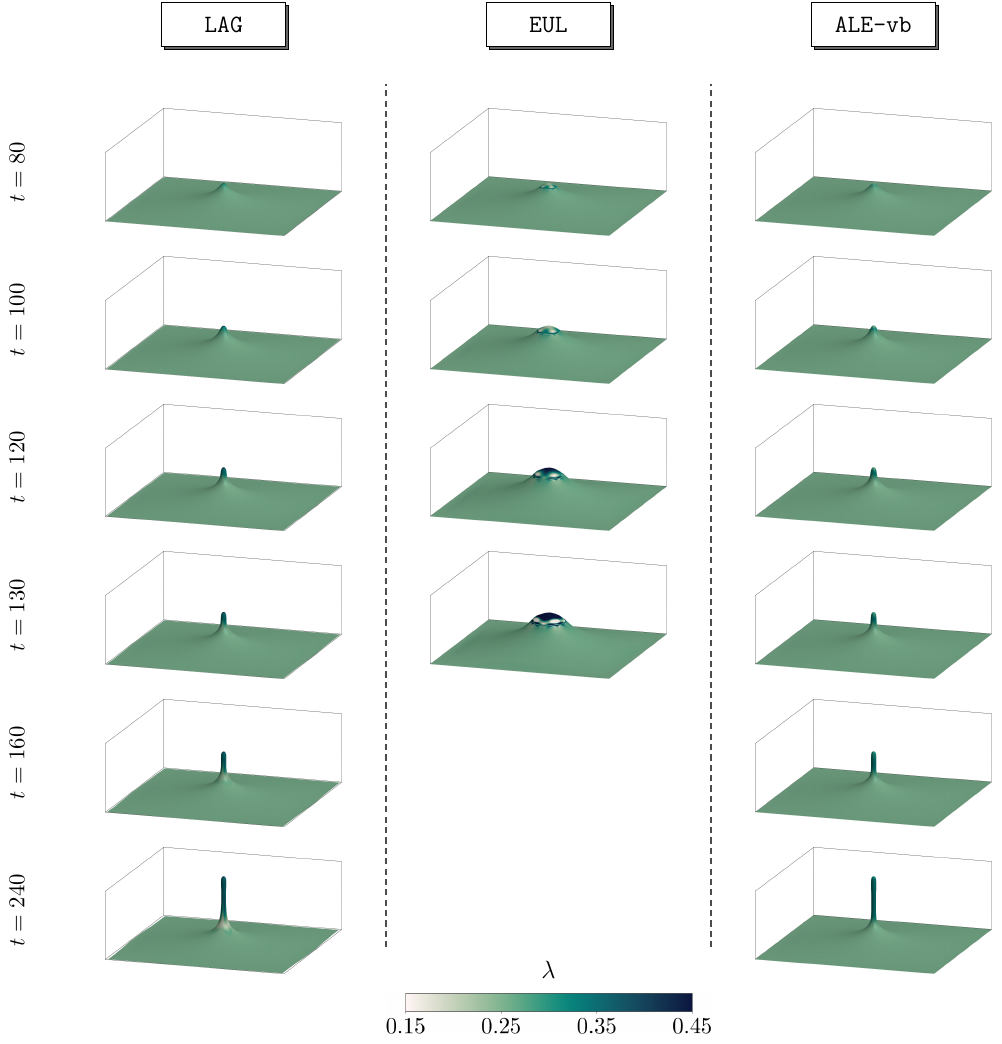}
	\vspace{-1pt}
	\caption{%
		Snapshots from tether pulling simulations with a $ 65 \times 65 $ mesh, in
		which three different mesh motions are employed.
		The color bar indicates the surface tension, in units of $ \kb / \rc^2 $,
		for all snapshots.
		Times are measured in units of $ \zeta \rc^2 / \kb $.
		(left) Lagrangian simulations (\lag) successfully generate a tether.
		The morphological shape change from a tent to a tube occurs around time
		$ t = 120 $.
		Since the membrane is area-incompressible, the patch boundary must be pulled
		inwards to accommodate the tether surface area.
		(center) The Eulerian mesh motion (\eul) is not able to form a tube.
		Rather, the tent morphology persists and bulges outward until the method
		fails around time
		$ t = 143 $.
		The material velocity degrees of freedom of the central
		element are constrained to move upwards; no such constraint is placed on the
		mesh velocity degrees of freedom.
		(right) An ALE mesh motion that is viscous and resists bending (\alevb)
		successfully forms a tether.
		Both material and mesh velocities of the central element are constrained to
		move upwards.
		Since the mesh is area-compressible, the patch boundary can be constrained
		to remain stationary as the tether develops.
		For all three mesh motions,
		$ \Delta t = 0.5 $,
		$ \SL = 0.1 $,
		and
		$ \itGamma = 1024 $,
		for which
		$ \ell = 64 \, \rc $
		and
		$ \vp = 0.1 \, \kb / (\zeta \rc ) $
		according to Eqs.\ \eqref{eq_sim_pull_fvk_radius} and
		\eqref{eq_sim_pull_sl}.
		See also Movies \hyperref[movie_pull_lag]{1}--\hyperref[movie_pull_ale]{3}.%
	}
	\label{fig_sim_pull_schemes}
\end{figure}

\begin{figure}[p]
	\centering
	~\hfill\includegraphics[width=0.89\textwidth]{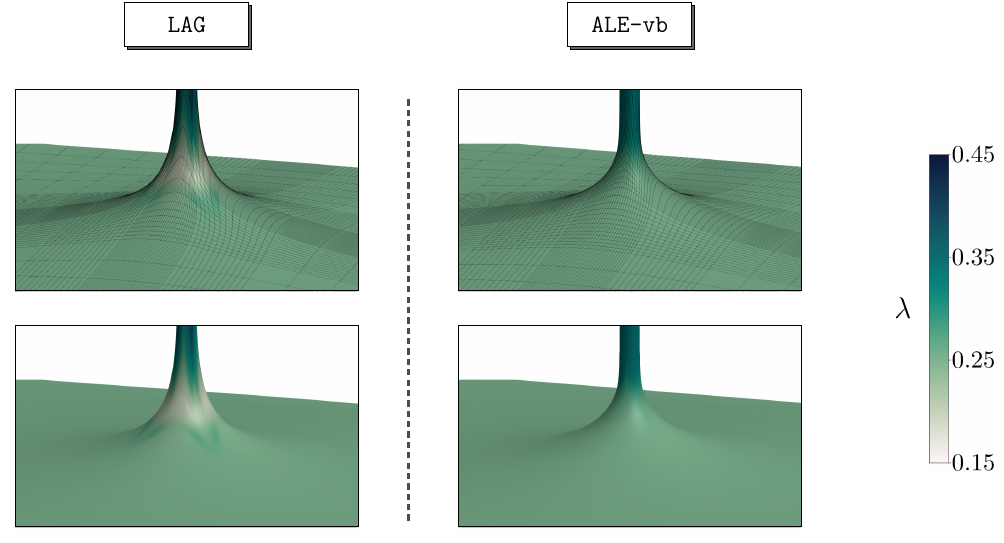}
	\vspace{3pt}
	\caption{%
		Zoomed-in views of the
		$ t = 240 \, \zeta \rc^2 / \kb $
		snapshots from Fig.\ \ref{fig_sim_pull_schemes}, with the underlying mesh
		shown (top) and hidden (bottom).
		In the Lagrangian simulations (left), the mesh is drawn into the tube along
		with the lipids due to the areal incompressibility of the membrane.
		Mesh elements close to the diagonal of the square patch, where
		$ y = \pm \mk x $, become highly distorted---which leads to an artificially
		low surface tension in the transition region between the tether and
		surrounding membrane.
		Numerical artifacts are visible in the striation of the tension.
		The ALE-viscous-bending result (right), in contrast, shows a less distorted
		mesh because the choice of mesh constitution resists both shear and
		dilation.
		No surface tension artifacts are visible, and a smooth tension gradient is
		observed from the flat patch into the tether.%
	}
	\label{fig_sim_pull_zoom}
\end{figure}

\begin{figure}[p]
	\centering
	\includegraphics[width=0.40\textwidth]{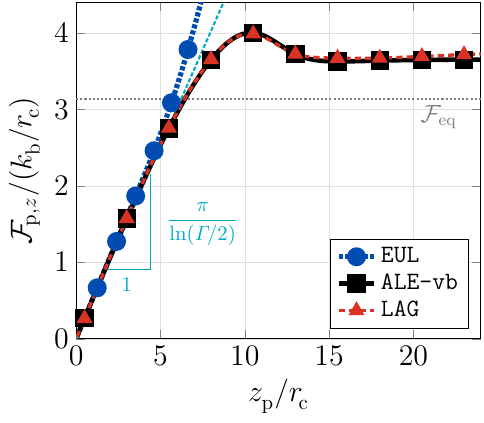}
	\vspace{3pt}
	\caption{%
		The $ z $-component of the pull force ($ \mcfpz $) as a function of the
		$ z $-displacement.
		The dashed cyan line is the result from the linear theory, as presented in
		Eq.\ \eqref{eq_sim_pull_force_linear}.
		The Eulerian simulation (blue circles, dotted line) is unable to form a
		tether, and is unphysical.
		The Lagrangian (red triangles, dashed line) and ALE (black squares, solid
		line) simulations capture the tent-to-tube transition, after which the
		Lagrangian steady-state pull force $ \mcfss $ is slightly larger.
		Both overshoot the equilibrium pull force $ \mcfeq $ 
		\eqref{eq_sim_pull_force_eq} due to dynamical effects from tether pulling
		(see \ref{sec_sim_pull_geodyn}).
		Numerical parameters are those specified in Fig.\
		\ref{fig_sim_pull_schemes}.%
	}
	\label{fig_sim_pull_force_scheme}
\end{figure}

We begin by pulling a tether in a scenario with \FvK\ number
$ \itGamma = 1024 $
and Scriven--Love number
$ \SL = 0.1 $.
Given these dimensionless numbers,
$ \ell = 64 \mk \rc $
and
$ \vp = 0.1 \mk \kb / (\zeta \mk \rc) $
according to Eqs.\ \eqref{eq_sim_pull_fvk_radius} and \eqref{eq_sim_pull_sl}.
Snapshots from tether-pulling simulations with each mesh motion are shown in
Fig.\ \ref{fig_sim_pull_schemes}, with corresponding videos in the \memalefem\
package repository \cite{mem-ale-fem}.
A zoomed-in view of the late-time snapshots, with and without the underlying
mesh, are shown in Fig.\ \ref{fig_sim_pull_zoom} to emphasize the advantages of
the ALE scheme.
In addition, Fig.\ \ref{fig_sim_pull_force_scheme} shows how the $ z $-component
of the pull force,
$ \mcfpz := \bmmcfp \bmcdot \bme_z $,
varies as a function of the vertical displacement $ \zp $.
We note that in the quasi-static limit
($ \SL \rightarrow 0 $),
the steady-state pull force of a perfect cylinder is given by
\cite{evans-cpl-1994}%
\footnote{%
	Some studies define
	$ \kappa := \kb / 2 $
	to be the bending modulus, in which case
	$ \mcfeq = 2 \pi \sqrtl{ 2 \mk \kappa \mk \lambdaz } $.%
}
\begin{equation} \label{eq_sim_pull_force_eq}
	\mcfeq
	\, := \, \dfrac{\pi \mk \kb}{\rc}
	\, = \, 2 \mk \pi \sqrtl{ \kb \mk \lambdaz }
	~.
\end{equation}
Since a tether pulled from a flat patch deviates slightly from a cylinder
\cite{powers-pre-2002}, Eq.\ \eqref{eq_sim_pull_force_eq} serves as an excellent
approximation for $ \mcfpz $ when
$ \SL = 0 $.

In what follows, we comment on the efficacy of the three mesh motions:
Lagrangian (\lag), Eulerian (\eul), and ALE-viscous-bending (\alevb).
Note that additional variables are used to solve for membrane behavior in the
latter two schemes: the fundamental variables are
$ \bmv $ and $ \lambda $ (\lag);
$ \bmv $, $ \bmvm $, and $ \lambda $ (\eul);
and
$ \bmv $, $ \bmvm $, $ \lambda $, and $ \ptm $ (\alevb).
We present the number of degrees of freedom for each motion, as well as the wall
clock run time, in Table \ref{tab_sim_pull_benchmark}.
When appropriate, our results are compared to those of prior theoretical and
numerical developments.


\bigskip

\noindent
\textbf{Lagrangian scheme.---}%
We first consider results from simulations with a Lagrangian mesh motion, as
shown in the left column of Figs.\ \ref{fig_sim_pull_schemes} and
\ref{fig_sim_pull_zoom} (see also Movie \hyperref[movie_pull_lag]{1}).
A tent-like structure develops and grows until the vertical displacement
$ \zp := \vp \mk t \approx 8 \mk \rc $,
for which
$ t \approx 80 \, \zeta \mk \rc^{\, 2} / \kb $.
Around this point, the linear relationship between $ \mcfpz $ and $ \zp $ breaks
down, as shown in Fig.\ \ref{fig_sim_pull_force_scheme}.
The membrane undergoes a morphological transition between
$ t \approx 80 \, \zeta \mk \rc^{\, 2} / \kb $
and
$ t \approx 120 \, \zeta \mk \rc^{\, 2} / \kb $,
during which $ \mcfpz $ reaches its largest value.
After the shape transition, a tether continues to be drawn at an approximately
constant force, which we refer to as the steady-state force $ \mcfss $ (see
Fig.\ \ref{fig_sim_pull_force_scheme}).
However, a pronounced area of low surface tension develops in the region where
the tether meets the flat patch (see Figs.\ \ref{fig_sim_pull_schemes} and
\ref{fig_sim_pull_zoom}).  We comment on the unphysical nature of this
development following a presentation of our ALE results.

The majority of our findings from Lagrangian tether-pulling simulations, as well
as the Lagrangian implementation itself, are not new.
\textsc{T.R.\ Powers} \textit{et al.} \cite{powers-pre-2002} provided a detailed
account of the axisymmetric, quasi-static membrane shape in response to vertical
displacements, and along with Ref.\ \cite{derenyi-prl-2002} numerically
calculated the force versus displacement curve in the quasi-static limit.
More recently, rate effects were included in axisymmetric simulations that
focused primarily on the slip between two monolayer leaflets
\cite{rahimi-pre-2012}---an effect not considered in the present work.
General non-axisymmetric Lagrangian formulations were developed since then.
The implementation in Ref.\ \cite{rodrigues-jcp-2015} included rate effects when
pulling a tether, but the force was not provided as a function of displacement
and it is unclear if the membrane underwent the morphological
transition from a tent to a tube in simulations.
In contrast, Ref.\ \cite{mandadapu-jcp-2017} pulled a tether from the center of
an initially axisymmetric mesh and illustrated the tent-to-tube transition.
While this implementation seems to be capable of capturing rate effects, the
reported tether-pulling results employed a numerical stabilization scheme that
did not capture the coupling between in-plane lipid flows and out-of-plane
forces.
Thus, to the best of our knowledge, we present the first non-axisymmetric
Lagrangian simulation capturing the dynamics of a tether pulled from a membrane
patch---though we do not consider the results to be novel, as Refs.\
\cite{rodrigues-jcp-2015} and \cite{mandadapu-jcp-2017} may have been able to do
the same.


\bigskip

\noindent
\textbf{Eulerian scheme.---}%
Let us next examine results from an Eulerian mesh motion.
As shown in Movie \hyperref[movie_pull_eul]{2} and the center column of Fig.\
\ref{fig_sim_pull_schemes}, the numerical implementation fails to form a tether.
In addition, unphysical gradients develop in the surface tension, and the code
fails at time
$ t = 143 \, \zeta \mk \rc^{\, 2} / \kb $.
In what follows, we discuss why such a failure is to be expected, as a tether
cannot form when the mesh motion is Eulerian.

\begin{table}[!t]
	\centering
		\caption{%
			Number of degrees of freedom (\texttt{dofs}) and wall clock run time for
			the three different mesh motions, corresponding to the results presented
			in Fig.\ \ref{fig_sim_pull_schemes}.
			All computations were carried out on a single node of the
			\texttt{Perlmutter} system at the National Energy Research Scientific
			Computing Center, with area element calculations distributed across 32
			cores.%
		}
		\setlength{\tabcolsep}{14.5pt}
		\renewcommand{\arraystretch}{1.5}
		\begin{tabular}{c c c}
			\hline
			\hline
			mesh motion
			&
			~\# \texttt{dofs}~
			&
			wall clock time
			\\
			\hline
			~ & ~ & ~
			\\[-16pt]
			\texttt{LAG}    & 17{,}401 & 271 minutes~
			\\
			\texttt{EUL}    & 30{,}833 & 525 minutes$^\text{a}$
			\\
			\texttt{ALE-vb} & 34{,}282 & 636 minutes~
			\\[3pt]
			\hline
			\hline
		\end{tabular}
		\\[2pt]
		{\footnotesize$^\text{a}$ Scaled to the total number of time steps, as the
		simulation failed.}
		\label{tab_sim_pull_benchmark}
\end{table}

We begin by introducing $ J $ and $ \Jm $ as the relative area dilations of the
membrane and mesh, respectively.
These dilations are related to the corresponding flow fields according to
\cite[Ch.V\,\S1(c)]{sahu-thesis}
\begin{equation} \label{eq_sim_pull_eul_J_dot}
	\dfrac{\dot{J}}{J}
	\, = \, v^\calpha_{; \calpha}
	\, - \, 2 \mk v H
	\qquad
	\text{and}
	\qquad
	\dfrac{\dot{J}^\tm}{\Jm}
	\, = \, (v^\calpha_{\tm})_{ ; \, \calpha}
	\, - \, 2 \mk \vm \mkn H
	~.
\end{equation}
Since the membrane is incompressible,
$ \dot{J} / J = 0 $
and
$ v^\calpha_{; \calpha} = 2 \mk v H $.
To understand how the mesh dilates or compresses, we recognize that
$ \vm = v $
and
$ \vam = 0 $
according to Eq.\ \eqref{eq_gov_mesh_eul_vm_strong}.
We thus find
\begin{equation} \label{eq_sim_pull_eul_mesh_compr}
	\dfrac{\dot{J}^\tm}{\Jm}
	\, = \, - 2 \mk v H
	\, \ne \, 0
	~.
\end{equation}
At this point, we make three observations regarding the geometry and dynamics of
the initial tent formation:
($i$) $ v \ge 0 $ everywhere,
($i\mkn i$) $ H < 0 $ in a central region where the surface is concave down, and
($i\mkn i\mkn i$) $ H \ge 0 $ elsewhere.
Accordingly,
$ \dot{J}^\tm / \Jm > 0 $
in the center of the tent, and the mesh continuously dilates.
In this manner, the mesh expands laterally and under-resolves the region where
the morphological transition would occur.

\begin{figure}[t!]
	\centering
	\begin{subfigure}[b]{0.385\columnwidth}
		\centering
		\caption{\hfill~}
		\vspace{-10pt}
		\includegraphics[width=\textwidth]{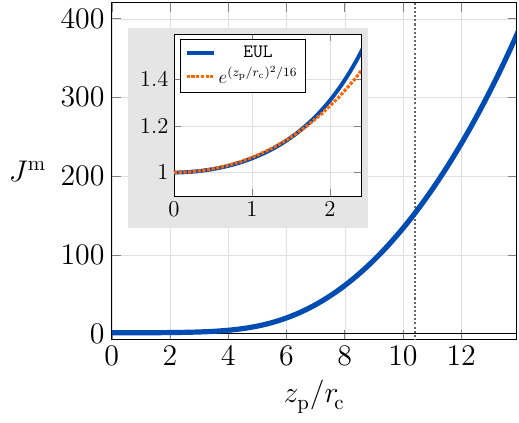}
		\label{fig_sim_pull_eul_axis}
	\end{subfigure}
	\hspace{60pt}
	\begin{subfigure}[b]{0.38\columnwidth}
		\centering
		\caption{\hfill~}
		\vspace{-12pt}
		\includegraphics[width=\textwidth]{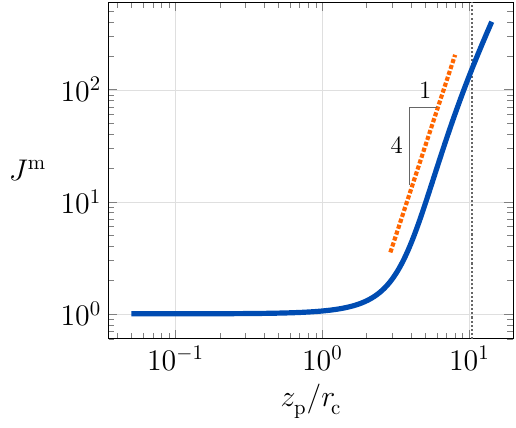}
		\label{fig_sim_pull_eul_log}
	\end{subfigure}
	\vspace{-16pt}
	\caption{%
		Mesh dilation of the Eulerian simulation shown in Fig.\
		\ref{fig_sim_pull_schemes}.
		(a) The relative area change at the patch center, $ \Jm $, is plotted as a
		function of the displacement $ \zp $.
		The dashed vertical line is the displacement at which a tent is expected to
		transition to a tether, according to Lagrangian and ALE data in Fig.\
		\ref{fig_sim_pull_force_scheme}.
		(inset) At small displacements, simulation data agrees with an approximate
		solution to Eq.\ \eqref{eq_sim_pull_eul_mesh_compr}.
		Here
		$ v = \dot{z}_{\tp} $;
		if the mean curvature is linear in $ \zp $ then
		$ H \sim - \zp / \rc^{\, 2} $
		and integrating Eq.\ \eqref{eq_sim_pull_eul_mesh_compr} yields
		$ \ln \Jm \sim (\zp / \rc)^2 $.
		The number 16 in the analytical expression is a fitting parameter.
		(b) The same data in a log--log plot suggests a power-law growth of the
		dilation at large displacements, prior to the expected morphological
		transition (dotted vertical line)---though the data does not span even a
		single decade.
	}%
	\label{fig_sim_pull_eul}
\end{figure}

Mesh dilation at the patch center, where
$ \zeta^\cone = 0.5 $
and
$ \zeta^\ctwo = 0.5 $
[see Eq.\ \eqref{eq_sim_pull_setup_ic_3}], is quantified in Fig.\
\ref{fig_sim_pull_eul}.
Here, the relative area change of the mesh $ \Jm $ is plotted as a function of
the membrane displacement $ \zp $.
The dilation by two orders of magnitude suggests an Eulerian simulation with a
similar increase in the number of elements.
Such a patch, even after the large dilation, may be sufficiently resolved to
undergo the morphological transition to a tube.
However, even if a tether was formed, it likely could not be extended with an
Eulerian mesh motion: lipid flows are primarily in-plane during tether
elongation, so the Eulerian mesh would remain stationary and the tether would be
poorly resolved.
We thus believe the inability to pull a tether is a general failure of Eulerian
methods, including those implemented previously \cite{sahu-jcp-2020,
reuther-jfm-2020}.
Moreover, since the ALE implementation of Ref.\ \cite{torres-jfm-2019} employed
a mesh motion that was close to Eulerian, we are unsure if it would be able to
successfully pull a tether.

\bigskip

\noindent
\textbf{ALE scheme.---}%
We close by discussing the results of our ALE mesh motion.
As shown in Figs.\ \ref{fig_sim_pull_schemes}--\ref{fig_sim_pull_force_scheme}
and Movie \hyperref[movie_pull_ale]{3}, a tether was successfully pulled with
the \alevb\ scheme, and the surface tension in the region where the tether meets
the flat patch was approximately constant.
This latter point is quantified by considering the range of surface tension
values at time
$ t = 240 \, \zeta \mk \rc^{\, 2} / \kb $.
In the ALE simulation, the lowest (dimensionless) value of the surface tension
is $ 0.249 $---approximately equal to the magnitude of the in-plane force per
length
$ \lvert \bmFbar_\parallel \rvert = \lambdaz = 0.25 $
maintained on the boundary.
Larger tension values, in this case up to $ 0.345 $, arise in the tether to draw
in lipids during the dynamic process of pulling.
In contrast, the minimum value of the tension in Lagrangian simulations is
$ 0.191 $: well below $ \lvert \bmFbar_\parallel \rvert $ and also less than the
larger values (up to $ 0.446 $) attained on the tether.
Since lipids flow from regions of low to high tension, there does not seem to be
a smooth flow of lipids from the flat patch into the tether.
We thus find the ALE-viscous-bending mesh motion to be superior to its
Lagrangian counterpart, and report only \alevb\ results for the remainder of our
tether-pulling analysis.

It is important to note that while the \alevb\ scheme successfully pulls a
tether, the purely viscous ALE mesh motion---employed in \ref{sec_sim_bend}---%
does not: as the center of the patch is translated upwards, the tether tapers to
a point.
We believe an angular shape arises in the \alev\ simulation because of
incompatible boundary conditions for the material and mesh.
More specifically, since all nodes over a single element are translated upwards,
a zero-slope condition is implicitly prescribed at the element boundary.
Both the membrane and \alevb\ mesh can sustain such a requirement due to their
inherent bending stiffness.
In contrast, the \alev\ mesh has no bending rigidity and so cannot maintain zero
slope on the element boundary.
The tether resulting from the \alev\ scheme is consequently unphysical.

%
%

\subsubsection{The geometry and dynamics of tether pulling}
\label{sec_sim_pull_geodyn}

In comparing different mesh motions in \ref{sec_sim_pull_compare}, all
simulations were carried out at a single choice of $ \itGamma $ and $ \SL $.
We now investigate the effects of altering the patch size relative to the tether
radius, as well as changing the speed of tether pulling---which respectively
modify the \FvK\ and Scriven--Love numbers.
We confirm that $ \itGamma $ dictates the initial slope of the force versus
displacement curve, as previously observed \cite{powers-pre-2002,
derenyi-prl-2002, mandadapu-jcp-2017}, while $ \SL $ captures the overshoot
of of the tether pull force relative to the equilibrium (or quasi-static)
result.

\begin{figure}[p]
	\centering
	\begin{subfigure}[b]{0.48\columnwidth}
		\centering
		\caption{\hfill~}
		\includegraphics[width=0.90\textwidth]{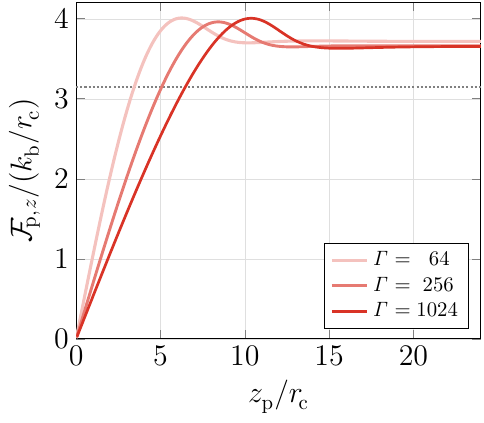}
		\label{fig_sim_pull_gamma}
	\end{subfigure}
	\hfill
	\begin{subfigure}[b]{0.48\columnwidth}
		\centering
		\caption{\hfill~}
		\includegraphics[width=0.90\textwidth]{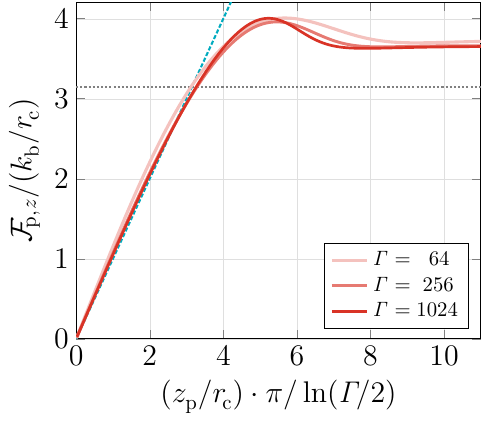}
		\label{fig_sim_pull_gamma_collapse}
	\end{subfigure}
	\\[20pt]
	\begin{subfigure}[b]{0.48\columnwidth}
		\centering
		\caption{\hfill~}
		\includegraphics[width=0.90\textwidth]{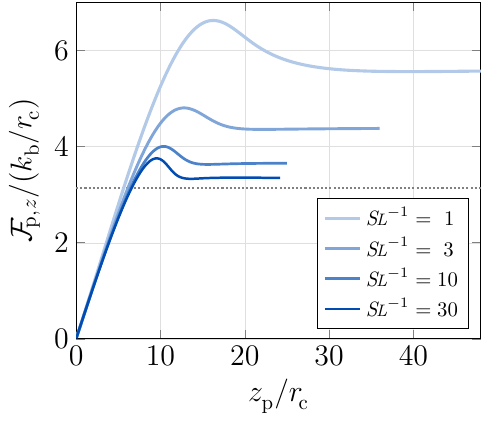}
		\label{fig_sim_pull_sl}
	\end{subfigure}
	\hfill
	\begin{subfigure}[b]{0.48\columnwidth}
		\centering
		\caption{\hfill~}
		\includegraphics[width=0.95\textwidth]{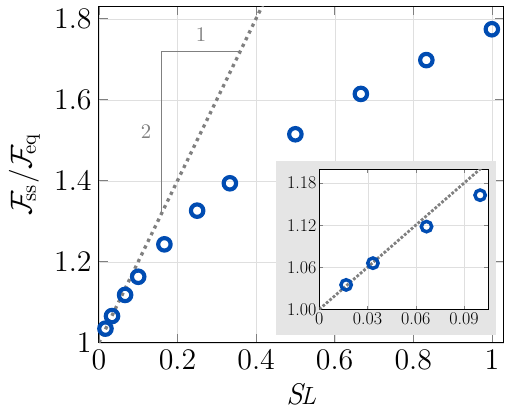} \hfill ~
		\\[-10pt]
		~
		\label{fig_sim_pull_sl_ss}
	\end{subfigure}
	\caption{%
		The $ z $-component of the pull force ($ \mcfpz $) as a function of $ \zp $
		for different values of $ \itGamma $ and $ \SL $.
		(a) When
		$ \SL = 0.1 $ and $ \itGamma $ is varied, the initial slope of the force-vs%
		-displacement curve is altered according to the linear theory [see Eq.\
		\eqref{eq_sim_pull_force_linear}].
		(b) By scaling the $ z $-displacement appropriately, the data collapses---%
		with the steady-state pull force independent of $ \itGamma $ after tether
		formation.
		The cyan line has slope one.
		(c) When
		$ \itGamma = 1024 $
		and $ \SL $ is varied, the initial slope of the force-vs-displacement curve
		is unchanged.
		The tent-to-tube transition occurs at larger displacements, and the long-%
		time pull force increases with $ \SL $.
		(d) Long-time pull force, $ \mcfss $, as a function of $ \SL $ for
		$ \itGamma = 1024 $.
		The dotted grey line is the linear prediction from Eq.\
		\eqref{eq_sim_pull_force_sl_small}, which agrees with simulation results
		when
		$ \SL \ll 1 $%
		---as highlighted by the zoomed-in inset.
		The nonlinear dependence of $ \mcfss $ on $ \SL $ arises from the coupling
		between in-plane flows and out-of-plane shape deformations.%
	}
	\label{fig_sim_pull_geodyn}
\end{figure}

We first investigate the dependence of the pull force on the patch geometry.
Figure \ref{fig_sim_pull_gamma} plots $ \mcfpz $ as a function of $ \zp $ for
different values of $ \itGamma $, at fixed
$ \SL = 0.1 $.
We observe that $ \itGamma $ alters the initial slope of the force versus
displacement curve, but does not affect the steady-state pull force after the
tether is formed---the latter of which is expected to be independent of the
patch size (see Eq.\ \eqref{eq_sim_pull_force_eq} and Refs.\
\cite{powers-pre-2002, derenyi-prl-2002}).
To calculate the initial dependence of $ \mcfpz $ on $ \zp $, we analyze the
tent-like membrane shape when deformations are small.
In this limit, the shape equation \eqref{eq_gov_mem_lin_mom_shape} is decoupled
from in-plane lipid flows, and identical to its quasi-static counterpart
\cite{sahu-pre-2020}.
We thus take the small-deformation, quasi-static membrane tent result from Ref.\
\cite{powers-pre-2002} and calculate the pull force to be given by [cf.\ Eq.\
\eqref{eq_sim_pull_force_eq}]
\begin{equation} \label{eq_sim_pull_force_linear}
	\dfrac{\mcfpz}{\kb / \rc}
	\ = \ \dfrac{\pi}{\ln (\itGamma \! / 2)} \cdot \dfrac{\zp}{\rc}
	~,
	\qquad\quad
	\text{or equivalently}
	\qquad\quad
	\dfrac{\mcfpz}{\mcfeq}
	\ = \ \dfrac{1}{\ln (\itGamma \! / 2)} \cdot \dfrac{\zp}{\rc}
	~,
\end{equation}
which is shown as the dashed cyan line in Fig.\ \ref{fig_sim_pull_force_scheme}.
Our calculation of Eq.\ \eqref{eq_sim_pull_force_linear} is provided in Appendix
\ref{sec_a_flat}, and the collapse of force versus displacement curves is shown
in Fig.\ \ref{fig_sim_pull_gamma_collapse}.

The dependence of the pull force on the speed of tether pulling is investigated
next.
Results from simulations with variable $ \SL $ and fixed
$ \itGamma = 1024 $
are plotted in Fig.\ \ref{fig_sim_pull_sl}.
The observed increase in pull force with increasing Scriven--Love number can be
justified as follows.
During tether pulling, lipids in the surrounding region flow inwards towards the
tether.
A mass balance, with the assumption of axisymmetry, indicates that the flow
speed is approximately $ \vp \mk \rc / r $, where $ r $ is the
distance from the $ z $-axis.
Since the in-plane velocity grows as we approach the tether from its
surroundings, a surface tension gradient is required to sustain the flow of
lipids.
As the tether is pulled more quickly, larger velocities and thus larger tension
gradients ensue.
A greater surface tension in the tether results, and leads to the larger pull
force observed in simulations.

At present, we are unable to determine a general functional form for the long-%
time, steady-state pull force as a function of $ \SL $ due to the high degree of
nonlinearity in the governing equations.
The main difficulty is that the surface tension and membrane geometry enter both
the in-plane and shape equations.
Moreover, the viscous--curvature coupling forces---which arise due to the flow
of lipids---lead to $ \mco(\SL) $ changes of the membrane shape in the tent-like
region.
Since we are unable to approximate how the steady-state pull force depends on
the Scriven--Love number, we choose not to collapse the data contained in Fig.\
\ref{fig_sim_pull_sl}.
Instead, we plot the steady-state pull force $ \mcfss $ as a function of
Scriven--Love number in Fig.\ \ref{fig_sim_pull_sl_ss}.
We expect
$ \mcfss \approx \mcfeq $ when
$ \SL \rightarrow 0 $.
In the limit where
$ \SL \ll 1 $, we approximate dynamical effects by calculating the change in
surface tension under the assumption of no membrane shape changes.
We find
\begin{equation} \label{eq_sim_pull_force_sl_small}
	\mcfss (\SL)
	\, = \, \big(
		1
		+ 2 \mk \SL
	\big) \, \mcfeq
	\qquad
	\text{for}
	\quad
	\SL
	\, \ll \, 1
	~,
\end{equation}
which is shown as the dotted grey line in Fig.\ \ref{fig_sim_pull_sl_ss}.
Evidently, Eq.\ \eqref{eq_sim_pull_force_sl_small} is a reasonable approximation
when
$ \SL < 0.1 $.

\begin{figure}[p]
	\centering
	\includegraphics[height=0.89\textheight]{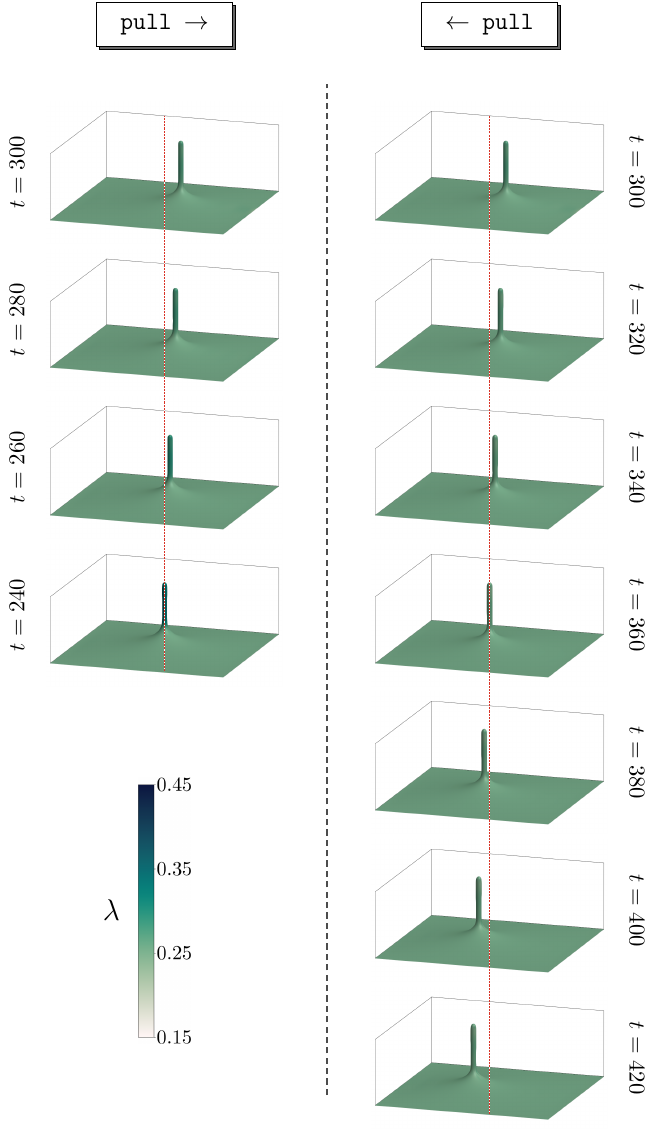}
	\vspace{-1pt}
	\caption{%
		Tether translation in the $ + \bme_x $ (left column) and then $ - \bme_x $
		(right column) directions.
		Times are measured in units of $ \zeta \mk \rc^{\, 2} / \kb $, the pull
		velocity is specified in Eq.\ \eqref{eq_sim_pull_lat_vp}, and all other
		parameters are identical to those in Fig.\ \ref{fig_sim_pull_schemes}.
		The vertical red lines are a visual aid to highlight the lateral translation
		of the tether.
		For a video of the simulation results, see Movie
		\hyperref[movie_translate_ale]{4}.
	}
	\label{fig_sim_pull_lateral}
\end{figure}

%
%

\subsubsection{The lateral translation of a membrane tether}
\label{sec_sim_pull_lat}

Once a tube is pulled from a patch of membrane, a lateral force applied at the
end of the tether causes it to translate relative to the surrounding region.
Lipids quickly flow and readjust to accommodate the translating tether due to
the in-plane fluidity of the membrane.
While tether translation via a lateral force is observed in \textit{in vitro}
studies \cite{datar-bpj-2015, perez-sa-2022, shi-sa-2022}, the physics of tether
translation remains poorly understood.
A theoretical description of tether translation is difficult because the system
is no longer axisymmetric, and the greatly simplified axisymmetric equations
\cite{powers-pre-2002, derenyi-prl-2002, agrawal-bmmb-2008, agrawal-zamp-2011,
omar-bpj-2020} no longer apply.
Tether translation also cannot be captured with general, non-axisymmetric
Lagrangian numerical methods---as laterally translating the tether induces a
rigid body translation of the entire patch (see Appendix
\ref{sec_a_lag_translate}).
ALE finite element methods are thus required to simulate tether translation.

In Fig.\ \ref{fig_sim_pull_lateral}, we present snapshots from a simulation of
tether translation, which employed the \alevb\ mesh motion.
Starting with the
$ t = 240 \, \zeta \mk \rc^{\, 2} / \kb $
configuration shown in Figs.\ \ref{fig_sim_pull_schemes} and
\ref{fig_sim_pull_zoom}, we prescribe a lateral velocity in the $ x $-direction
given by
\begin{equation} \label{eq_sim_pull_lat_vp}
	\bmvp (t)
	\ = \, \begin{cases}
		\, + \mk \vp \mk \bmex & ~ \text{ for } ~ 240 < t \cdot \kb / (\zeta \mk \rc^{\, 2}) < 300 \\[4pt]
		\, - \mk \vp \mk \bmex & ~ \text{ for } ~ 300 < t \cdot \kb / (\zeta \mk \rc^{\, 2}) < 420 
		~,
	\end{cases}
\end{equation}
where in both cases
$ \vp = 0.1 \mk \kb / (\zeta \mk \rc) $
to be consistent with our choice
$ \SL = 0.1 $.
In this simulation,
$ \itGamma = 1024 $
is unchanged.
We observe that the surface tension does not appreciably change, as the tether
is translated slowly relative to the fundamental timescale of lipid
rearrangements.
Our results demonstrate that ALE methods can be used in scenarios where
Lagrangian methods fail, and set the stage for future investigations of the
forces, geometry, and dynamics of tether translation.
A video of the laterally-pulled tether simulation is provided in the software
repository \cite{mem-ale-fem}.

%
%

\section{Conclusions and future work} \label{sec_concl}

In the present work, we ($i$) developed a robust ALE numerical method for lipid
membranes and ($i\mkn i$) applied the method to a scenario where established
Lagrangian and Eulerian schemes fail.
In our development, the mesh is treated as a material that is independent from
the membrane---with the mesh equations of motion and corresponding boundary
conditions arbitrarily prescribed by the user.
Mesh and material surfaces are constrained to overlap with a Lagrange multiplier
field, which enters the mesh dynamics as a force per area in the normal
direction.
By choosing for the mesh to resist shear and dilation (through the mesh
viscosity $ \zetam $) and bending (through the mesh bending moduli $ \kbm $ and
$ \kgm $), we successfully pulled a tether from a flat patch and then translated
it laterally across the membrane surface.
In contrast, Lagrangian and Eulerian simulations are respectively unable to
translate and pull a tether.
Our results thus mark the first numerical demonstration of lateral tether
translation.
We also analyzed the dynamics of tether pulling by determining how the pull
force increases monotonically with increasing pull speed.
Our findings were presented in terms of the \FvK\ number $ \itGamma $ and Scriven--Love number $ \SL $,
which define the tether-pulling scenario.

We close by highlighting that our numerical implementation \cite{mem-ale-fem}
is structured such that one can easily choose different constitutive relations
and boundary conditions for the mesh motion.
In this manner, the mesh itself can resist shear, dilation, or their rates of
change---irrespective of the material behavior.
Only purely viscous and viscous-with-bending mesh motions were considered in the
present work;
we aim to investigate different choices of mesh motion in a future study.
As a consequence of the modular structure of the numerical implementation, it is
straightforward to adapt the code to simulate 2D materials with different
constitution.
We hope to support the open-source community in doing so.
We also intend to extend our method to describe additional phenomena governing
biological membranes---including the coupling between lipid flows and the
hydrodynamics of the surrounding fluid \cite{narsimhan-jfm-2015, gross-jcp-2018,
faizi-pnas-2024}, membrane permeability \cite{alkadri-arxiv-2024}, the effects
of embedded particles \cite{stone-jfm-2010, stone-jfm-2015, sabass-prl-2016},
and in-plane diffusion and phase transitions in multi-component membrane systems
\cite{subramaniam-nm-2013, yu-arxiv-2023, venkatesh-jfm-2025,
venkatesh-arxiv-2024}.

	%
	%

	\vspace{13pt}

\section*{Acknowledgments}

We are indebted to Prof.\
\href{https://www.cchem.berkeley.edu/~kranthi/}{Kranthi Mandadapu}
for many scientific discussions on membrane dynamics and finite element methods
within the setting of differential geometry.
We are also grateful to Prof.\
\href{https://me.berkeley.edu/people/panayiotis-papadopoulos/}{Panos Papadopoulos}
and Dr.\
\href{https://scholar.google.com/citations?hl=en&user=fUuBj2sAAAAJ&view_op=list_works&sortby=pubdate}{Yannick Omar}
for bringing important details of finite element analysis to our attention,
including the Dohrmann--Bochev method \cite{dohrmann-ijnmf-2004} and numerical
differentiation of the residual vector \cite{tanaka-cmame-2014}.
It is a pleasure to thank Dr.\
\href{https://scholar.google.com/citations?hl=en&user=UrwMUscAAAAJ&view_op=list_works&sortby=pubdate}{Jo\"el Tchoufag}
regarding conversations on fluid mechanics and membrane dynamics in different
geometries.

Simulations were carried out on the
\href{https://www.nersc.gov/systems/perlmutter/}{\texttt{Perlmutter}}
high-performance computing (HPC) system at the National Energy Research
Scientific Computing Center.
Our open-source package is written in the
\href{https://julialang.org/}{\texttt{Julia}}
programming language \cite{julia}.
Figures and videos of the simulation results were generated with the open-source
\texttt{Julia} package \href{https://docs.makie.org}{\texttt{Makie.jl}}
\cite{makie}.
This work was partially supported by the \href{https://welch1.org/}{Welch
Foundation} via Grant No.\ F-2208-20240404.

\begin{center}\textit{
	Declaration of Interests.
	The authors report no conflict of interest.
}\end{center}

	%
	%

	\begin{center}%
		\par\noindent\rule{0.94\textwidth}{0.4pt}%
	\end{center}

	\begin{appendices}
		\titleformat{\section}[block]%
		{\large\bfseries\filcenter}{Appendix \thetitle.\ }{0em}{}
		\numberwithin{equation}{section}
		\makeatletter
			\renewcommand{\p@section}{}
			\renewcommand{\p@subsection}{}
		\makeatother

		\addcontentsline{toc}{part}{Appendices}

%
%

\section{The finite element implementation} \label{sec_a_fem}

The main novelty of the present work is the treatment of the mesh as an
independent material with its own constitution, as discussed in
\ref{sec_gov_mesh_ale}.
However, none of the finite element techniques used here are new.
All fundamental unknowns, their arbitrary variations, and the surface position
itself are discretized with the same basis functions---as is standard in
isoparametric finite element methods.
Over any element
$ \Omegae \in \mcth $,
we have
\begin{equation} \label{a_fem_xh}
	\bmxh (\zetaalpha, t)
	\, = \, \sum_{k = 1}^{\nen} \Nke (\zetaalpha) \, \bmxke (t)
	\, = \, \mNeza \, \mxet
	\qquad
	\forall ~ \zetaalpha \in \Omegae
	~.
\end{equation}
In Eq.\ \eqref{a_fem_xh}, the matrix $ \mNeza $ contains the nonzero elemental
basis functions (see \ref{sec_fem_disc}) and the column vector $ \mxet $
collects the corresponding local degrees of freedom (or nodal positions) in the
usual manner of finite element analysis:
\begin{equation} \label{a_fem_mNeza}
	\mNeza
	\, := \, \Big[ \hspace{5pt}
		N_{1}^e (\zetaalpha) \, \mone \hspace{9pt}
		N_{2}^e (\zetaalpha) \, \mone \hspace{5pt}
		\ldots \hspace{5pt}
		N_{\nen}^e (\zetaalpha) \, \mone \hspace{5pt}
	\Big]
	~,
\end{equation}
where $ \mone $ is the $ 3 \times 3 $ identity matrix in Cartesian coordinates,
and
\begin{equation} \label{a_fem_mxet}
	\mxet
	\, := \,
	\begin{bmatrix}
		~ \\[-13pt]
		\, \bmx_1^e (t) \, \\
		\, \vdots \, \\[2pt]
		\, \bmx_{\nen}^e (t) \, \\[2pt]
	\end{bmatrix}
	~.
\end{equation}
The velocity and mesh velocity are similarly decomposed as
\begin{align}
	\bmvh (\zetaalpha, t)
	\, &= \, \mNeza \, \mvet
	\label{a_fem_bmvh}
	\shortintertext{and}
	\bmvmh (\zetaalpha, t)
	\, &= \, \mNeza \, \mvmet
	\label{a_fem_bmvmh}
\end{align}
for all
$ \zetaalpha \in \Omegae $.
Since the membrane tension and mesh pressure are scalar quantities, they are
respectively discretized as
\begin{align}
	\lambdah (\zetaalpha, t)
	\, &= \, \mNveceza \, \mlambdaet
	\label{a_fem_lambdah}
	\shortintertext{and}
	\pmh (\zetaalpha, t)
	\, &= \, \mNveceza \, \mpmet
	\label{a_fem_pmh}
\end{align}
for all
$ \zetaalpha \in \Omegae $, where $ \mNveceza $ is the row vector of nonzero
elemental basis functions given by [cf.\ Eq.\ \eqref{a_fem_mNeza}]
\begin{equation} \label{a_fem_mNveceza}
	\mNveceza
	\, := \, \Big[ \hspace{5pt}
		N_{1}^e (\zetaalpha) \hspace{9pt}
		N_{2}^e (\zetaalpha) \hspace{5pt}
		\ldots \hspace{5pt}
		N_{\nen}^e (\zetaalpha) \hspace{5pt}
	\Big]
	~.
\end{equation}
Assuming a known solution at time $ t $, the temporal evolution of the membrane
is obtained via the backward Euler method.
In particular, the membrane surface is updated according to [cf.\ Eq.\
\eqref{eq_gov_mesh_velocity},\eqref{eq_gov_mesh_position}]
\begin{align}
	\bmx (\zetaalpha, t + \Delta t)
	\, &= \, \bmx (\zetaalpha, t)
	\, + \, \Delta t \, \bmvm (\zetaalpha, t + \Delta t)
	~,
	\label{a_fem_be_bmx}
	\shortintertext{or equivalently}
	[ \mathbfx^e (t + \Delta t) ]
	\, &= \, \mxet
	\, + \, \Delta t \, [ \mathbfv^{\tm, e} (t + \Delta t) ]
	~.
	\label{a_fem_be_mxe}
\end{align}
As discussed in \ref{sec_fem_summary}, the mesh velocity and all other unknowns
satisfy the residual vector equation
\begin{equation}
	\big[ \mathbfr ( [ \mathbfu (t + \Delta t) ] ) \big]
	\, = \, \mzero
	~,
\end{equation}
which is solved via the Newton--Raphson method---in which $ \mut $ is used as
the initial guess.
We thus close our discussion of the finite element implementation by presenting
the contributions to the residual vector; all details can be found in the
software documentation \cite{mem-ale-fem}.

%
%

\subsection{The contributions to the residual vector} \label{sec_a_fem_res}

Our finite element implementation allows one to choose whether the mesh motion
is Lagrangian, Eulerian, or ALE---and, in the latter case, whether the mesh
dynamics are purely viscous or viscous with a bending resistance.
The direct Galerkin expressions in Eqs.\ \eqref{eq_gov_mesh_lag_mcg},
\eqref{eq_gov_mesh_eul_mcg}, \eqref{eq_gov_mesh_visc_mcg}, and
\eqref{eq_gov_mesh_vb_mcg} dictate the residual vector for each mesh motion.
In what follows, we present the local residual vector contributions---%
corresponding to a single finite element
$ \Omegae \in \mcth $%
---for each fundamental unknown.

%
%

\subsubsection{The surface tension contribution} \label{sec_a_fem_res_lambda}

As detailed in Ref.\ \cite{sahu-jcp-2020}, the residual vector corresponding to
Eq.\ \eqref{eq_gov_inc_mcgl} is given by
\begin{equation} \label{a_fem_res_mrel}
	\mrel
	\ = \int_{\Omegae} \mNvece^\tT \big(
		\bma^\calpha \bmcdot \bmv_{, \calpha}
	\big) \, \JO ~ \td \Omega
	\ - \ \dfrac{\alphaDB}{\zeta} \bigg\{
		\int_{\Omegae}
			\mNvece^\tT \mk \lambda
		~ \td \Omega
		\ - \ \mGe^\tT \mk \mHe^{-1} \mGe \, \mlambdae
	\bigg\}
	~,
\end{equation}
where all quantities in curly braces arise from the Dohrmann--Bochev method
\cite{dohrmann-ijnmf-2004}.
The matrices $ \mGe $ and $ \mHe $ in Eq.\ \eqref{a_fem_res_mrel} are
constructed from the basis functions $ \brNek (\zetaalpha) $ to the space
$ \brL $ onto which surface tensions are projected.
Since functions in $ \brL $ are linear and form a plane over $ \Omegae $ [see
Eq.\ \eqref{eq_fem_disc_brL}], they can be expressed as
$ a + b \mk \zetaone + c \mk \zetatwo = 0 $
for some constants
$ a, b, c \in \mathbb{R} $.
In our numerical implementation, the planar basis functions over a single
element are chosen to be
$ ~ \hspace{-0pt} \brNe{1} (\zetaalpha) = 1 $,
$ \brNe{2} (\zetaalpha) = \xi (\zetaone) $,
and
$ \brNe{3} (\zetaalpha) = \eta (\zetatwo) $,
where $ \xi $ and $ \eta $ parametrize the reference square
$ [-1, 1] \times [-1, 1] $
onto which $ \Omegae $ is mapped.
Accordingly, over any element
$ \Omegae \in \mcth $
we define the row vector
\begin{equation} \label{a_fem_res_mbrNeza}
	\mbrNeza
	\, := \, \Big[ \hspace{5pt}
		\brNe{1} (\zetaalpha) \hspace{9pt}
		\brNe{2} (\zetaalpha) \hspace{9pt}
		\brNe{3} (\zetaalpha) \hspace{5pt}
	\Big]
	~,
\end{equation}
with which the matrices $ \mGe $ and $ \mHe $ in Eq.\ \eqref{a_fem_res_mrel}
are expressed as
\begin{equation}
	\mGe
	\ := \, \int_{\Omegae} \! \mbrNe^\tT \mNvece \ \td \Omega
	\qquad
	\text{and}
	\qquad
	\mHe
	\ := \, \int_{\Omegae} \! \mbrNe^\tT \mbrNe \ \td \Omega
	~.
\end{equation}

%
%

\subsubsection{The material velocity contribution} \label{sec_a_fem_res_bmv}

It is straightforward to determine the contributions to the residual vector from
Eq.\ \eqref{eq_gov_mem_mcgv}.
We take advantage of the symmetry of $ \sigmaab $ and $ \Mab $ to obtain
\begin{equation} \label{a_fem_res_mrev}
	\begin{split}
		\mrev
		\ &:= \, \int_{\Omegae}
				\mNe^\tT_{, \calpha} \, \bmab \, \sigmaab
		\, \JO ~ \td \Omega
		\ + \, \int_{\Omegae}
			\mNe^\tT_{; \calpha \cbeta} \, \bmn \, \Mab
		\, \JO ~ \td \Omega
		\ - \, \int_{\Omegae} \mNe^\tT \, \bmf \, \JO ~ \td \Omega
		\\[5pt]
		&\hspace{20pt}
		- \ \sum_{j = 1}^3 \mk \int_{\partial \Omegae \cap \GammaFj} \hspace{-17pt}
			\mNvece^\tT \, \barFj
		\, \JG ~ \td \Gamma
		\ - \, \int_{\partial \Omegae \cap \GammaM} \hspace{-21pt}
			\mNe^\tT_{, \calpha} \, \nu^\calpha \, \bmn \, \barM
		\, \JG ~ \td \Gamma
		~.
	\end{split}
\end{equation}
In Eq.\ \eqref{a_fem_res_mrev}, it is understood that the $ x $, $ y $, and
$ z $ components of the force terms are written to appropriate entries of the
residual vector.

%
%

\subsubsection{The Eulerian mesh velocity contribution}
\label{sec_a_fem_res_bmvm_E}

For the choice of an Eulerian mesh motion, the residual vector contribution
corresponding to Eq.\ \eqref{eq_gov_mesh_eul_vm_weak} is given by
\begin{equation} \label{a_fem_res_mremE}
	\mremE
	\ := \ \alphamE \int_{\Omegae}
		\mNe^\tT \, \Big(
			\bmvm
			\, - \, \big(
				\bmn \otimes \bmn
			\big) \mk \bmv
		\Big)
	\, \JO ~ \td \Omega
	~.
\end{equation}

%
%

\subsubsection{The ALE mesh pressure contribution} \label{sec_a_fem_res_pm}

When an ALE mesh motion is employed, the mesh pressure ensures the kinematic
constraint \eqref{eq_gov_geodyn_kinematic_constraint} is satisfied.
With the Dohrmann--Bochev method applied once again, the residual vector
resulting from Eq.\ \eqref{eq_gov_mesh_visc_mcgp} is given by [cf.\ Eq.\
\eqref{a_fem_res_mrel}]
\begin{equation} \label{a_fem_res_mrep}
	\begin{split}
		\mrep
		\ &= \ - \int_{\Omegae} \mNvece^\tT \, \Big(
			\bmn \bmcdot \big(
				\bmvm
				\, - \, \bmv
			\big)
		\Big) \, \JO ~ \td \Omega
		\\[3pt]
		&\hspace{24pt}
		\ - \ \dfrac{\alphaDB}{\zeta \mk \ell^2} \bigg\{
			\int_{\Omegae}
				\mNvece^\tT \mk \ptm
			~ \td \Omega
			\ - \ \mGe^\tT \mk \mHe^{-1} \mGe \, \mpme
		\bigg\}
		~.
	\end{split}
\end{equation}

%
%

\subsubsection{The ALE mesh velocity contribution}
\label{sec_a_fem_res_bmvm_A}

When an ALE mesh motion is employed, the residual vector contribution from the
mesh velocity looks similar to that from the material velocity [cf.\ Eq.\
\eqref{a_fem_res_mrev}]:
\begin{equation} \label{a_fem_res_mremAv}
	\mremA
	\ := \, \int_{\Omegae}
			\mNe^\tT_{, \calpha} \, \bmab \, \sigmaabm
	\, \JO ~ \td \Omega
	\ + \, \int_{\Omegae}
		\mNe^\tT_{; \calpha \cbeta} \, \bmn \, \Mabm
	\, \JO ~ \td \Omega
	\ - \, \int_{\Omegae} \mNe^\tT \, \ptm \mk \bmn \, \JO ~ \td \Omega
	~.
\end{equation}
In Eq.\ \eqref{a_fem_res_mremAv}, $ \sigmaabmv $ and $ \Mabmv $ (resp.\
$ \sigmaabmvb $ and $ \Mabmvb $) are substituted when the motion is ALE-viscous
(resp.\ ALE-viscous-bending).

%
%

\section{The static portion of a cylinder} \label{sec_a_cyl}

Here we consider a static membrane patch that is a portion of a cylinder---for
which
\begin{equation} \label{a_cyl_x_og}
	\bmx (\theta, z)
	\, = \, \rc \, \bmer (\theta)
	\, + \, z \mk \bmez
	~.
\end{equation}
In Eq.\ \eqref{a_cyl_x_og}, $ \theta $ and $ z $ are the canonical cylindrical
coordinates and $ \rc $ is the cylinder radius.
With our differential geometric formulation, we arbitrarily choose to
parametrize the surface as
\begin{equation} \label{a_cyl_param}
	\zetaone
	\, := \, \theta
	\, \in \, [\mk 0, \alpha \mk]
	\qquad
	\text{and}
	\qquad
	\zetatwo
	\, := \, \dfrac{z}{\ell}
	\, \in \, [\mk 0, 1 \mk]
	~,
\end{equation}
such that
\begin{equation} \label{a_cyl_x_dg}
	\bmx (\zetaalpha)
	\, = \, \rc \, \bmer (\zetaone)
	\, + \, \ell \mk \zetatwo \mk \bmez
	~.
\end{equation}
It is then straightforward to determine \cite[Ch.IX\,\S1]{sahu-thesis}
\begin{equation} \label{a_cyl_dg}
	\begin{split}
		\bma_\cone
		\, &= \, \rc \, \bmetheta (\zetaone)
		~,
		\hspace{33pt}
		\bma_\ctwo
		\, = \, \ell \mk \bmez
		~,
		\hspace{34pt}
		\bmn
		\, = \, \bmer
		~,
		\hspace{33pt}
		a_{\calpha \cbeta}
		\, = \, \text{diag} \big( \rc^{\, 2}, \ell^2 \big)
		~,
		\\[6pt]
		a^{\calpha \cbeta}
		\, &= \, \text{diag} \big( \rc^{\, -2}, \ell^{-2} \big)
		~,
		\hspace{25pt}
		b_{\calpha \cbeta}
		\, = \, \text{diag} \big( -\rc, 0 \big)
		~,
		\hspace{25pt}
		b^{\calpha \cbeta}
		\, = \, \text{diag} \big( -\rc^{\, -3}, 0 \big)
		~,
		\hspace{20pt}
	\end{split}
\end{equation}
for which the mean and Gaussian curvatures are respectively given by
\begin{equation} \label{a_cyl_curvatures}
	H
	\, = \, \dfrac{\, -1 \,~}{2 \mk \rc}
	\qquad
	\text{and}
	\qquad
	K
	\, = \, 0
	~.
\end{equation}
The couple-stress components $ \Mab $ and couple-free in-plane stress components
$ \sigmaab $ are calculated via Eqs.\ \eqref{eq_gov_mem_Mab} and
\eqref{eq_gov_mem_sigmaab} as
\begin{align}
	\Mab
	\ &= \ \begin{pmatrix}
		\
		\dfrac{\, -\kb \,~}{2 \mk \rc^{\, 3}}
		\
		& 0
		\\[11pt]
		0 &
		\
		\dfrac{\, -\kb \,~}{2 \mk \rc \mk \ell^{\mk 2}}
		\, - \, \dfrac{\kg}{\rc \mk \ell^{\mk 2}}
		\
	\end{pmatrix}
	\label{a_cyl_Mab}
	\shortintertext{and}
	\sigmaab
	\ &= \ \begin{pmatrix}
		\
		\dfrac{\lambdaz}{\rc^{\, 2}}
		\, - \, \dfrac{3 \mk \kb}{4 \mk \rc^{\, 4}}
		\
		& 0
		\\[3pt]
		0
		&
		\
		\dfrac{\lambdaz}{\ell^{\mk 2}}
		\, + \, \dfrac{\kb}{4 \mk \rc^{\, 2} \ell^{\mk 2}}
		\
	\end{pmatrix}
	~,
	\label{a_cyl_sigmaab}
\end{align}
where for a static patch there are no viscous stresses---for which
$ \piab = 0 $.
Here $ \lambdaz $ denotes the constant surface tension, which is set by the
balance of forces in the out-of-plane direction.
With some additional calculations, we find the stress vectors $ \bmTa $ to be
given by
\begin{equation} \label{a_cyl_stress_vector}
	\bmT^\cone
	\, = \, \bigg(
		\dfrac{\lambdaz}{\rc}
		\, - \, \dfrac{\kb}{4 \mk \rc^{\, 3}}
	\bigg) \, \bmetheta
	\qquad
	\text{and}
	\qquad
	\bmT^\ctwo
	\, = \, \bigg(
		\dfrac{\lambdaz}{\ell}
		\, + \, \dfrac{\kb}{4 \mk \rc^{\, 2} \mk \ell}
	\bigg) \, \bmer
	~.
\end{equation}
We now separately consider the top and bottom edges, where $ \zetatwo $ is
fixed, and the left and right edges, where $ \zetaone $ is fixed.

\bigskip

\noindent
\textbf{The top and bottom surfaces.---}%
At the top
($ \zetatwo = 1 $)
and bottom
($ \zetatwo = 0 $)
edges,
$ \bmnu = \pm \bmez $,
$ \nu^{}_\cone = 0 $,
$ \nu^{}_\ctwo = \pm \ell $,
$ \bmtau = \mp \bmetheta $,
$ \tau^{}_\cone = \mp \rc $,
and
$ \tau^{}_\ctwo = 0 $.
We thus determine (see \ref{sec_gov_mem_bc})
\begin{gather}
	M
	\, = \, \Mab \nu^{}_\calpha \mk \nu^{}_\cbeta
	\, = \, M^{\ctwo \ctwo} \nu^{}_\ctwo \mk \nu^{}_\ctwo
	\, = \, - \dfrac{\kb}{2 \mk \rc}
	\, - \, \dfrac{\kg}{\rc}
	\label{a_cyl_top_bottom_boundary_moment}
	\shortintertext{and}
	\bmF
	\, = \, \bmTa \nu^{}_\calpha
	\, - \, \big(
		\Mab \mk \nu^{}_\calpha \mk \tau^{}_\cbeta \, \bmn
	\big)_{\! , \cmu} \, \tau^\cmu
	\, = \, \bmT^\ctwo \nu^{}_\ctwo
	\, = \, \pm \bigg(
		\lambdaz
		\, + \, \dfrac{\kb}{4 \mk \rc^{\, 2}}
	\bigg) \, \bmez
	~.
	\label{a_cyl_top_bottom_boundary_force}
\end{gather}
The sign of the moment in Table \ref{tab_sim_bend_force_moment} is opposite that
of Eq.\ \eqref{a_cyl_top_bottom_boundary_moment} due to the difference in
orientation of the unit normal with respect to the surface.

\bigskip

\noindent
\textbf{The left and right surfaces.---}%
At the left
($ \zetaone = \alpha $)
and right
($ \zetaone = 0 $)
edges,
$ \bmnu = \pm \bmetheta $,
$ \nu^{}_\cone = \pm \rc $,
$ \nu^{}_\ctwo = 0 $,
$ \bmtau = \pm \bmez $,
$ \tau^{}_\cone = 0 $,
and
$ \tau^{}_\ctwo = \pm \ell $.
In the same manner, we calculate
\begin{gather}
	M
	\, = \, \Mab \nu^{}_\calpha \mk \nu^{}_\cbeta
	\, = \, M^{\cone \cone} \nu^{}_\cone \mk \nu^{}_\cone
	\, = \, - \dfrac{\kb}{2 \mk \rc}
	\label{a_cyl_left_right_boundary_moment}
	\shortintertext{and}
	\bmF
	\, = \, \bmTa \nu^{}_\calpha
	\, - \, \big(
		\Mab \mk \nu^{}_\calpha \mk \tau^{}_\cbeta \, \bmn
	\big)_{\! , \cmu} \, \tau^\cmu
	\, = \, \bmT^\cone \nu^{}_\cone
	\, = \, \pm \bigg(
		\lambdaz
		\, - \, \dfrac{\kb}{4 \mk \rc^{\, 2}}
	\bigg) \, \bmetheta
	~.
	\label{a_cyl_left_right_boundary_force}
\end{gather}
The moment in Eq.\ \eqref{a_cyl_left_right_boundary_moment} again differs from
the moment reported in Table \ref{tab_sim_bend_force_moment} due to the choice
of normal vector.

%
%

\section{The numerical calculation of the pull force} \label{sec_a_pull_num}

In this section, we determine how to set the membrane velocity $ \bmv $ to a
desired value $ \bmvp $ at the center of the membrane patch.
Given the use of non-interpolatory basis functions, there is no unique way to do
so.
We thus choose to set the membrane velocity over the entire finite element
containing the point of interest.
The pull force $ \bmmcfp (t) $ resulting from the imposed displacement is
calculated via variational arguments.

%
%

\subsection{The inability to uniquely displace a single point on the membrane}
\label{sec_a_pull_num_unique}

To begin, at a chosen point $ (\zetaonep, \zetatwop) $ we intend for
\begin{equation} \label{a_pull_num_set_single}
	\bmv (\zetaalphap, t)
	\, = \, \bmvp (t)
	~,
\end{equation}
where $ \bmvp (t) $ is a known function.
With the velocity discretization in Eq.\ \eqref{a_fem_bmvh}, Eq.\
\eqref{a_pull_num_set_single} can be expressed as
\begin{equation} \label{a_pull_num_v_constraint}
	\sum_{k = 1}^{\nen} \Nke (\zetaalphap) \, \bmvke (t)
	\ = \, \bmvp (t)
	~,
\end{equation}
where $ \{ \Nke (\zetaalphap) \} $ and $ \{ \bmvke \} $ are respectively the
local basis functions and nodal velocities of the finite element $ \Omegaep $
containing $ \zetaalphap $.
Importantly, the basis function values are set by the choice of $ \zetaalphap $,
and they are non-interpolatory \cite{piegl-tiller}---and thus all $ \nen $
nodal velocities contribute to Eq.\ \eqref{a_pull_num_v_constraint}.
Since
$ \bmvp \in \mathbb{R}^3 $
and
$ \bmvke \in \mathbb{R}^3 $
for all $ k $,
Eq.\ \eqref{a_pull_num_v_constraint} can be understood as a system of three
scalar equations involving $ 3 \cdot \nen $ scalar unknowns.
As
$ \nen = ( \poly + 1)^2 $
for B-splines of polynomial order \poly \cite{piegl-tiller}, there is no unique
way to specify the $ \{ \bmvke \} $ in Eq.\ \eqref{a_pull_num_v_constraint}.
In practice, we set
\begin{equation} \label{a_pull_num_v_uniform}
	\bmvke (t)
	\, = \, \bmvp (t)
	\qquad
	\text{for }
	k \in \{1, 2, \ldots, \nen\}
	~.
\end{equation}
Given the properties of B-spline functions \cite{piegl-tiller}, Eq.\
\eqref{a_pull_num_v_uniform} results in a uniform prescribed velocity over the
entire parametric element $ \Omegaep $---expressed as
\begin{equation} \label{a_pull_num_v_element}
	\bmv (\zetaalpha, t)
	\, = \, \bmvp (t)
	\qquad
	\text{for all }
	\zetaalpha \in \Omegaep
	~.
\end{equation}

%
%

\subsection{The pull force when a finite element is uniformly displaced}
\label{sec_a_pull_num_elem}

In this section, we calculate the force $ \bmmcfp $ required to pull an entire
finite element at a prescribed velocity as in Eq.\ \eqref{a_pull_num_v_element}.
From the strong form of the governing membrane equations
(\ref{eq_gov_mem_lin_mom_in_plane},\ref{eq_gov_mem_lin_mom_shape}) we
recognize the pull force is the area integral of the net body force per area
$ \bmf $ on the membrane---expressed as
\begin{equation} \label{a_pull_elem_fp_traction}
	\bmmcfp (t)
	\ = \, \int_{\Omegaep}
		\bmf (\zetaalpha, t) \mk \JO (\zetaalpha, t)
	~ \td \Omega
	~.
\end{equation}
Since the prescribed velocity in Eq.\ \eqref{a_pull_num_v_element} can be viewed
as a constraint, $ \bmf $ is then understood as the associated Lagrange
multiplier field over the element $ \Omegaep $.

Our task now is to determine how to calculate $ \bmmcfp $ numerically, which is
accomplished by considering how the weak form would be modified if Eq.\
\eqref{a_pull_num_v_element} was not satisfied directly.
The principle of virtual power would then necessitate the quantity
\begin{equation} \label{a_pull_elem_constraint_weak}
	\begin{split}
		&\delta \mk \bigg\{
			\int_{\Omegaep} \! \bmf (\zetaalpha, t) \bmcdot \Big(
				\bmv (\zetaalpha, t)
				\, - \, \bmvp (t)
			\Big)
			\, \JO (\zetaalpha, t) ~ \td \Omega
		\bigg\}
		\\[3pt]
		& \hspace{20pt}
		   = \, \int_{\Omegaep} \!
			\delta \bmv \bmcdot \bmf
		\, \JO ~ \td \Omega
		\  + \, \int_{\Omegaep} \!
			\delta \bmf \bmcdot \big(
				\bmv
				\ - \ \bmvp
			\big)
		\, \JO ~ \td \Omega
		\  + \, \int_{\Omegaep} \!
			\bmf \bmcdot \big(
				\bmv
				- \bmvp
			\big)
		\, \delta \JO ~ \td \Omega
	\end{split}
\end{equation}
be subtracted from the direct Galerkin expression, where
$
	\delta \JO
	= \Delta t \mk \JO \, \bma^{\calpha} \bmcdot \delta \bmvm_{, \calpha}
$
(see Appendix C.2.1 of Ref.\ \cite{sahu-jcp-2020}).
At this point, the fundamental unknowns and arbitrary variations are discretized
as
\begin{equation} \label{a_pull_elem_unk_discretization}
	\begin{split}
		\bmv (\zetaalpha, t)
		\, &= \, \mNep \mk \mvept
		~,
		\hspace{20pt}
		\bmvm (\zetaalpha, t)
		\,  = \, \mNep \mk \mvmept
		~,
		\hspace{20pt}
		\bmf (\zetaalpha, t)
		\,  = \, \mNep \mk \mfept
		~,
		\\[6pt]
		\delta \bmv (\zetaalpha)
		\, &= \, \mNep \mk \mdeltavep
		~,
		\hspace{31pt}
		\delta \bmvm (\zetaalpha)
		\,  = \, \mNep \mk \mdeltavmep
		~,
		\hspace{30.5pt}
		\delta \bmf (\zetaalpha)
		\,  = \, \mNep \mk \mdeltafep
		~.
	\end{split}
\end{equation}
In Eq.\ \eqref{a_pull_elem_unk_discretization}, we introduced the shorthand
$ \mNep := [ \mathbfN^e (\zetaalpha) ] $
for all
$ \zetaalpha \in \Omegaep $.
Upon substituting Eq.\ \eqref{a_pull_elem_unk_discretization} into the second
line of Eq.\ \eqref{a_pull_elem_constraint_weak} and defining the elemental mass
matrix $ \mMep $ as
\begin{equation} \label{a_pull_elem_mass_matrix}
	\mMep
	\  := \, \int_{\Omegaep} \mNep^\tT \mk \mNep \, \JO ~ \td \Omega
	~,
\end{equation}
we find the quantity
\begin{equation} \label{a_pull_elem_constraint_weak_discrete}
	\begin{split}
		& \mdeltavep^\tT \mk \mMep \, \mfept
		\, + \, \mdeltafep^\tT \mk \bigg\{
			\mMep \, \mvept
			\, - \, \bigg(
				\int_{\Omegaep} \mNep^\tT \mk \JO \ \td \Omega
			\bigg) \, \bmvp (t)
		\bigg\}
		\\[2pt]
		& \hspace{20pt}
		\, + \, \mdeltavmep^\tT \int_{\Omegaep}
			\mNep^\tT_{, \calpha} \bmcdot \big(
				\bma^\calpha \otimes \bmf
			\big) \, \big(
				\bmv
				- \bmvp
			\big)
		\, \Delta t \, \JO ~ \td \Omega
	\end{split}
\end{equation}
is to be subtracted from the discretized weak Galerkin expression $ \mcgh $.
We now separately consider the portions of Eq.\
\eqref{a_pull_elem_constraint_weak_discrete} arising from variations in the pull
force, material velocity, and mesh velocity.

%
%

\subsubsection{The pull force contribution} \label{sec_a_pull_elem_gp}

The portion of the direct Galerkin expression associated with the net force per
unit area $ \bmf $ can be expressed as
\begin{equation} \label{a_pull_elem_direct_galerkin_p}
	\mcgf
	\, = \, \mdeltafep^\tT \mk \bigg(
		\int_{\Omegaep} \mNep^\tT \mk \JO \ \td \Omega
	\bigg) \, \bmvp (t)
	\, - \, \mdeltafep^\tT \mk \mMep \, \mvept
	\, = \, 0
	\qquad
	\forall ~ \mdeltafep
	~.
\end{equation}
Importantly, the mass matrix $ \mMep $ is invertible and the variation
$ \mdeltafep $ is arbitrary, so the nodal velocity degrees of freedom are found
to be given by
\begin{equation} \label{a_pull_elem_velocity_dof_soln}
	\mvept
	\, = \, \mMep^{-1} \mk \bigg(
		\int_{\Omegaep} \mNep^\tT \mk \JO \ \td \Omega
	\bigg) \, \bmvp (t)
	~.
\end{equation}
It is a well-known property of B-splines \cite{piegl-tiller} that if the
constraint
$ \bmv (\zetaalpha, t) = \bmvp (t) $
is enforced over the entire element, then the unique solution for the nodal
degrees of freedom is given by
\begin{equation} \label{a_pull_elem_velocity_dofs}
	\mvept
	\ = \,
	\begin{bmatrix}
		~ \\[-13pt]
		\, \bmvp (t) \, \\
		\, \vdots \, \\[2pt]
		\, \bmvp (t) \, \\[2pt]
	\end{bmatrix}
	\ = \, \begin{bmatrix}
		~ \\[-12pt]
		\, \mone \, \\
		\, \vdots \, \\[1pt]
		\, \mone \, \\[2pt]
	\end{bmatrix} \, \bmvp (t)
	~.
\end{equation}
By comparing Eqs.\ \eqref{a_pull_elem_velocity_dof_soln} and
\eqref{a_pull_elem_velocity_dofs}, we recognize
\begin{equation} \label{a_pull_elem_identity}
	\mMep^{-1} \mk \int_{\Omegaep} \mNep^\tT \mk \JO \ \td \Omega
	\, = \, \begin{bmatrix}
		~ \\[-12pt]
		\, \mone \, \\
		\, \vdots \, \\[1pt]
		\, \mone \, \\[2pt]
	\end{bmatrix}
	~,
\end{equation}
which will be useful in our subsequent developments.

%
%

\subsubsection{The material velocity contribution} \label{sec_a_pull_elem_gv}

The material velocity portion of the discretized direct Galerkin expression
$ \mcgh $ is given by
\begin{equation} \label{a_pull_elem_direct_galerkin_v}
	\mcgv
	\, = \, \mdeltav^\tT \mk \mrv
	\, - \, \mdeltavep^\tT \mk \mMep \, \mfept
	\, = \, 0
	\qquad
	\forall ~ \, ~ \mdeltav
	~,
\end{equation}
where $ \mdeltav $ is the global variation of the nodal velocities,
$ \mdeltavep $ is the velocity variation of the \nen\ nodes associated with
$\Omegaep$, and $ \mrv $ is the velocity portion of the global residual vector
in the absence of a pull force.
Let us imagine reordering the global velocity degrees of freedom such that those
associated with $ \Omegaep $ appear last, and those not associated with the pull
force appear first.
The global velocity unknowns, their arbitrary variation, and residual vector in
the absence of a pull force are respectively expressed as
\begin{equation} \label{a_pull_elem_dc_ordering}
	\mvt
	\, = \, \begin{bmatrix}
		~ \\[-13pt]
		\, \mvnpt \, \\[4pt]
		\, \mvept \, \\[2pt]
	\end{bmatrix}
	~,
	\qquad
	\mdeltav
	\, = \, \begin{bmatrix}
		~ \\[-13pt]
		\, \mdeltavnp \, \\[4pt]
		\, \mdeltavep \, \\[2pt]
	\end{bmatrix}
	~,
	\qquad
	\text{and}
	\qquad
	\mrv
	\, = \, \begin{bmatrix}
		~ \\[-13pt]
		\, \mrvnp \, \\[4pt]
		\, \mrvep \, \\[2pt]
	\end{bmatrix}
	~.
\end{equation}
Here, the subscript `$\bar{\tp}$' is used to signify ``\texttt{not} p,'' i.e.\
degrees of freedom not associated with the pulled element $ \Omegaep $.
Substituting Eq.\ \eqref{a_pull_elem_dc_ordering} into
Eq.\ \eqref{a_pull_elem_direct_galerkin_v} yields
\begin{equation} \label{a_pull_elem_dc_direct_galerkin}
	\mdeltavnp^\tT \mk \mrvnp
	\, + \, \mdeltavep^\tT \mk \mrvep
	\, - \, \mdeltavep^\tT \mk \mMep \, \mfept
	\, = \, 0
	\qquad
	\forall ~ \, ~ \mdeltavnp, \, \mdeltavep
	~.
\end{equation}
Since the arbitrary variations $ \mdeltavnp $ and $ \mdeltavep $ are independent
of one another, Eq.\ \eqref{a_pull_elem_dc_direct_galerkin} requires
$ \mrvnp = \mzero $%
---which is the equation one obtains when the nodal velocities over $ \Omegaep $
are set directly.
Additionally, the nodal body force degrees of freedom are found to be given by
\begin{equation} \label{a_pull_elem_traction_dofs}
	\mfept
	\, = \, \mMep^{-1} \mk \mrvep
	~.
\end{equation}
By substituting Eqs.\ \eqref{a_pull_elem_unk_discretization} and then
\eqref{a_pull_elem_traction_dofs} into Eq.\ \eqref{a_pull_elem_fp_traction}, we
calculate the pull force as
\begin{equation} \label{a_pull_elem_force_calc}
	\bmmcfp (t)
	\  = \, \bigg(
		\int_{\Omegaep} \mNep^\tT \mk \JO \ \td \Omega
	\bigg) \, \mfept
	\ = \, \bigg(
		\int_{\Omegaep} \mNep^\tT \mk \JO \ \td \Omega
	\bigg) \, \mMep^{-1} \mk \mrvep
	~.
\end{equation}
Finally, recognizing the mass matrix is symmetric and substituting the transpose
of Eq.\ \eqref{a_pull_elem_identity} into Eq.\ \eqref{a_pull_elem_force_calc}
yields
\begin{equation} \label{a_pull_elem_force}
	\bmmcfp (t)
	\, = \, \Big[ \hspace{5pt}
		\mone \hspace{9pt}
		\mone \hspace{5pt}
		\ldots \hspace{5pt}
		\mone \hspace{5pt}
	\Big]
	\, \mrvep
	~,
\end{equation}
which is straightforward to calculate within finite element subroutines.
Equation \eqref{a_pull_elem_force} is the main result of this section.

%
%

\subsubsection{The mesh velocity contribution} \label{sec_a_pull_elem_gm}

The mesh velocity portion of $ \mcgh $ can be written as
\begin{equation} \label{a_pull_elem_direct_galerkin_m}
	\mcgm
	\, = \, \mdeltavm^\tT \mk \mrm
	\, - \, \mdeltavmep^\tT \mk \int_{\Omegaep}
			\mNep^\tT_{, \calpha} \bmcdot \big(
				\bma^\calpha \otimes \bmf
			\big) \, \big(
				\bmv
				- \bmvp
			\big)
		\, \Delta t \, \JO ~ \td \Omega
	\ = \ 0
\end{equation}
for any arbitrary variation $ \mdeltavm $.
By reordering global velocity degrees of freedom in the same manner as Eq.\
\eqref{a_pull_elem_dc_ordering}, we find
\begin{equation} \label{a_pull_elem_dc_direct_galerkin_m}
	\mdeltavmnp^\tT \mk \mrmnp
	\, + \, \mdeltavmep^\tT \mk \mrmep
	\, - \, \mdeltavmep^\tT \mk \int_{\Omegaep}
			\mNep^\tT_{, \calpha} \bmcdot \big(
				\bma^\calpha \otimes \bmf
			\big) \, \big(
				\bmv
				- \bmvp
			\big)
		\, \Delta t \, \JO ~ \td \Omega
	\ = \ 0
\end{equation}
for all independent variations $ \mdeltavnp $ and $ \mdeltavep $.
Equation \eqref{a_pull_elem_dc_direct_galerkin_m} requires
$ \mrmnp = \mzero $,
which solves for all mesh velocity degrees of freedom not associated with the
finite element $ \Omegaep $.
In addition, we find
\begin{equation} \label{a_pull_elem_vm_dofs}
	\mrmep
	\ = \mk \int_{\Omegaep}
		\mNep^\tT_{, \calpha} \bmcdot \big(
			\bma^\calpha \otimes \bmf
		\big) \, \big(
			\bmv
			- \bmvp
		\big)
	\, \Delta t \, \JO ~ \td \Omega
	~,
\end{equation}
where $ \mrmep $ is the mesh velocity portion of the residual vector
corresponding to $ \Omegaep $ in the absence of a pull force.
The following discussion explains why Eq.\ \eqref{a_pull_elem_vm_dofs} is not
used in our code.

%
%

\subsection{The numerical implementation} \label{sec_a_pull_elem_numerics}

In our numerical implementation, we calculate the pull force $ \bmmcfp (t) $
directly---rather than with a Lagrange multiplier field.
To do so, we set the $ \nen $ nodal material velocity values
$ \{ \bmvke (t) \} $ associated with $ \Omegaep $ directly according to Eq.\
\eqref{a_pull_elem_velocity_dofs}.
Though these nodes are removed from the degree-of-freedom list, the residual
$ \mrvep $ is still calculated, from which the pull force is determined
according to Eq.\ \eqref{a_pull_elem_force}.
We also directly set the $ \nen $ mesh velocity degrees of freedom
$ \{ \bmvmke (t) \} $ associated with $ \Omegaep $ to be $ \bmvp (t) $ in a
similar fashion.
Since the nodal mesh velocities on $ \Omegaep $ are known, they are removed from
the degree-of-freedom list, and so the integral term in Eq.\
\eqref{a_pull_elem_vm_dofs} is not evaluated in practice.

\section{The pull force at small deformations} \label{sec_a_flat}

Consider a membrane patch which, prior to any deformation, is at the constant
surface tension $ \lambdaz $ and spans the region between two concentric circles
in the $ x $--$ y $ plane.
We denote $ \ell $ as the diameter of the outer circle and $ 2 \mk \rp $ as the
diameter of the inner circle.
Eventually, we will take the limit as
$ \rp \rightarrow 0 $.
When a deformation is applied quasi-statically and the membrane height
$ h = h (r, \theta) $
above the $ x $--$ y $ plane is small
(i.e.\ $ h \ll \ell $),
the membrane shape is known to satisfy \cite[Ch.\ VII]{sahu-thesis}
\begin{equation} \label{a_flat_shape}
	\lambdaz \mk \nabla^2 h
	\, - \, \dfrac{1}{2} \, \kb \mk \nabla^4 h
	\, = \, 0
	~,
\end{equation}
where
\begin{equation} \label{a_flat_laplacian}
	\nabla^2 \pdp
	\, = \, \dfrac{1}{r} \, \pp{}{r} \, \bigg(
		r \, \pp{\pdp}{r}
	\bigg)
	\, + \, \dfrac{1}{r^2} \, \pp{\pdp}{\theta}
\end{equation}
is the 2D Laplacian expressed in terms of the canonical cylindrical coordinates
$ r $ and $ \theta $.
In what follows, all lengths are non-dimensionalized by $ \ell / 2 $, for which
\begin{equation} \label{a_flat_nd}
	\hnd
	\, := \, \dfrac{2 \mk h}{\ell}
	~,
	\qquad
	\rnd
	\, := \, \dfrac{2 \mk r}{\ell}
	~,
	\qquad
	\text{and}
	\qquad
	\nabland^2
	\, := \, \dfrac{\ell^2}{4} \, \nabla^2
	~.
\end{equation}
With Eqs.\ \eqref{eq_sim_pull_fvk} and \eqref{a_flat_nd}, the shape equation
\eqref{a_flat_shape} is presented in dimensionless form as
\begin{equation} \label{a_flat_shape_nd}
	\nabland^2 \hnd
	\, - \, \dfrac{\itGamma}{2} \, \nabland^4 \hnd
	\, = \, 0
	~.
\end{equation}
In comparing Eq.\ \eqref{a_flat_shape_nd} with the description in Ref.\
\cite{powers-pre-2002}, we recognize the small parameter $ \epsilon $ in the
latter is given by $ 2 / \itGamma $ in our notation.
Thus, for an axisymmetric membrane with
($i$) prescribed displacement $ \zp $ at $ r = \rp $,
($i \mkn i$) zero slope at $ r = \rp $,
($i \mkn i \mkn i$) no displacement at $ r = \ell / 2 $, and
($i \mkn v$) no moment at $ r = \ell / 2 $,
we reproduce the Powers \textit{et al.} solution \cite{powers-pre-2002} as%
\footnote{%
	In our numerics, we require zero slope rather than zero moment
	on the outer boundary.
	However, when membrane deformations are small, we do not expect this choice of
	boundary condition to affect the pull force calculation.%
}
\begin{equation} \label{a_flat_h_solution}
	\hnd (\rnd)
	\ = \ \zpnd \cdot \dfrac{\,
		\alpha \mk K_1 (\alpha) \, \ln \rnd
		\, + \, K_0 ( \alpha \mk \rnd \! / \rpnd \,)
	\,}{\,
		\alpha \mk K_1 (\alpha) \, \ln \rpnd
		\, + \, K_0 (\alpha)
	\,}
	~,
\end{equation}
where
\begin{equation} \label{a_flat_alpha}
	\alpha
	\, := \, \rpnd \, \sqrt{\dfrac{\itGamma}{2} \, }
\end{equation}
is a constant parameter defined for notational convenience.
In Eqs.\ \eqref{a_flat_h_solution} and \eqref{a_flat_alpha}, $ K_0 $ and $ K_1 $
are modified Bessel functions of the second kind, and
$ \zpnd := 2 \mk \zp / \ell $
is the dimensionless displacement at the inner membrane boundary---which is
located at
$ \rnd = \rpnd := 2 \mk \rp / \ell $.

\begin{figure}[p]
	\centering
	\includegraphics[width=0.8\textwidth]{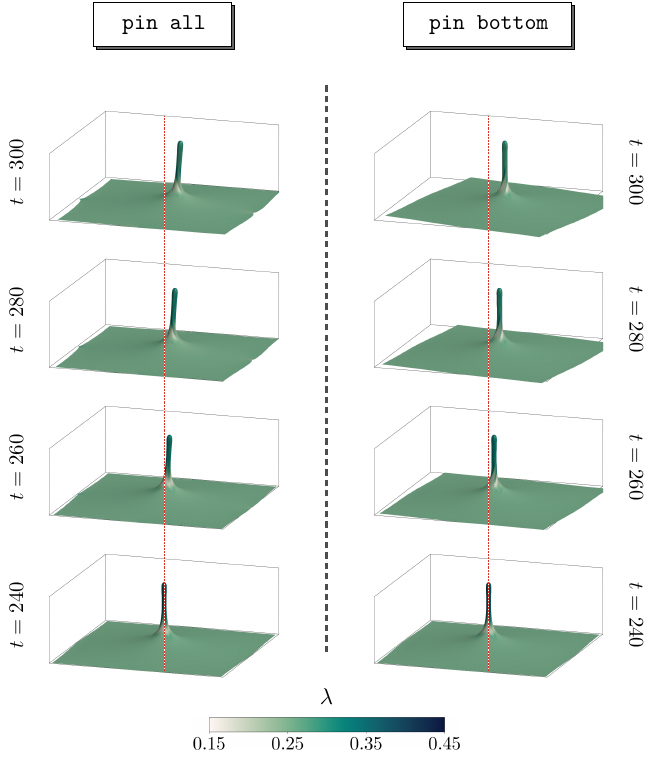}
	\vspace{-1pt}
	\caption{%
		Tether translation in the $ \bme_x $ direction with a Lagrangian mesh
		motion.
		In the left column, simulation boundary conditions are those specified in
		the first two rows of Table \ref{tab_sim_pull_bcs}.
		Twisting and tilting of the tether, along with further striation of the
		surface tension, ensue.
		In the right column, an attempt is made to remove unphysical constraints on
		the membrane by only pinning the center of the bottom boundary.
		Upon lateral pulling, the mesh rotates about the center of the bottom edge;
		twisting and tension striation are once again observed.
		Both simulations fail to capture the expected behavior shown in the left
		column of Fig.\ \ref{fig_sim_pull_lateral}.
		As in the main text, times are measured in units of
		$ \zeta \mk \rc^{\, 2} / \kb $, the pull velocity is specified in the first
		row of Eq.\ \eqref{eq_sim_pull_lat_vp}, and all other parameters are
		identical to those in Fig.\ \ref{fig_sim_pull_schemes}.
		The vertical red lines are a visual aid to highlight the motion of the
		tether.
	}
	\label{fig_a_lag_translate}
\end{figure}

With the solution for the membrane shape at a given displacement in Eq.\
\eqref{a_flat_h_solution}, we seek to determine the magnitude of the pull force
in the $ \bmez $ direction: $ \mcfpz $.
The pull force is related to $ F_z $, the force per length in the $ \bmez $
direction at the inner boundary, via
\begin{equation} \label{a_flat_f_pull}
	\mcfpz
	\, = \, \lim_{\rp \mk \rightarrow \, 0} \ 2 \mk \pi \mk \rp \mk F_z
	~.
\end{equation}
Since the membrane patch is parallel to the $ x $--$ y $ plane at the inner
boundary, we recognize \cite[Ch.V\,\S6(e)]{sahu-thesis}
\begin{equation} \label{a_flat_f_z}
	F_z
	\, = \, \kb \, \pp{H}{r} \bigg\rvert_{\rp}
	\, = \, \dfrac{\kb}{2} \, \bigg(
		\pp{}{r} \big(
			\nabla^2 h
		\big)
	\bigg)\bigg\rvert_{\rp}
	~,
\end{equation}
where the mean curvature
$ H = \tfrac{1}{2} \mk \nabla^2 h $
in the limit of small deformations.
Moreover, since the membrane height is axisymmetric and has zero slope at the
inner boundary, we apply Eq.\ \eqref{a_flat_laplacian} and find
\begin{equation} \label{a_flat_f_pull_nd}
	\mcfpz
	\, = \, \lim_{\rpnd \mk \rightarrow \, 0} \
	\dfrac{2 \mk \pi \mk \kb}{\ell} \, \Bigg(
		\rpnd \, \ddth{\hnd}{(\rnd)} \bigg\rvert_{\rpnd}
		\, + \, \ddt{\hnd}{(\rnd)} \bigg\rvert_{\rpnd}
	\,\Bigg)
	~.
\end{equation}
With the height solution in Eq.\ \eqref{a_flat_h_solution} and properties of
Bessel functions \cite{abramowitz-stegun}, the terms in parenthesis in Eq.\
\eqref{a_flat_f_pull_nd} are found to be
\begin{align}
	\ddt{\hnd}{(\rnd)} \bigg\rvert_{\rpnd}
	&= \ \dfrac{
		\zpnd \, (\itGamma \! / 2 ) \, K_0 (\alpha)
	}{\,
		\alpha \mk K_1 (\alpha) \, \ln \rpnd
		\, + \, K_0 (\alpha)
	\,}
	\label{a_flat_ddt}
	\shortintertext{and}
	\rpnd \, \ddth{\hnd}{(\rnd)} \bigg\rvert_{\rpnd}
	&= \ \dfrac{\,
		- \zpnd \, (\itGamma \! / 2 ) \, \big[
			K_0 (\alpha)
			\, + \, \alpha \mk K_1 (\alpha)
		\,]
	\,}{
		\alpha \mk K_1 (\alpha) \, \ln \rpnd
		\, + \, K_0 (\alpha)
	}
	~.
	\label{a_flat_ddth}
\end{align}
Substituting Eqs.\ \eqref{a_flat_ddt} and \eqref{a_flat_ddth} into Eq.\
\eqref{a_flat_f_pull_nd} yields
\begin{equation} \label{a_flat_f_pull_step_2}
	\mcfpz
	\, = \, \lim_{\rpnd \mk \rightarrow \, 0} \
	\dfrac{\zpnd \mk \pi \mk \kb \mk \itGamma}{\ell} \, \Bigg(
		\dfrac{
			- \alpha \mk K_1 (\alpha)
		}{
			\alpha \mk K_1 (\alpha) \, \ln \rpnd
			\, + \, K_0 (\alpha)
		}
	\,\Bigg)
	~.
\end{equation}
At this point, we evaluate the limit by recognizing
$ \alpha \rightarrow 0 $
as
$ \rpnd \rightarrow 0 $.
In addition, we use the Bessel function relations \cite{abramowitz-stegun}
\begin{equation} \label{a_flat_bessel_limits}
	\lim_{\alpha \, \rightarrow \, 0} \ \alpha \mk K_1 (\alpha)
	\, = \, 1
	\qquad
	\text{and}
	\qquad
	\lim_{\alpha \, \rightarrow \, 0} \ K_0 (\alpha)
	\, = \, - \ln (\alpha)
	~.
\end{equation}
By substituting Eqs.\ \eqref{eq_sim_pull_fvk_radius}, \eqref{a_flat_alpha}, and
\eqref{a_flat_bessel_limits} into Eq.\ \eqref{a_flat_f_pull_step_2} and
rearranging terms, we obtain the small-deformation pull force expression in
Eq.\ \eqref{eq_sim_pull_force_linear}---presented here as
\begin{equation} \label{a_flat_f_pull_final}
	\mcfpz
	\, = \, \dfrac{\pi \mk \kb \mk \zp}{\rc^{\, 2} \mk \ln (\itGamma \! / 2)}
	~.
\end{equation}

%
%

\section{The translation of a tether with a Lagrangian mesh motion}
\label{sec_a_lag_translate}

When a pulled tether is translated laterally across a membrane surface, lipids
in the surrounding patch flow in-plane to accommodate the large out-of-plane
shape deformations.
For a given prescribed lateral tether velocity as in Eq.\
\eqref{eq_sim_pull_lat_vp}, we do not know the corresponding velocity field
$ \bmv (\zetaalpha, t) $ over the membrane patch \textit{a priori.}
Thus, any boundary conditions that pin certain nodes yield unphysical results,
as shown in Fig.\ \ref{fig_a_lag_translate}.
If no nodes are pinned, however, a rigid body translation results---and the
dynamics of lateral tether motion are not obtained.
Lagrangian simulations are thus unable to capture the physics of tether
translation.

\vfill

%
%

\section*{Supplemental movies} 

All supplemental movies can be viewed on the home page of the software package
repository, where they are included in the \texttt{README.md} file:
\href{https://github.com/sahu-lab/MembraneAleFem.jl}{\small\texttt{github.com/sahu-lab/MembraneAleFem.jl}}

\paragraph{Movie 1:}\label{movie_pull_lag}
\texttt{pull-lag.mov}---Lagrangian simulation of tether pulling, corresponding
to the left columns of Figs.\ \ref{fig_sim_pull_schemes} and 
\ref{fig_sim_pull_zoom}.
Available for download at the software repository:
\\
\href{https://github.com/sahu-lab/MembraneAleFem.jl/raw/refs/heads/main/assets/pull-lag.mov}{\small\texttt{github.com/sahu-lab/MembraneAleFem.jl/raw/refs/heads/main/assets/pull-lag.mov}}

\paragraph{Movie 2:}\label{movie_pull_eul}
\texttt{pull-eul.mov}---Lagrangian--Eulerian simulation of tether pulling,
corresponding to the center column of Fig.\ \ref{fig_sim_pull_schemes}.
Available for download at the software repository:
\\
\href{https://github.com/sahu-lab/MembraneAleFem.jl/raw/refs/heads/main/assets/pull-eul.mov}{\small\texttt{github.com/sahu-lab/MembraneAleFem.jl/raw/refs/heads/main/assets/pull-eul.mov}}

\paragraph{Movie 3:}\label{movie_pull_ale}
\texttt{pull-ale.mov}---Arbitrary Lagrangian--Eulerian simulation of tether
pulling, corresponding to the right columns of Figs.\ \ref{fig_sim_pull_schemes}
and \ref{fig_sim_pull_zoom}.
Available for download at the software repository:
\href{https://github.com/sahu-lab/MembraneAleFem.jl/raw/refs/heads/main/assets/pull-ale.mov}{\small\texttt{github.com/sahu-lab/MembraneAleFem.jl/raw/refs/heads/main/assets/pull-ale.mov}}

\paragraph{Movie 4:}\label{movie_translate_ale}
\texttt{translate-ale.mov}---Arbitrary Lagrangian--Eulerian simulation of
lateral translation of a tether, corresponding to Fig.\
\ref{fig_sim_pull_lateral}.
Available for download at the software repository:
\\
\href{https://github.com/sahu-lab/MembraneAleFem.jl/raw/refs/heads/main/assets/translate-ale.mov}{\small\texttt{github.com/sahu-lab/MembraneAleFem.jl/raw/refs/heads/main/assets/translate-ale.mov}}

	\end{appendices}

	%
	%

	\newpage
	\small
	\addcontentsline{toc}{section}{References}
	\bibliographystyle{bibStyle}
	\bibliography{refs}

\end{document}